\documentclass[11pt,leqno]{article}
\usepackage{latexsym}
\usepackage{epsfig}
\usepackage{color}
\usepackage{amssymb,amsmath}

\usepackage[noline,nofillcomment,noend,noresetcount,boxed,figure]{algorithm2e}

\numberwithin{equation}{section}

\newtheorem{theorem}{Theorem}[section]
\newtheorem{lemma}[theorem]{Lemma}
\newtheorem{proposition}[theorem]{Proposition}

\newtheorem{definition}{Definition}[section]
\newtheorem{corollary}[theorem]{Corollary}
\newtheorem{example}{Example}[section]
\def\remark #1{\noindent{\bf Remark:} #1\\}
\long\def\remarks #1{\noindent{\bf Remarks:} #1\\}
\long\def\claim #1 #2{\bigskip\noindent{\bf Claim {#1}} {\it #2}\bigskip}
\def\xclaim #1 #2{\noindent{\bf Claim {#1}} {\it #2}\bigskip}
\newenvironment{proof}{\noindent{\bf Proof:}}{\hfill $\Box $\\}

\newcommand{\bproof}{\noindent{\bf Proof: }}

\newcommand{\ecproof}{\hfill $\diamondsuit$\\}

\renewcommand{\thetheorem}{\arabic{section}.\arabic{theorem}}
\def\sqr#1#2{{\vcenter{\vbox{\hrule height .#2pt
              \hbox{\vrule width .#2pt height#1pt \kern#1pt
              \vrule width .#2pt} \hrule height .#2pt}}}}

\def\ncas #1 {\noindent {\bf Case #1.}\ }

\def\bipart #1 #2{\bigskip \noindent {\bf #1} {\it #2}}
\def\xbipart #1 #2{\noindent {\bf #1} {\it #2}}
\def\iipart #1 #2{\bigskip \noindent {\it #1} {\it #2}}
\def\xiipart #1 #2{\noindent {\it #1} {\it #2}}
\def\brpart #1 #2{\bigskip \noindent {\bf #1} {#2}}
\def\xbrpart #1 #2{\noindent {\bf #1} {#2}}
\def\irpart #1 #2{\bigskip \noindent {\it #1} {#2}}
\def\xirpart #1 #2{\noindent {\it #1} {#2}}

\def\o {\overline}

\def\mod{{\rm{\ mod\;}}}

\def\case #1{\bigskip\noindent{{\bf Case} {\em #1}:}}
\def\subcase #1{\bigskip\noindent{{\bf Subcase} {\em #1}:}}
\def\numcase #1 #2{\bigskip\noindent{{\bf Case #1} {\em #2}:}}

\def\bclaim{\bigskip \noindent{\bf Claim: }}

\def\nclaim #1 {\noindent{\bf Claim #1: }} 

\newcommand{\bcproof}{\smallskip\noindent{\bf Proof: }}

\def\obs #1 {\bigskip\noindent{\bf Observation #1: }} 

\begin{document}

\ifcase 0  

\title{Data Structures for Weighted Matching 
and Extensions to $b$-matching and $f$-factors%
\thanks{This paper is a combination of two conference papers:
A preliminary version of the data structures part 
appeared in 
{\em Proc.~1st Annual ACM-SIAM Symp.~on Disc.~Algorithms}, 
1990 
\cite{G90}. A preliminary version of the extensions part, based on
reduction to matching, appeared in
{\em Proc.~15th Annual ACM Symp.~on Theory of Comp.}, 1983
\cite{G83}.}
\author{%
Harold N.~Gabow
\thanks{Department of Computer Science, University of Colorado at Boulder,
Boulder, Colorado 80309-0430, USA. 
Research supported in part by NSF Grant No. CCR-8815636.
E-mail: {\tt hal@cs.colorado.edu} 
}
}
}

\date{December 18, 2014; revised August 28, 2016}

\maketitle
\def\today{\ifcase\month\or
January\or February\or March\or April\or May\or June\or
July\or August\or September\or October\or November\or December\fi
\ \number\day, \number\year}
\def\date#1.#2.{\ifcase#1\or
January\or February\or March\or April\or May\or June\or
July\or August\or September\or October\or November\or December\fi
\ #2, \number\year}
\def\ydate#1.#2.#3.{\ifcase#1\or
January\or February\or March\or April\or May\or June\or
July\or August\or September\or October\or November\or December\fi
\ #2, 199#3}
\def\nydate#1.#2.{\ifcase#1\or
January\or February\or March\or April\or May\or June\or
July\or August\or September\or October\or November\or December\fi
\ #2}
\def\doublespace{\multiply\baselineskip by3\divide\baselineskip by2%
                 \def\doublespace{}}
\def\bigdoublespace{\multiply\baselineskip by2%
                 \def\bigdoublespace{}}
\def\imp{\ifmmode {\ \Longrightarrow \ }\else{$\ \Longrightarrow \ $}\fi}
\def\rimp{\ifmmode {\ \Longleftarrow \ }\else{$\ \Longleftarrow \ $}\fi}
\def\ximp{\ifmmode {\Longrightarrow\ }\else{$\Longrightarrow\ $}\fi}
\def\xrimp{\ifmmode {\Longleftarrow\ }\else{$\Longleftarrow\ $}\fi}
\def\iff{\ifmmode {\ \Longleftrightarrow \ }\else{$\ \Longleftrightarrow \ $}\fi}
\def\xiff{\ifmmode {\Longleftrightarrow\ }\else{$\Longleftrightarrow\ $}\fi}
\def\tru{\ {\bf true}\ }
\def\fal{\ {\bf false}\ }
\def\wrt{\ {\it wrt}\ }
\def\endskip{\medskip}
\def\qed{$\Box$}
\def\qedn{\ \vrule width4pt depth-1pt height7pt }
\def\rqed{\hfill\hbox to 24 pt{\vrule width4pt depth-1pt
height7pt\hfil}\bigskip}
\def\rqedn{\hfill\hbox to 24 pt{\vrule width4pt depth-1pt height7pt\hfil}}
\def\log{\ifmmode \,{ \rm log}\,\else{\it log }\fi}
\def\con {\subseteq}
\def\pcon{\subset}
\def\firstnumstp#1 {\bigskip \noindent{\it Step} #1.\newquad}
\def\numstp#1 {\endskip\noindent{\it Step} #1.\newquad}
\def\newquad{\hskip1ex}
\def\stp#1.{\endskip
\noindent{\it #1 Step.}\newquad}
\def\firststp#1.{\bigskip
\penalty-1000
\noindent{\it #1 Step.}\newquad}
\def\cas#1 {\smallskip\noindent{\bf Case} #1.\ } 
%
%
%
\long\def\sec#1{\bigskip
\penalty-2000%
\noindent{\twelvebf #1}\par\ignorespaces\noindent\ignorespaces}
\def\aorbsec#1{\noindent{\twelvebf #1}}
\def\nsec#1{\penalty-2000%
\noindent{\bf #1\hfill\break}
\hbox to \parindent{\hfill}\ignorespaces}
\long\def\res #1. #2{\bigskip
\penalty-1000
\noindent {\bf #1.}\newquad%
#2 \bigskip}
\long\def\nres #1. #2{\bigskip
\noindent {\bf #1.}\newquad%
#2}
\def\pf{\noindent {\bf Proof.}\newquad}
\def\cont{\ifmmode\star\else$\star$\fi}
\def\+{\tabalign} 
\def\nskp{\def\bigskip{}}
\def\i{($i$) } \def\xi{($i$)}
\def\ii{($ii$) } \def\xii{($ii$)}
\def\iii{($iii$) } \def\xiii{($iii$)}
\def\iv{($iv$) } \def\xiv{($iv$)}
\def\pa{({\it a}) } \def\xpa{({\it a})} 
\def\pb{({\it b}) } \def\xpb{({\it b})}
\def\pc{({\it c}) } \def\xpc{({\it c})}
\def\hi{\hskip20pt\i} \def\hii{\hskip20pt\ii} \def\hiii{\hskip20pt\iii}
\def\ha{\hskip20pt\pa} \def\hb{\hskip20pt\pb} \def\hc{\hskip20pt\pc}
\def\tran{{\buildrel*\over\to}}
\def\n{\rlap{$\>/$}}
\def\({{\rm(}} \def\){{\rm)}}
\def\c#1{\lceil {#1} \rceil}
\def\f#1{\lfloor {#1} \rfloor}
\long\def\boxit#1{\vtop{\hrule
\hbox{\vrule\quad\vtop{\vskip5pt\hbox{#1}\vskip5pt}\quad\vrule}
\hrule}} 
\def\iboxit#1{\vtop{\hrule
\hbox{\vrule\quad\vtop{\vskip5pt\hbox{{\it #1}}\vskip5pt}\quad\vrule}
\hrule}} 
\def\x{\iffalse}
\def\b{\bigskip}
\def\set #1#2{\{ #1:#2 \}}
\def\pset #1#2{( #1:#2 )}
\def\h{\hskip20pt}
\def\hi{\advance\parindent by 20pt}

\def\o{\overline} 
\def\u{\underline}
\def\opn{\hangindent=40pt\hangafter=1}
\def\h{{\hskip 20pt}}
\def\v{\vfill}
\def\hi{\advance \parindent by 20pt}
\def\d{\cdot}
\def \il #1{\log^{(#1)} }
\def\al.{{\it add\_leaf}}
\def\alm{{\it add\_leaf}$\,$}
\def\O{o\hbox{-}smallest}
\def\os.{\ifmmode{ \o{\cal S} }\else{$\o {\cal S}$}\fi}
\def\oP.{\ifmmode{ \o{\cal P} }\else{$\o {\cal P}$}\fi}
\def\ot.{\mathy{ \o{\cal T} }}
\def\oG{\o G}
\def\oB{\o B}
\def\oE.{\mathy{\overline E}}
\def\p(#1,#2){\ifmmode p(#1,#2) \else{$p(#1,#2)$}\fi}
\def\op(#1,#2){\ifmmode \o{p}(#1,#2) \else{$\o{p}(#1,#2)$}\fi}
\def\lb{\ifmmode \,{ \rm log}_\beta \else{\it log XX }\fi}
\def\wh{\widehat}
\def\wx.{\ifmmode \wh x \else$\wh x$\fi}
\def\wy.{\ifmmode \wh y \else$\wh y$\fi}
\def\wz.{\ifmmode \wh z \else$\wh z$\fi}
\def\wv.{\ifmmode \wh v \else$\wh v$\fi}
\def\Px.{\ifmmode \wh x \else$\wh x$\fi}
\def\Py.{\ifmmode \wh y \else$\wh y$\fi}
\def\Pz.{\ifmmode \wh z \else$\wh z$\fi}
\def\Pv.{\ifmmode \wh v \else$\wh v$\fi}
\def\Pr.{\ifmmode \wh r \else$\wh r$\fi}
\def\Pr.{\ifmmode \wh r \else$\wh r$\fi}
\def\A.{\mathy{{\cal A}}}
\def\B.{\mathy{{\cal B}}}
\def\E.{\ifmmode {{\cal E}}\else{{$\cal E$}}\fi}
\def\F.{\mathy{\cal F}}
\def\H.{\mathy{\cal H}}
\def\M.{\mathy{\cal M}}
\def\P.{\mathy{\cal P}}
\def\S.{\ifmmode {{\cal S}}\else{{$\cal S$}}\fi}
\def\mathy #1{\ifmmode {#1}\else{$#1$}\fi}
\def\pkt(#1,#2){\mathy{packet(#1,#2)}}

\def\rt{\mathy{\rho}}
\def\pt{\pi}
\long\def\example #1. #2{\bigskip \noindent{\bf Example #1.} 
{#2}\bigskip} 
\long\def\xexample #1 {\bigskip \noindent{\bf Example.} 
{#1}\bigskip}
\def\goin{\hspace{17pt}}

\begin{abstract}
This paper shows the weighted matching problem on general graphs
can be solved in time $O(n(m + n\log n))$ for $n$ and $m$ the
number of vertices and edges, respectively. This was previously known
only for bipartite graphs. 
The crux is a data structure for blossom creation. It uses 
a dynamic nearest-common-ancestor algorithm
to simplify blossom steps, so they
involve only back edges rather than arbitrary nontree edges.

The rest of the paper presents direct extensions of Edmonds'
blossom algorithm to weighted $b$-matching and $f$-factors. Again
the  time bound 
is the one previously
known for
bipartite graphs: for $b$-matching the time is $O(\min\{b(V),n\log
n\}(m + n\log n))$ 
and for $f$-factors the time is $O(\min\{f(V),m\log n\}( m + n\log n) )$,
where $b(V)$ and $f(V)$ denote the sum of all degree constraints.  
Several immediate applications of the $f$-factor algorithm 
are given:
The generalized shortest path structure of \cite{GS13}, i.e., the analog 
of the shortest path tree for
conservative undirected graphs, is shown to be a version of 
the blossom structure for $f$-factors. 
This structure
is found in time
$O(|N|(m+n\log n))$ for $N$ the set of negative edges ($0<|N|<n$).
A shortest $T$-join is found in time $O(n(m+n\log n))$, or
$O(|T|(m+n\log n))$ when all costs are  nonnegative. These bounds are all
slight improvements of previously known ones, and are simply achieved
by proper initialization of the $f$-factor algorithm.
\end{abstract}
\fi 

\def\switch{0}
\ifcase \switch 
\iftrue
\section {Introduction} 
\label{IntroSec}

This paper solves a well-known problem in data structures 
to achieve 
an efficient algorithm for weighted matching. It also extends the results
to the most general weighted matching problems.
This section defines the problems and states the results.

A {\it matching}  on a graph is a set of vertex-disjoint edges.
A matching is {\em perfect} if it covers every vertex. More generally it
is {\em maximum cardinality} if it has the greatest possible number of edges,
and {\em cardinality $k$} if it has exactly $k$ edges.
Let each edge $e$ have a real-valued
{\it weight} $w(e)$. 
%
%
The weight $w(S)$ of a set of edges $S$
is the sum of the individual edge weights.
Each of the above variants has a maximum weight version, e.g.,
a {\em maximum weight matching} has the greatest possible weight,
a {\em maximum weight perfect matching} has maximum weight
subject to the contraint that it is {\em perfect}, etc.
Alternatively edges may have real-valued costs,
and we define {\em minimum cost matching}, etc.
The {\em weighted matching problem} is to find a matching of
one of these types,
e.g., find a maximum weight perfect matching on a given graph, etc.
All these variants are essentially equivalent from an algorithmic viewpoint.
For definiteness this paper concentrates on maximum weight perfect matching.

In stating resource bounds for graph algorithms
we assume throughout this paper that the given graph 
has $n$ vertices and $m$ edges. For notational simplicity  we assume
$m\ge n/2$. In the weighted matching problem this can always be achieved
by discarding isolated vertices.

Weighted matching is a classic problem 
in network optimization; detailed discussions are in \cite{L, LP, PS, S, CCPS}.
Edmonds gave the first polynomial-time algorithm for weighted matching \cite{E}.
Several implementations of Edmonds' algorithm have 
been given  with increasingly fast running times: 
$O(n^3)$ \cite{G73, L}, $O(mn\log n)$  \cite{BD, GMG},
$O(n ( m \log  \log  \log_{2+m/n} n  +  n \log n))$ 
\cite{GGS}.
Edmonds' algorithm is a generalization of the Hungarian algorithm, due to
Kuhn, for  weighted matching on bipartite graphs \cite{K55, K56}.
Fredman and Tarjan implement the Hungarian algorithm in 
$O(n ( m + n\log n) )$ time using Fibonacci heaps \cite{FT}.
They ask if general matching can be done in this time.
We answer affirmatively: We show that a
search in Edmonds' algorithm can be implemented in time $O(m+n\log n)$.
This implies that the weighted matching problem can be solved
in time $O(n ( m +  n \log n) )$. The space is $O(m)$.
Our implementation of a search is in some sense optimal: 
As shown by Fredman and Tarjan \cite{FT} for Dijkstra's algorithm, 
one search of 
Edmonds' algorithm can be used to sort $n$ numbers. Thus a
 search requires time
$\Omega(m+n\log n)$ in an appropriate model of computation.

Weighted matching algorithms based on cost-scaling
have a better asymptotic time bound
when costs are small integers \cite{GT89}.
However our result remains of interest for at least
two reasons: First, Edmonds' algorithm is theoretically attractive because
its time bound is strongly polynomial.
Second,  for a number of matching and related problems, the 
best known solution
amounts to performing one search of Edmonds' algorithm, e.g.,
most forms of sensitivity analysis for weighted matching
\cite{BD, CM, G85b, W}.
Thus our implementation of a search  in time $O(m+n\log n)$ gives
the best-known algorithm for these problems.

The paper continues by presenting versions of Edmonds' blossom
algorithm for weighted $b$-matching and weighted $f$-factors.  These
problems generalize ordinary matching 
to larger degree-constrained subgraphs
and are defined as follows.  For
an undirected multigraph $G=(V,E)$ with function $f:V\to \mathbb Z_+$,
an {\em $f$-factor} is a subgraph where each vertex $v\in V$ has
degree exactly $f(v)$.  For an undirected graph $G=(V,E)$ where $E$
may contain loops, with function $b:V\to \mathbb Z_+$, a (perfect)
{\em $b$-matching} is a function $x:E \to \mathbb{Z}_+$ where each
vertex $v\in V$ has $\sum_{w: vw\in E} x(vw) = b(v)$.  Given in
addition a weight function $w:E\to \mathbb R$, a {\em maximum
  $b$-matching} is a (perfect) $b$-matching with the greatest weight
possible; similarly for {\em maximum $f$-factor}.  We find  maximum
$b$-matchings and $f$-factors in
the same time bound as  was known for bipartite graphs:
for $b$-matching the time is
 $O(\min\{b(V),n\log n\} ( m + n\log n) )$ 
where $b(V)$ is the sum of all degree constraints;
for $f$-factors the time is
 $O(\min\{f(V),m\log n\} ( m + n\log n) )$ 
where $f(V)$ is the sum of all degree constraints.
A blossom algorithm for $b$-matching is given in Pulleyblank's thesis
\cite{Pu} (\cite{Pu12} gives a very high level description, different from our algorithm).
The pseudo-polynomial parts of the above bounds 
(i.e., the bounds using $b(V)$ and $f(V)$) 
can also be achieved using the current paper's 
algorithm for ordinary matching plus
the reduction to matching presented in the original version of the current
 paper \cite{G83}.

Here we prefer  direct implementations of the general matching algorithms,
 to avoid practical inefficiencies
and to illuminate the properties of blossoms. As an example of the latter, 
the algorithm's blossom structure is shown to be 
exactly the generalized shortest path structure of Gabow and Sankowski \cite{GS13}, 
i.e., the analog of the shortest path tree
for conservative undirected graphs.
(The paths in blossoms
that are used to augment the matching give the shortest paths to
the fixed source in a conservative undirected graph.)
Our discussion of blossoms 
also leads to (and requires) 
simple proofs of (previously known) min-max formulas
for the maximum size of a $b$-matching, equation
(\ref{MaxBMatchEqn}), or partial $f$-factor, (\ref{fFactorSizeEqn}).
Lastly our algorithm shows that  $b$-matching blossoms
have the same structure as ordinary  
matching blossoms (unlike $f$-factors there are no ``$I(B)$-sets'', i.e.,
pendant edges, and no ''heavy blossoms'', only ``light'' ones, at least
 in the final output).

We find the generalized shortest path structure 
in time $O(|N|(m+n\log n))$
for $N$ the set of negative edges
($0<|N|<n$ for conservative costs)
and a shortest $T$-join in time
$O(n(m+n\log n))$, or
$O(|T|(m+n\log n))$
for nonnegative costs.
These bounds are slight improvements
of previously known ones and  are achieved simply by
proper initialization of the $f$-factor algorithm.
(The  strong polynomial bound
of Gabow and Sankowski \cite[Section 10] {GS13} can be modified to achieve
the same  time as ours for the matching part, plus
additional time
$O(m\log n)$ for  post-processing.)
Good implementations of the $T$-join algorithm of Edmonds
use time $O(n^3)$ for general costs and the same time as ours plus
$O(|T|^3)$ for nonnegative costs, both cubic terms coming from
finding a minimum cost matching on a complete graph \cite[p.486 and p.488]{S}.)

The paper is organized as follows.  
This section concludes with some terminology and assumptions.
Section \ref{ReviewSec} reviews Edmonds'
algorithm and defines the ``blossom-merging problem'' -- the last
ingredient needed to obtain the time bound we seek.  Section \ref{TSec}
specializes this problem to ``tree-blossom-merging'' and solves it.
Section \ref{bMatchingSec} gives our
$b$-matching algorithm and Section \ref{fFactorSec} gives the $f$-factor algorithm.
Appendix A gives further details of Edmonds' matching algorithm.
Appendix B gives some
further details for $b$-matching and $f$-factors.  Appendix C gives an
efficient implementation of
the grow and expand steps of Edmonds' algorithm.
Gabow \cite{G85b} gives a faster algorithm, but 
Appendix C is simpler
and suffices to achieve our overall time bound.
Appendix C also gives the details for grow and expand steps 
for $b$-matching and $f$-factors. 
The latter is more involved but follows the same outline.

Our algorithm for tree-blossom-merging requires an algorithm that computes
nearest common ancestors in trees that grow by addition of new leaves.
The conference version 
of this paper \cite{G90} 
presented the required algorithm, as well as extensions.
For reasons of succinctness 
the data structure for nearest common ancestors is now given
separately in \cite{G17}.

Portions of this paper may be read independently.
Our implementation of Edmonds' matching algorithm is in Section \ref{TSec};
readers familiar with Edmonds'  algorithm can skip the review in
and go directly to Section \ref{BMergingSec}. 
Those interested in generalized versions of matching should concentrate on 
$f$-factors, our most general algorithm (Section 5), 
although some basic lemmas are proved in Section 4.

\paragraph*{History of this paper}
The conference paper \cite{G90}
presented a preliminary version of
the tree-blossom-merging algorithm. 
The current paper 
simplifies that algorithm, e.g., there is no need
to refine the strategy for sparse graphs ($m=o(n\log^2 n$).
The tree-blossom-merging algorithm
uses an algorithm that computes
nearest common ancestors in trees that grow by addition of new leaves.
\cite{G90}
presented the required algorithm, as well as extensions.
As mentioned this is now given in 
\cite{G17}. Subsequent to \cite{G90} Cole and Hariharan \cite{CH}
used a similar approach to allow other operations; they also achieve
time bounds
that are worst-case rather than amortized.

The results on $b$-matching and $f$-factors 
evolve from the conference paper \cite{G83}. That paper
achieved similar time bounds to those presented here by reducing the problems to
matching. The current paper gives direct approaches to the problems,
thus illuminating the structure of blossoms (see Sections 
\ref{bBlossomSec} and \ref{fBlossomSec}) and
avoiding the blow-up in problem size incurred
by reduction.

\paragraph*{Terminology}
We often omit set braces from singleton sets, denoting $\{v\}$ as
$v$.
We use interval notation for sets of integers: for $i,j\in \mathbb{Z}$, 
$[i..j]=\set {k^{\in \mathbb{Z}}} {i\le k\le j}$. 
We use a common summing notation: If $x$ is a function on elements
and $S$ a set of elements then $x(S)$ denotes $\sum_{s\in S}x(s)$.
$\log n$ denotes logarithm to the base two.
Assume that for a given integer $s\in [1..n]$ the value
$\f{\log s}$ can be computed in $O(1)$ time.
This can be done if we precompute these $n$
values and store them in a table. The precomputation time is $O(n)$.

For a graph $G$, $V(G)$ denotes its vertices and $E(G)$ its edges.
For vertices $x,y$ an {\em $xy$-path} has ends $x$ and $y$.
For a set of vertices $S\subseteq V$ and a subgraph $H$ of $G$,
$\delta(S,H)$ ($\gamma(S,H)$) denotes the set of edges with exactly
one (respectively two) endpoint in $S$. (Loops are in $\gamma$ but not
$\delta$.)
$d(v,H)$ denotes the degree of vertex $v$ in $H$.  
When
referring to the given graph $G$ we often omit the last argument and write, e.g.,
$\delta(S)$. 
(For example a vertex $v$ has $d(v)=|\delta(v)|+2|\gamma(v)|$.)

Fix a matching $M$ on the given graph.
A vertex is {\it free} if it is not on any matched edge.
An {\it alternating  path} is a vertex-simple path 
whose edges are alternately matched and unmatched.
(Paths of 0 or 1 edges are considered alternating.)
An {\it augmenting path  P} is an alternating path joining two distinct free
vertices.
To {\it augment the matching along $P$} means to enlarge the matching $M$ 
to $ M \oplus P$ (the symmetric difference of $M$ and $P$).
This gives a matching with one more edge.
\section{Edmonds' algorithm and its implementation}
\label{ReviewSec} 

\def\fm.{$min\_edge$}
\def\fms.{$min\_edges$}

This section summarizes Edmonds' algorithm  and 
known results on its implementation.
Sections \ref{BlossomSec}--\ref{EdAlgSec}
sketch the high level algorithm.
They include all 
the details needed for our implementation but
do not give a complete development.
For the latter see, e.g., 
\cite{E, G73, L, PS}.
Section \ref{BMergingSec} 
reviews the portions of the algorithm
for which efficient implementations are known, and 
the outstanding problem of efficient ``blossom merging''.

\begin{figure}[th]
\centering
\input{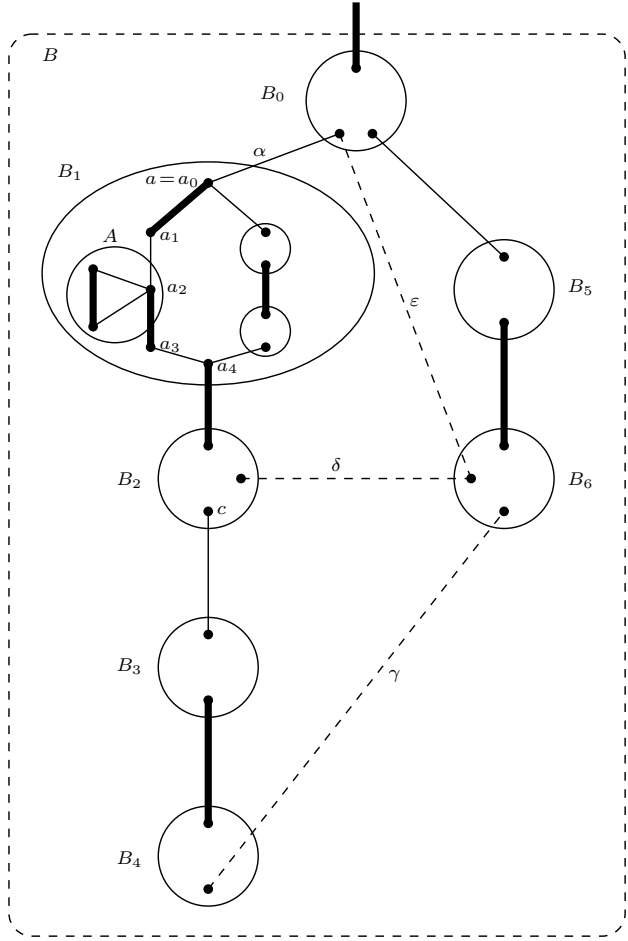}
 \caption{Blossoms in a search of Edmonds' algorithm.
Blossom $B$ is formed from subblossoms $B_0,\ldots,B_6$.
Heavy edges are matched.}
 \label{1Fig}
 \end{figure}

\subsection{Blossoms}
\label{BlossomSec} 
Throughout this section the notation $P(x,y)$ denotes an $xy$-path.
This includes the possibility that $x=y$, i.e., $P(x,x)=(x)$.

Edmonds' algorithm is based on the notion of blossom.
We start with a data-structure-oriented definition,
illustrated in  Fig.~\ref{1Fig}.
Begin by considering two even-length alternating paths $P(x_i,y)$,
 $i=0,1$,
with $x_0\ne x_1$ and $y$ the only common vertex. 
Each path begins  with the matched edge at $x_i$, unless $x_i=y$. 
These paths plus an edge $x_0x_1$ form a simple example of a blossom.
Edmonds' algorithm contracts blossoms. This leads to the general, recursive
 definition:

\begin{definition}
\label{BlossomDefn}
Let $G$ be a graph with a distinguished matching.
A {\em blossom} is a subgraph defined by rules (a) and (b):
 
(a) Any single vertex $b$ is a blossom. 

(b) Let $\oG$ be a graph formed from $G$ by contracting zero or more
vertex-disjoint blossoms.
Let $X_0,X_1,Y$ be $\oG$-vertices, $X_0\ne X_1$.
For
$i=0,1$ let
 $P(X_i,Y)$ be an even-length alternating path that starts with a 
matched
edge or has $X_i=Y$, 
with $Y$ the only common 
$\oG$-vertex.
These paths plus an edge $X_0X_1$ form a blossom. 
\end{definition}

In Fig.~\ref{1Fig} blossom $B$ is formed from
paths $P(B_4,B_0)$, $P(B_6,B_0)$ and edge $\gamma$.
We use the term ``vertex'' to refer to a vertex of the given graph $G$.
For a blossom $B$, $V(B)$ is
the set of vertices of $G$ contained in
any blossom in either path $P(X_i,Y)$;
we sometimes refer to them as {\em the vertices of $B$}.
The maximal blossoms in the paths $P(X_i,Y)$
are {\em the subblossoms of $B$}.

We use some properties of blossoms 
that are easily established by induction.
Any blossom has a {\em base vertex}: In 
Definition \ref{BlossomDefn}
a blossom of type (a) has base vertex $b$.
A blossom of type (b) has the same base vertex as $Y$.
We usually denote the base of $B$ as $\beta(B)$
(or $\beta$ if the blossom is clear). $\beta(B)$
is the unique vertex of $B$ that is not
matched to another vertex of $B$.
$\beta(B)$ is either free or matched to another vertex
not in $B$.
We call $B$ a {\em free blossom} or {\em matched blossom}
accordingly.

Consider a blossom $B$ with base vertex $\beta$.
Any vertex $x\in V(B)$ has an even-length alternating path
$P(x,\beta)$ that starts with the matched edge at $x$.
($P(\beta,\beta)$ has no edges.)
For example in Fig.\ref{1Fig} in blossom $B$, $P(a,\beta(B))$
starts with $P(a,\beta(B_1)) =(a_0,a_1,a_2,a_3,a_4)$
followed by edge $(\beta(B_1),\beta(B_2))$ and the reverse of 
$P(c,\beta(B_2))$,
and continuing along paths in $B_3,B_4,B_6,B_5,B_0$.

To define  $P(x,\beta)$
in general
consider the two paths
for $B$ in Definition \ref{BlossomDefn}.
Among the blossoms in these  paths
let $x$ belong to a blossom designated as $B_0$.
The edges of $B$ (i.e., the edges of $P(X_i,Y)$
plus $X_0X_1$) contain a unique even-length alternating path $A$
from $B_0$ to $Y$. Here $A$ is a path in $\o G$. 

First suppose
$B_0\ne Y$. $A$
starts with the matched edge at $B_0$.
$P(x,\beta)$ passes through the same blossoms as $A$.
To be precise let the $\o G$-vertices of
$A$ be $B_i, i=0,\ldots,k$, $B_k=Y$, with $k>0$ even.
Let $\beta_i$ be the base vertex of $B_i$.
So there are vertices $x_i\in V(B_i)$ such that 
the edges of $A$ are 
\[
\beta_0\beta_1,  x_1x_2 ,   \beta_2\beta_3 ,  x_3x_4 ,\ldots,  x_{k-1}x_k .  
\]
Here the $\beta_i\beta_{i+1}$ edges are matched 
and the $x_ix_{i+1}$ edges are unmatched.
Recursively define  $P(x,\beta)$ as the concatenation of $k+1$ subpaths 
\begin{equation}
\label{PvbEqn}
P(x,\beta)= P(x,\beta_0), P(\beta_1,x_1),P(x_2,\beta_2),P(\beta_3,x_3),\ldots,
P(x_k,\beta_k).
\end{equation}
For odd $i$, $P(\beta_i,x_i)$ is the reverse of path $P(x_i,\beta_i)$.

Now consider the base case
$B_0=Y$. If $Y$ is a vertex then $Y=x=\beta$
and $P(x,\beta)=(x)$. Otherwise $P(x,\beta)$ in blossom $B$ is identical
to $P(x,\beta)$ in $B_0$.%
\footnote{The $P(x,\beta)$ paths may intersect in nontrivial ways.
For instance in Fig.~\ref{1Fig}, $P(\beta(B_3),\beta(B))$ and $P(\beta(B_5),\beta(B))$
traverse edge $\gamma$ in opposite directions. 
So the paths have common subpaths, e.g., the subpath
of $P(\beta(B_5),\beta(B))$ joining $\gamma$ and edge $(\beta(B_3),\beta(B_4))$,
and disjoint subpaths, e.g., the subpath of $P(\beta(B_5),\beta(B))$
joining $\beta(B_3)$ and edge $\alpha$. This intersection pattern can continue
inside blossom $B_0$.
 So in general for two
vertices $x_0,x_1$ in a blossom $B$ with base $\beta$, the paths
$P(x_i,\beta)$ can have arbitrarily many subpaths that are alternately
common and disjoint.}

Edmonds' algorithm finds an augmenting path 
$\o P$ in the graph $\oG$ that has
every blossom contracted. $\o P$ corresponds to
an augmenting path $P$ in the given graph $G$:
For any contracted blossom $B$ on an unmatched edge $xy$ ($x\in V(B)$) 
of  $\o P$,
$P$ traverses the path $P(x,\beta(B))$. If we augment the matching of $G$
along $P$,
every blossom becomes a blossom in the new matched graph:
For instance the above vertex $x$ becomes the new base of $B$. In Fig.\ref{1Fig}
if $\o P$ contains an unmatched edge $\alpha'$ that enters
$B$ at vertex $a$,
$P$ contains the subpath $P(a,\beta(B))$. The augment makes $a$ the base of $B$
as well as $B_1$;
in the contracted graph $\alpha'$ is the matched edge incident to $B$.

\subsection{Edmonds' weighted matching algorithm}
\label{EdAlgSec} 
For definiteness consider the problem of finding a maximum weight 
perfect matching.
The algorithm is easily modified for all the other variants of weighted matching. Without loss of generality assume a perfect matching exists.

The algorithm
is a primal-dual strategy
based on Edmonds' formulation of weighted matching as a linear program.
It repeatedly finds a maximum weight
augmenting path and augments the
matching. The procedure to find one augmenting path is  a {\it
search}. If the search is successful, i.e., it finds an augmenting path
$P$, then an {\it augment step} is done. It augments the matching along
$P$. 
The entire algorithm consists of $n/2$ searches and augment steps.
At any point  in the algorithm
$V(G)$ is  partitioned into blossoms.
Initially every vertex is a
singleton blossom.

The following pseudocode gives a precise specification of the
search for an augmenting path; a more detailed discussion with examples follows.
Assume the graph has a perfect matching so this path exists.
For any vertex $v$, $B_v$ denotes the maximal blossom containing $v$.

\begin{algorithm}
\DontPrintSemicolon

make every free vertex or free blossom the (outer) root of an \os.-tree\;

{\bf loop}

\Indp

\If{$\exists$ tight edge $e=xy$, $x$ outer and $y\notin \S.$}
{\tcc{grow step}
let $\beta$ be the base of $B_y$, with $\beta\beta'\in M$\;
add $xy, B_y, \beta\beta', B_{\beta'}$ to \S.\;
}

\ElseIf{$\exists$ tight edge $e=xy$, $x,y$ outer in the same search tree, $B_x\ne B_y$}
{\tcc{blossom step}
merge all blossoms in the fundamental cycle of $e$ in \os.\;
}

\ElseIf{$\exists$ tight edge $e=xy$, $x,y$ outer in different search trees}
{\tcc{augment step}
\tcc{$xy$ plus the \os.-paths to $x$ and  $y$ form
an augmenting path $P$}
augment the matching along $P$ and end the search\;
}

\ElseIf{$\exists$ a nonsingleton inner blossom $B$ with $z(B)=0$}
{\tcc{expand step}


let \os. contain edges $xy$ and $\beta\beta'$ incident to $B$ where
$\beta$ is the base of $B$, $x\in V(B)$

let $B$ have subblossoms $B_i$

in \os. replace $B$ by the even length alternating path of subblossoms
$B_0,\ldots, B_k$ 

\goin that has $x\in B_0$, $\beta\in B_k$

\tcc{the remaining subblossoms of $B$ are no longer in \S.}

}

\lElse {adjust duals}

\bigskip

\caption{Pseudocode for a search in Edmonds' algorithm.}
\label{MAlg}
\end{algorithm}

A search constructs a subgraph $\cal S$. $\cal S$ is initialized
to contain every free blossom.
It is enlarged by executing three types of steps, called {\it grow},
{\it blossom},  and {\it expand steps} in 
\cite{G85b}. In addition the search changes the linear programming
dual variables in {\it dual adjustment steps}.
After a dual adjustment step, one or more of the other steps can be performed.
Steps are repeated until $\cal S$ contains the desired augmenting path.

$\cal S$ consists of blossoms forming a forest.
More precisely
if each blossom is contracted to a vertex, $\cal S$ becomes a
forest \os. whose roots are the free blossoms.
A blossom of $\cal S$ that is an even (odd)
distance from a root of \os. is {\it outer} ({\it inner}).
(The third possibility is a blossom not in $\cal S$.)
A vertex of $\cal S$
is outer or inner depending on the blossom that contains it. 
Any path from a root to a node in its tree of \os. is alternating.
So the matched edge incident to a nonroot outer blossom 
goes to its parent; the matched edge incident to an inner
blossom goes to its unique child.

In Fig.~1 suppose $B_0,\ldots,B_6$ are maximal blossoms,
edges $\gamma,\delta,\varepsilon$ are not part of \S.,
and blossom $B$ 
has not yet formed.
If $B_0$ is outer, the outer blossoms are the $B_i$ with $i$ even.
If a blossom step forms blossom $B$, the $B_i$ with $i$ odd change from
inner to outer.
Now we discuss the three steps that build up $\cal S$
(see also Fig. \ref{MAlg}).

\b

\noindent
{\bf Grow steps} 
A grow step enlarges \S..
It is executed for an edge $vx$ where vertex $v$ is outer and $x$ is
not in \S.. Let $x$ be in the maximal blossom $B$ with base vertex $\beta$.
$B$ is a matched blossom (by the initialization of \S.).
Let  $\beta\beta'$ be the corresponding matched edge.
$\beta'$ is the base vertex of a maximal blossom $B'$.
The grow step adds edge $vx$, 
$B$, edge $\beta\beta'$ and $B'$ to \S..
$B$ and
 $B'$ 
are new inner and
outer blossoms respectively.

In Fig.~1 if $\cal S$ contains outer blossom $B_0$ but no other $B_i$,
a grow step for edge $\alpha$ adds 
$\alpha, B_1$,  $(\beta(B_1),\beta(B_2))$ and $B_2$ to $\cal S$.
Two more grow steps add the
other $B_i$, $i=3,\ldots,6$.

\b

\noindent
{\bf Blossom steps} 
A blossom step is executed for an edge $e$ that joins two distinct
outer blossoms in the same tree of \os..
It combines all blossoms along the fundamental cycle
of $e$  to form a new
blossom $B$. Note that $B$ is outer: In proof let $A$ be the blossom
closest to the root in $e$'s fundamental cycle. $B$ is outer if $A$ is.
If $e$ is incident to $A$ then 
$A$ is outer, by definition. If $e$ is not incident to $A$ then
the ends of $e$ descend from 2 distinct children of $A$.
$A$ must be outer, since as previously noted any inner blossom has only
one child. 

In Fig.~1 suppose $\cal S$ contains blossoms $B_i$, $i=0,\ldots,6$,
with $B_i$ outer for $i$ even.
A blssom step for $\gamma$ would form blossom $B$.
Alternatively the search
might do a blossom step for $\varepsilon$, then later one for $\delta$,
and still later one for $\gamma$. Other sequences are possible, and the one
executed 
depends on the costs of these edges.

\b

\noindent
{\bf Expand steps} 
An expand step replaces an inner blossom $B$ of $\cal S$ by some of its
subblossoms. Specifically let
$B$, with
base vertex $\beta$, be  incident to
edges $xy$ and 
$\beta\beta'$
in \os., with $x\in V(B)$.
Let $x$ ($\beta$) be in the maximal subblossom
$B_0$ ($B_r$) of $B$, respectively, and let $P(B_0,B_r)$ be
the even-length alternating path formed from  edges of $B$.
Then $B$ is replaced by  $P(B_0,B_r)$ in \os.. The remaining
subblossoms of $B$ are no longer in \os. -- they are now eligible
to enter \os. in grow steps.
\os. remains a forest.
 
In Fig.~1 when $\cal S$ contains all blossoms $B_i$,
an expand step for $B_1$ replaces it by
vertices $a_0,a_1,$ blossom $A$, $a_3$, and  $a_4$.
The other two subblossoms of $B_1$ leave \os..

\b

This completes the description of the three steps that construct \S..
Note that once a vertex becomes outer it remains 
in $\cal S$ and outer for the rest of the search.
In contrast vertices
can alternate between being inner  and 
not being in $\cal S$ (perhaps ultimately becoming outer in a grow
or blossom step).
This alternation can occur $\Theta(n)$ times for a given vertex $v$
in one search
(the upper bound  $n$  holds since each time $v$ is
involved in an expand step, the size of the maximal blossom containing $v$
decreases). 

The sequence of steps executed by the search depends on
the costs of edges and the values of the dual variables. The algorithm
executes a given step when a corresponding edge becomes {\em tight},
i.e., its dual variables satisfy
the complementary slackness conditions of linear
programming. For example in Fig.~1 the blossom step for $\gamma$ is done
when $\gamma$ becomes tight. The dual adjustment step modifies dual
variables so that additional edges become tight and corresponding grow,
blossom or expand steps can be done.
The dual adjustment step involves finding a candidate edge 
that is closest to being tight, and then changing the dual variables to make it
tight. It is given in detail in Appendix \ref{EdAppendix}.

The search terminates in an augment step.  It is executed for an edge
$e$ that joins 2 different trees of \os..  Let $e$ join blossoms $X_0$
and $X_1$, with $X_i$ in the tree rooted at free blossom $B_i$.  $e$
plus the edges of \os. from $X_i$ to $B_i$ forms an augmenting path
$\o P$ in the current contracted graph $\oG$.  We augment the matching
along the path $P$ in $G$, as described above. 
This changes base vertices of various
blossoms on $\o P$ as well as $P$.

After an augment the algorithm halts if every vertex is matched.  
Otherwise the next search is initialized: Every free blossom is made a
tree root of \S.. Then we proceed as above.
Note that the initialization retains all the blossoms --
a maximal blossom that is not free is now a blossom not in \S..%
\footnote{This contrasts with maximum cardinality matching,
where initialization discards the blossoms.
As a result there are no inner blossoms.}

\b

\noindent
{\bf Data structure for blossoms}
The data structure consists of two  parts,
both  used in most implementations of Edmonds' algorithm. 
The maximal outer blossoms are tracked 
using
a data structure for set-merging.
It suffices to use 
an algorithm that executes $O(m)$  
$finds$ and $O(n)$ $unions$ in a universe of $n$ elements in
 $O(m+n\log n)$ total time and $O(n)$ space
(e.g., the  simple relabel-the-smaller-set algorithm ~\cite{AHU, CLRS}).  
The incremental-tree set-merging algorithm of 
Gabow and Tarjan \cite{GT85} improves this to linear time.
The set-merging data structure is used to manipulate the
vertices of \os., e.g., a blossom step does 
$finds$ 
to identify the outer vertices in
the fundamental cycle in $\o G$
and $unions$ to  contract the cycle.
(The inner vertices are identified using the algorithm of Appendix \ref{GrowExpandSection},
as discussed at the start of Section \ref{BMergingSec}.)

The second part of the data structure
is based on  a forest representing
the laminar family of blossoms.
The main use is to compute
$P(x,\beta)$ paths. These paths are used to find  augmenting paths,
and to compute the supporting forest for tree-blossom-merging
(Section \ref{TSec}).
The data structure is also used 
in expand steps to update \os..

The laminar family is defined as follows.
Every maximal blossom 
$B^*$ has a corresponding tree $T(B^*)$.
The root of $T(B^*)$ is a node corresponding to $B^*$ and the leaves
correspond to $V(B^*)$.
The children of any node $B$ are the
subblossoms $B_i$ of  $B$.
$T(B^*)$ is an ordered tree,
with 
the children $B_i$ of $B$ ordered according to
the edges of $B$ that form a cycle spanning the  contracted
$B_i$ (in Definition \ref{BlossomDefn} these are the edges
of the two paths $P(X_i,Y)$ plus $X_0X_1$). 

The data structure for the laminar family has the following components. 
The root node (for $B^*$) records the vertex $\beta(B^*)$.
The children
of any node $B$ form a doubly linked ring. Each link also records the
edge $xy$ of $G$ ($x,y\in V(G)$) that joins  the two subblossoms.
Finally  $T(B^*)$ has parent pointers.
Additionally we assume the edges of the matching $M$ are marked as such.

This data structure has size $O(|V(B^*)|)=O(n)$.
In proof the
leaves of the tree form the set $V(B^*)$.
Each interior node is a nonsingleton blossom, so
there are fewer than $|V(B^*)|$ interior nodes.

The data structure is initially constructed in the blossom step.
The expand step involves replacing the inner node $B^*$ with
a path formed from the children of $B^*$. The other applications are
based on  $P(x,\beta)$ paths.
For completeness we give a routine for computing these paths.

We start with
some simple primitive operations.
Given $x\in V(B^*)$ the blossoms containing $x$ are identified by
tracing the  path from the leaf $x$ to the root of $ T(B^*)$.
An interior node $B\ne B^*$ with a link corresponding to
edge $xy\in M$, $x\in V(B)$
has base vertex $x$. 
If neither link from $B$ is matched then 
the base vertex of $B$ is  that of its first ancestor 
having an incident matched edge.

We turn to computing the $P(x,\beta)$ paths.
The main algorithm requires
$P(x,\beta)$ in the form of a list of edges, say $L$.
The supporting forest requires this list to be ordered
as an $x\beta$-path.
The following recursive routine uses a global list $L$.
Each recursive invocation enlarges $L$ with the appropriate edges.
 
Consider any blossom $B$ (maximal or not) and a vertex $x\in V(B)$.
The recursive routine adds the edges of $P(x,\beta(B))$
or its reverse to $L$.
The routine is called with 4 
arguments: 

\bigskip

{\narrower

{\parindent=0pt

vertex $x$ and the child $B_0$ of $B$ that contains $x$;

the base vertex $\beta(B)$; 

a bit $r$ equal to 1 if the reverse path
 $P(\beta(B),x)$ is desired, else 0.

}
}

\bigskip

The following procedure is used if $r=0$:
First consider the general case $B_0\ne B_k$.
Starting at $B_0$ follow links to the siblings $B_i$ corresponding
to the blossoms of (\ref{PvbEqn}).
The last blossom $B_k$ is identified as the node whose two links
both have unmatched edges.
For each $B_i$ use a recursive call to find the 
corresponding subpath of (\ref{PvbEqn}), and then for $i<k$ add the edge
corresponding to the link leaving $B_i$
(i.e., $\beta_i \beta_{i+1}$ or $x_i x_{i+1}$) to $L$.
The arguments for each recursive call 
$P(x_i,\beta(B_i))$ are easily found using the above primitives,
with the value of $r$  corresponding to (\ref{PvbEqn}).
(The argument $\beta(B_i)$ is known from the matched edge incident
to $B_i$, for any $i<k$. For $i=k$ it is known from the 
matched edge incident to an
ancestor of $B$, unless $\beta(B)=\beta(B^*)$. That base is explicitly
recorded in the data structure.) 
The base case $B_0=B_k$ is handled as in the definition of
$P(x,\beta)$, i.e., immediately return if
$B_0$ is a vertex else  recurse as above.

Now suppose the routine is called with $r=1$.
The base case is unchanged. For the general case first find
which of the two links from $B_k$ corresponds to an edge on the desired path.
Do this by
following  links from $B_0$ to $B_k$, starting with the appropriate link from
$B_0$.  Then follow links from $B_k$ to $B_0$, starting with the appropriate link. As in the $r=0$ case, issue recursive calls, and for $i>0$ add the
edge leaving $B_i$.
The recursive calls have $r$  set to
the complement of the bit corresponding to (\ref{PvbEqn}).

The arguments for the initial call to find $P(x,\beta(B))$
are found as above.

Regarding efficiency first observe that 
the procedure is easily implemented to use time $O(|V(B)|)$, as follows.
For any recursive call
$P(x_i,\beta(B_i))$ the path-tracing to find the subblossom containing
$x_i$ is only done the first time it is needed.
Once the path is known it is treated like a stack to get
subsequent subblossoms containing $x_i$.

The time can be improved to $O(|P(x,\beta(B))|)$.
This requires eliminating part of the path-tracing to find the blossoms
containing $x_i$. Specifically we must skip over the smallest blossoms 
containing $x_i$ as base vertex, since they do not contain any edge of
the output  $P(x,\beta(B))$.
To accomplish this we augment the data structure 
so each matched edge corresponding to a link in $T(B^*)$ 
has a
pointer to that link.
The path-tracing 
for $x_i$ starts at the link for the matched edge incident to $x_i$.
This bypasses all the unwanted blossoms
with base vertex $x_i$.

\subsection{The blossom-merging problem}
\label{BMergingSec} 
This completes the sketch of Edmonds' algorithm.
Our task is to implement a search in time $O(m+n\log n)$.
It is known how to implement most parts
within the desired time bound. For dual adjustment steps
a Fibonacci heap $\cal F$ is used. It contains the candidate edges
for tightness
mentioned above. The heap minimum gives the next edge to be made
tight. This is analogous to Fredman and Tarjan's
implementation of Dijkstra's algorithm
by Fibonacci heaps 
and uses the same time per search, 
$O(m+n\log n)$. 
Dual variables are maintained in time $O(n)$ per search using
offset quantities (e.g., \cite{GMG}).
The processing associated with grow and expand steps can be done in time 
$O(m\alpha(m,n))$ using a data structure for list splitting given in 
Gabow \cite{G85b}.%
\footnote{After the conference version of this paper
the time for  list splitting was improved 
by Pettie \cite{Pettie} and Thorup \cite{Thorup}.}
A simpler algorithm is presented in Appendix \ref{GrowExpandSection}; it suffices for our time bound and makes this paper self-contained.
The current blossom containing a given vertex is found using
the set-merging data structure of last section
for  outer vertices and the grow/expand algorithm for nonouter vertices.
Finally note after a successful search,
the corresponding augment step can be done in time $O(n)$ using the $P(x,\beta)$ paths
 \cite{G76}. 
This leaves only the blossom steps: implementing the blossom steps of a 
search in time $O(m+n\log n)$ gives the desired result.
(This observation is also made by Gabow, Galil, and Spencer\cite{GGS}.)

The problem of implementing the blossom steps of a search
can be stated precisely as the {\it blossom-merging problem} 
which we now define. (\cite{GGS} defines a similar problem called 
on-line restricted component merging, solving it
 in time  $O( m \log  \log  \log_{2+m/n} n  +  n \log n)$.)
The universe consists of a graph with vertex set $\cal O$ and edge set $\cal E$,
both initially empty.
($\cal O$ will model the set of outer vertices, $\cal E$ the set of edges joining
two $\cal O$-vertices and thus candidates for blossoms.)
At any point in time $\cal O$ is partitioned into
subsets called blossoms.
The problem is to process (on-line) a sequence of the following types
of operations:

\b

\long\def\twolines #1 #2{\vtop{\hbox{#1}\hbox{#2}}}

{

{

$make\_blossom(A)$ -- \twolines
{add the set of vertices $A$ to $\cal O$ and make $A$ a blossom}
{(this assumes $A\cap {\cal O}=\emptyset$ before the operation);}

\smallskip

$merge(A,B)$ -- \twolines {combine blossoms $A$ and $B$ into a new blossom} 
{(this destroys the old blossoms $A$ and $B$);}

\smallskip

$make\_edge(vw, t)$ -- 
{add edge $vw$, with cost $t$, to $\cal E$}
(this assumes $v, w\in O$);

\smallskip

\fm. -- \twolines {return an edge $vw$ of $\cal E$ that has minimum cost
subject to the constraint} 
{that $v$ and $w$ are (currently)
in distinct blossoms.}

}}

\b

Let us sketch how these four operations are used to implement 
a search. Grow, expand 
and blossom steps each create new outer blossoms. They perform
$make\_blossom$ operations to add the new outer vertices to $\cal O$. They also
perform $make\_edge$ operations for the new edges that join two 
outer vertices. For example in Fig. \ref{1Fig}
if $B_0$, $B_5$ and $B_6$ are in $\cal S$
and a grow step is done for edge $\alpha$ then
$make\_blossom(B_2)$ is done; also
$make\_edge$ is done for edge $\delta$. Note that in $make\_edge(vw,t)$, 
$t$ is not the given cost $c(vw)$. 
Rather $t$ is $c(vw)$ modified by dual values; 
this modification allows the algorithm to make dual adjustments efficiently
(see e.g., Gabow, Micali, and Galil\cite{GMG}). The value of $t$
is unknown until the time of the $make\_edge$ operation. From now
on, since we are only concerned with the blossom-merging problem,
the ``cost'' of an edge of $\cal E$ refers to this value $t$, not the
cost input to the matching algorithm.

A blossom step performs $merges$ to construct the new
blossom. In Fig.~\ref{1Fig} the operations
{\it merge}$(B_i,B_0)$, $i=1,\ldots, 6$ construct $B$.
Note that information giving the
edge structure of blossoms is maintained and used in the
outer part of the algorithm -- 
it is not relevant to the blossom-merging problem. For this problem
a blossom $B$ is identical to its vertex set $V(B)$;
the $merge$ operation need only update the information about the
partition of $\cal O$ induced by blossoms. Also in the blossom-merging problem 
``blossom'' refers to a set of the vertex partition, i.e., the result
of a $make\_blossom$ or $merge$ operation. 
The latter may be only a
piece of a blossom in Edmonds' algorithm (as in the 6 merges above)
 but this is not relevant.

A \fm. operation
is done at the end of each of the three search steps.
 The returned edge,
say $e$,
is used in the above-mentioned Fibonacci heap $\cal F$
that selects the next step of the search.
Specifically $\cal F$ has one entry that maintains the smallest
cost edge of $\cal E$. If that entry already contains $e$
nothing is done. If the entry contains an edge of larger cost than $e$,
the entry is updated to $e$ and a corresponding $decrease\_key$ is done.
The smallest key in $\cal F$ (which may or may not be the key for $e$)
is used for the next dual adjustment and the next step of the search.

To illustrate this process in 
Fig.\ \ref{1Fig} suppose $\cal O$ consists of 
$B_i$ for $i=0,1,2,5,6$ and $\cal E=\{\delta,\varepsilon\}$.
Furthermore  the entry in $\cal F$ contains edge $\delta$.
A grow step for $B_3$ and $B_4$ adds $V(B_4)$ to $\cal O$ and
 $make\_edge$  
is done for $\gamma$. If $\gamma$ is now the smallest edge in
$\cal E$, 
the entry in \F. for the next blossom step changes from
$\delta$ to $\gamma$ and a corresponding {\it decrease\_key} is performed.

Our task is to implement a sequence of these 
operations: $make\_blossoms$ adding a total of $\le n$ vertices,
$\le m$ $make\_edges$, $\le n$ $merges$ and $\le n$ \fms.,
in time $O(m+n\log n)$. The bound on \fms. follows since
\fm. need not be done after an expand step that does not create a
new outer vertex (e.g., in Fig.~\ref{1Fig} with blossom $B_1$ already
expanded,
expanding $A$ does not add outer vertices).
Every other step creates a new outer vertex
(a blossom step changes an inner vertex to outer). So
there are at most $n$ such steps and $n$ corresponding \fms..

The difficulty in solving the blossom-merging problem is
illustrated by Fig.~\ref{1Fig}. When each $B_i$, $i=0,\ldots, 6$ is a
blossom, edges  $\gamma, \delta, \epsilon$ are candidates
for \fm.. If a blossom step for $\gamma$ is done, 
$\delta$ and $\epsilon$
become irrelevant -- they  no longer join distinct blossoms.
If we store the edges of $\cal E$ in a priority queue
 useless edges like $\delta, \epsilon$ 
can end up in the queue.  
These edges 
eventually get deleted from the queue but the deletions 
accomplish no useful work. The time bound 
becomes $\Omega(m\log n)$.
(This also indicates why 
there is no need for a {\it delete\_min} operation in the 
blossom-merging problem: If edge $\gamma$  gets returned by
\fm. and as above
a blossom step forms $B$, edge $\gamma$ becomes  irrelevant.)

\section{Tree-blossom-merging}
\label{TSec}
Tree-blossom merging incorporates 
the topology  of the search graph
into general blossom-merging, in two ways.  This section starts by defining the
tree-blossom-merging problem and showing
how it can be used to implement the blossom
steps of Edmonds' algorithm. Then it presents our 
tree-blossom-merging algorithm.

The first goal is to maintain a representation of the search graph $\cal S$ by a
forest that changes as little as
possible.  We cannot avoid adding nodes, e.g., in grow steps.
But we can define a forest that does not change in expand steps.
Consider a search tree, i.e., a tree \ot. in the forest \os..
 
\begin{definition}
\label{STreeDefn}
A tree $T$ {\em  supports the search tree \ot.}
if each blossom $B$ of \ot. has a corresponding
subtree $T_B$ in $T$, these subtrees partition 
the vertices of $T$ and are joined by  edges of \ot., and 
for each blossom $B$:

\case{$B$ is outer} Let $B$ have  base vertex $\beta$.  $V(B)=V(T_B)$. 
If $B$ is incident to the matched edge $\beta\beta'$ in \ot.
 then $\beta'$ is the parent of $\beta$
in $T$. If $B$ is a free vertex then $\beta$ is the root of $T$.

\case{$B$ is inner} Let $B$ be  incident to edges $vx, \beta\beta'$ 
in \ot., where $x,\beta\in V(B)$ and $\beta\beta'$ is matched. Then
$T_B$ is the path $P(x,\beta)$ and $v$ is the parent of $x$ in $T$.
\end{definition}

Take any vertex $v$ in an outer blossom $B$ of \ot..
$v$ has a path to the root in both \ot. and $T$,
say $p_{\ot.}(v)$ and $p_T(v)$ respectively. 
Let
$p_{\ot.}(v)=(B_0=B,B_1,\ldots,B_k)$.
$p_T(v)$ consists of subpaths through each subtree $T_{B_i}$.
For even $i$ the subpath 
contains the base vertex
$\beta(B_i)$ and perhaps other $B_i$-vertices. 
For odd $i$ 
the subpath is 
the entire path $T_{B_i}$.
This correspondence will allow us to track potential blossom steps,
as well as
execute them,
in the supporting forest. 

We
will maintain
the supporting tree $T$ 
using this operation:


\b

{


\alm$(x,y)$ -- add a new leaf $y$, with parent $x$, to  $T$.

}

\b

\noindent
Here we  assume
$x$ is a node already in $T$ and $y$
is a new node not in $T$. We also assume the data structure for $T$
records parent pointers created by \alm.

We now show how $T$ is maintained
as the search algorithm executes
grow, blossom and expand steps. 
Assume the partition of $V(T)$
into outer blossoms and individual 
vertices in inner blossoms
is maintained by a set-merging algorithm.

Suppose a grow step enlarges $\cal S$ by adding
unmatched edge $vx$, inner blossom $B$, matched edge 
$\beta\beta'$ and 
outer blossom $B'$,
where vertices $v\in V(\ot.)$, $x, \beta\in B$ and $\beta'\in B'$.
First compute $P(x,\beta)$ and write it as 
$x_i,\ i=0,\ldots, k$ where $x_0=x$, $x_k=\beta$.
Enlarge $T$ by performing 
\alm$(v,x_0)$, \alm$(x_i,x_{i+1})$ for $i=0,\ldots, k-1$,
\alm$(x_k,\beta')$ and finally
\alm$(\beta',w)$ for every $w\in B'-\beta'$.
Merge the vertices of $B'$ into one set.
(The search algorithm accesses the vertices of $B'$ from
the blossom data structure 
at the end of Section \ref{EdAlgSec}.)

Consider a blossom step for edge $vw$.  It combines the blossoms on
the fundamental cycle $C$ of $vw$ in \ot..  Let blossom $A$ be the
nearest common ancestor of the blossoms containing $v$ and $w$ in
\ot..  $A$ is an outer blossom.  $C$ consists of the subpaths of
$p_{\ot.}(v)$ and $p_{\ot.}(w )$ ending at $A$.  
In $T$, merge every
outer blossom in $C$ into the blossom $A$.  
For each inner blossom $B$ in $C$, 
do \alm$(v,u)$ for every vertex $u\in B-V(T)$.
Then merge every vertex of $B$ into $A$. 
The new blossom $A$ has the correct subgraph $T_A$ 
so the updated $T$ supports \ot..

Lastly consider an expand step. 
The expand step in the search algorithm
replaces an inner blossom $B$ by the subblossoms along the
path $P(x,\beta)$, say subblossoms $B_i$, $i=0,\ldots,k$,
where $x\in B_0$ and $\beta\in B_k$.
By definition $T_B$ is 
the path $P(x,\beta)$.
For odd $i$, $B_i$ is a new outer blossom of \ot.. 
Perform \alm$(\beta(B_i),v)$
for every vertex $v\in B_i-V(T)$.
Merge the vertices of $B_i$.

For correctness note that
for even $i$, $B_i$ is a new inner blossom of
\ot.. 
Equation
(\ref{PvbEqn}) gives the subpaths of $P(x,\beta)$ through the $B_i$.
$T$ contains the path $P(x_i,\beta(B_i))$, and $x_{i-1}$ is the parent
of $x_i$. 
So $T_{B_i}$ satisfies Definition \ref{STreeDefn}  as required.

This completes the algorithm to maintain $T$.
It is easy to see the total time in maintaining $T$ is $O(n)$, since
$P(x,\beta)$ paths are computed in time
linear in their size.

As mentioned the correspondence between paths $p_T$ and $p_{\ot.}$
allows us to track 
potential blossom steps in the supporting forest.
This still appears to be challenging.
Our second
simplification of the problem is to assume
$make\_edge$ adds only  back edges, i.e., edges joining
a vertex to some ancestor.
Clearly
 a blossom step for an edge $vw$
is equivalent to blossom steps for the two
edges $va$ and $wa$, for $a$ the nearest common ancestor of
$v$ and $w$ in $T$. So we can
replace $vw$ by these two back edges.

To accomplish this reduction we use a data structure for dynamic
nearest common ancestors. Specifically the algorithm maintains a tree 
subject to two operations.
The tree initially consists of a dummy root \rt, and
it grows using \alm\ operations. The second operation is

\b

{

{

$nca(x,y)$ -- return the nearest common ancestor of vertices $x$ and $y$.

}
}

\b

In summary we implement  Edmonds' algorithm as follows.
The search algorithm constructs 
supporting trees for the trees of \os.
using \al. operations. (Each supporting tree is rooted at a child of
\rt.)
When the search discovers an edge $vw\in E$
joining two outer vertices, it
performs $nca(v,w)$ to find the nearest common ancestor $a$.
If $a=\rt$ an augmenting path has been found.
Otherwise the search algorithm
executes the blossom-merging operations
$make\_edge(va,t)$ and $make\_edge(wa,t)$ for appropriate $t$. 

Each grow, blossom, and expand step performs all the appropriate 
$make\_edges$, and concludes with a 
\fm. operation. Assume this operation returns
back edge $va$, corresponding to edge $vw\in E$.
Assume neither $va$ nor  $wa$ has been 
previously returned.
As mentioned above, 
the Fibonacci heap \F. records edge $vw$ as the smallest candidate 
for a blossom step. If $vw$ is selected for the next step of the search algorithm, the corresponding blossom step is performed. Also blossom-merging $merge$ operations are
executed to form the new blossom in the supporting tree. These operations
place
$v$ and $w$ in the same blossom of the supporting tree. So 
$wa$ will never be returned by future 
\fms. (by definition
of that operation).

Let us estimate
the extra time that these operations add to a search
of Edmonds' algorithm.
The algorithm for maintaining the  supporting tree $T$
uses the incremental-tree set-merging algorithm of  Gabow and Tarjan \cite{GT85}
for $merge$ operations. It
maintains the partition of $V(T)$ into maximal blossoms
in linear time.
The dynamic nca algorithm of \cite{G90, G17} uses  $O(n)$ time for
$n$ \al. operations and $O(m)$ time for $m$ $nca$ operations.%
\footnote{After the conference version of this paper
Cole and Hariharan \cite{CH}
used a similar approach to allow these and other dynamic nca 
operations. In addition their time bounds are
worst-case 
rather than amortized.}
So excluding the time for $make\_edges$ and \fms.
the extra time is $O(m+n)$.

We have reduced our task to solving the
{\it tree-blossom-merging problem}. It is defined on a tree $T$
that is constructed incrementally by \al. operations.
Wlog assume the operation
\alm$(x,y)$ 
computes $d(y)$, the depth of vertex $y$ in its supporting tree.
There are three other operations, $make\_edge$, $merge$ and \fm.,
defined as in the blossom-merging problem with the restriction that
all edges of $\cal E$, i.e., the arguments to $make\_edge$, are back edges.

There is no $make\_blossom$ operation -- we assume Edmonds'
algorithm does the appropriate 
\al. and $merge$ operations.
Note that \E. is a multiset:
A given edge $vw$ may be added to \E. many times
(with differing costs) since although an edge $uv\in E$ will become
eligible for a blossom step at most once, different $u$ vertices can
give rise to the same back edge $vw$.
Our notation assumes
the edges of \E. are directed towards the root, i.e., $vw\in \E.$ has
$d(v)>d(w)$. 

As before for any vertex $x$,  
$B_x$ denotes the blossom currently containing $x$.
Assume that the $merges$ for a blossom step
are performed in the natural bottom-up order, i.e.,
for $vw\in \E.$ the search algorithm
traverses the path in \os. from $B_v$ to $B_w$, repeatedly
merging (the current) $B_v$ and
its parent blossom.
In tree-blossom-merging
we call any set resulting from  a $merge$ operation
a ``blossom'' even though it need not be a blossom of Edmonds' search algorithm.

\subsection{The tree-blossom-merging algorithm}
\label{TBMAlgSec}
\def\s.{{\em s}}
\def\l.{{\em l}}
\def\d.{{\em d}}
\def\Iline (#1) #2.{\noindent(#1)\hskip20pt #2.}
\def\emi{{\em i}}
\def\emii{{\em ii}}
\def\emiii{{\em iii}}
This section solves the tree-blossom-merging problem in time $O(m+n\log n)$.
First it  presents the basic principles
for our algorithm.
Then it gives the data structure, the algorithm statement, and its analysis.

Two features of
the supporting forest are essentially irrelevant to our algorithm:
The only role played by inner vertices,
prior to becoming outer,
 is to contribute to
depths $d(v)$.
The fact that supporting trees are  constructed incrementally is of no consequence, our algorithm only ``sees'' new $merge$ and $make\_edge$ operations.

Our strategy is to charge time to blossoms as their size doubles.
We
use a notion of
``rank'' defined for edges and blossoms:
The {\it rank} of an 
edge $vw\in {\cal E}$ is defined by
\[r(vw)=\f{\log (d(v)-d(w))}.
\]
The rank of an edge is between 0 and $\f{\log(n-1)}$.
A blossom $B$ has {\it rank} 
\[r(B)=\f{\log |B|}.\]
(Recall that in this section a blossom is a set of vertices.)

These notions are recorded in
the algorithm's data structure as follows.
(A complete description of the data structure is given below.)
We use 
a simple representation of $T$, each vertex $x$ in $T$ recording its
parent 
and depth $d(x)$. 
Each  blossom $B$ records its size $|B|$
and rank $r(B)$.

There are $\le n$ $merge$s, so each can perform a constant number of
time $O(\log n)$ operations, like Fibonacci tree $delete\_mins$ or
moving $\log n$ words.  There are $\le 2m$ $make\_edge$s, so each can
perform $O(1)$ operations.  Each vertex $v$ always belongs to some
current blossom $B_v$, and $\le \log n$ $merge$ operations increase the rank
of $B_v$. So we can charge $O(1)$ time to a vertex $v$
every time its blossom increases in rank.

\begin{table}[t]
\centering
\begin{tabular}{|c|l|l|c|l|}
\hline
Type&\multicolumn{2}{c|}{Conditions}&$u$&\multicolumn{1}{c|}{$r$}\\
\hline
{\em l}&$r(e)>r(A_v)$&&$v$&$r(e)$\\ 
\hline 
{\em s}&$r(e)\le r(A_v)$,&$r(A_v)\le r(A_w)$
&$v$& $\max\{ r(A_v)+1,\, r(A_w)\}$  \\
\cline{1-1}\cline{3-5} 
{\em d}&$A_v\ne A_w$&$r(A_v)>r(A_w)$
&$w$&$r(A_v)$\\
\hline
\end{tabular}
\caption{Edge  $e=vw\in \E.$ has Type {\em l,s,} or {\em d}
defined by the Conditions. Merging  blossoms $B_v$ and $B_w$ gives a blossom of rank $\ge r>r(A_u)$ (Proposition
\ref{TypePropertiesProp}).}
\label{TypeDefnTable}
\end{table}

\begin{figure}[ht]
\centering
\input{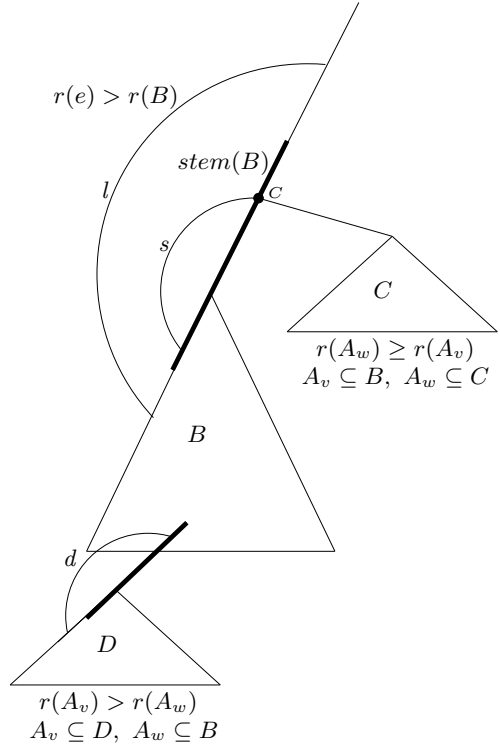}
 \caption{\l., \s., and \d. edges associated with
 blossom $B$, i.e., $A_u\con B$.}
 \label{ETypesFig}
 \end{figure}

\begin{table}[h]
\centering
\begin{tabular}{|c|c|c|c|c|}
\hline
State&$r(A_v)$&$r(A_w)$&$e:r$&reclassified\\
\hline 1&2&2&$e_1:3$&4\\
\hline 2&2&4&$e_2:4$&4\\
\hline 3&2&5&$e_3:5$&\\
\hline 4&4&6&$e_1:6$&\\
\hline
\end{tabular}
\caption{$s$-edges $e_1,e_2,e_3$, all copies of $vw$, and their
blossom parameters; $u=v$.
$e_1$ and $e_2$ are reclassified in state 4.}
\label{Ex1Table}
\end{table}

\begin{table}[h]
\centering
\begin{tabular}{|c|c|c|c|c|}
\hline
State&$r(A_v)$&$r(A_w)$&$e:r$&reclassified\\
\hline 1&3&2&$e_1:3$&4\\
\hline 2&3&2&$e_2:3$&4\\
\hline 3&5&2&$e_3:5$&\\
\hline 4&6&3&$e_1:6$&\\
\hline
\end{tabular}
\caption{$d$-edges $e_1,e_2,e_3$, all copies of $vw$, and their
blossom parameters; $u=w$.}
\label{Ex2Table}
\end{table}

We associate each edge $vw\in \E.$ with
one of its ends, say vertex $u\in \{v,w\}$, in such a way that 
a merge that places $v$ and $w$ in the same blossom increases the rank of
$B_u$.
To do this we assign  a {\em type} to every edge -- {\em long, short}
or {\em down}, respectively, or synonymously {\em l-edge, s-edge, d-edge}.
At any point in the execution of the algorithm an edge has a unique type,
but the type can change over time. 

Types are
defined in Table \ref{TypeDefnTable} and illustrated in 
Tables \ref{Ex1Table}--\ref{Ex2Table} and
Fig.\ref{ETypesFig}.
In Table \ref{TypeDefnTable}
$A_v$ and $A_w$ denote the blossoms containing 
$v$ and $w$, respectively, 
when the definition is applied to assign the edge's type.
Edges with a long span, type $l$,
are treated based on their
span; edges with a short span, type $s$ or $d$,
are treated according to the relative sizes
of the two blossoms containing their ends.
Clearly the Conditions column  specifies a unique type for
any edge $vw$ with $A_v\ne A_w$.

The edge type of $e=vw$ determines 
the end that $e$ is associated with -- this is vertex
$u\in \{v,w\}$ as specified in 
Table \ref{TypeDefnTable}.
As the algorithm progresses and $B_v$ and $B_w$ grow the type
of $e$ and other values in Table \ref{TypeDefnTable} may change.
(Indeed eventually we may have $B_v=B_w$ and $e$ has no type at all.)
The algorithm will not track every change in the type of $e$, instead
it examines $e$ from time to time, each time assigning a new type
according to the table.
This is illustrated in Fig.\ref{ETypesFig}
and the following example.

\example 1.
{Tables \ref{Ex1Table}--\ref{Ex2Table}
give classifications performed in a hypothetical execution of  the algorithm.
Table \ref{Ex1Table} shows type $s$
classifications of three edges, all of which are
copies of the same edge $vw$ with $r(vw)=2$.
$make\_edge(e_2)$ is executed after $make\_edge(e_1)$ and
so $e_2$ is classified after $e_1$, as illustrated in states 1--2.
$e_3$ is classified later still, in state 3.
At this point all three edges have different classifications.
In state 4, $e_1$ gets reclassified (as noted in the last column of the table). The algorithm also examines $e_2$,
detects it as a copy of $e_1$ with greater cost,
and so  permanently discards
$e_2$.
 $e_3$ is not reclassified at this point.
Further details of this scenario are given below in 
Example 2.
Table \ref{Ex2Table} gives a similar executions for three $d$-edges $vw$.}

We call the values specified in Table \ref{TypeDefnTable}
(specifically the edge type, blossoms $A_v$ and $A_w$, $r$, and $u$) the 
{\em type parameters}
of the classification. At a given point in the algorithm,
an edge $e=vw\in \E.$ will have two pairs of blossoms that are of interest:
the edge's current
 type parameters $A_v$ and $A_w$,
and the 
current blossoms $B_v$ and $B_w$. 
Clearly $A_v\con B_v$ and $A_w\con B_w$.

As mentioned above a merge making $B_v=B_w$
increases $r(B_u)$ to $r$ or more.
The following proposition makes this precise.

\begin{proposition}
\label{TypePropertiesProp}
Let $e=vw$ have type parameters $A_v,A_w,r,u$.

\i $r> r(A_u)$.

\ii A blossom $Z$ containing $v$ and $w$ has $r(Z)\ge r$.
\end{proposition}

\begin{proof}
\i An \l.-edge has $r=r(e)>r(A_v)$ and $v=u$.
An \s.-edge has
$r\ge r(A_v)+1$ and $v=u$.
A \d.-edge has $r=r(A_v)>r(A_w)$ and $w=u$.

\ii 
Suppose $e$ is an \l.-edge.
 $Z$ contains the path
from $v$ to $w$ so
$|Z|>d(v)-d(w)\ge 2^{r(e)}=2^r$.
Thus $r(Z)\ge r$.

If $e$ is a \d.-edge then $A_v \con Z$.
So $r(Z)\ge r(A_v)=r$.

If $e$ is an \s.-edge then $A_v, A_w\con Z$ 
and $r(A_w)\ge r(A_v)$ imply
$|Z|\ge |A_v|+|A_w|\ge \max \{2^{r(A_v)+1},\, 2^{r(A_w)}\}$.
So $r(Z) \ge r$.
\end{proof}

A crucial property for
the accounting scheme is that
the $s$ and $d$ edges are limited in number.
To make this precise recall that the vertices of a blossom
form a subtree of the supporting tree, so we can refer to a
blossom's root vertex.
(This differs from the notion of base vertex.
For instance recall that a tree-merging blossom
needn't be a complete blossom of Edmonds' algorithm.)
For any blossom $B$ define $stem(B)$
to consist of the
first 
$2|B|$ ancestors of the root of $B$ 
(or as many of these ancestors that exist).

\begin{proposition}
\label{StemLemma}
Let $e=vw$ where a blossom $B$ contains $v$ but not $w$ and
$r(e)\le r(B)$. Then $w\in stem(B)$.
\end{proposition}

\remark{The proposition shows 
an $s$ or $d$ edge $e=vw$ 
has $w\in stem(A_v)$, since
the definition of  Table \ref{TypeDefnTable}
 has $r(e)\le r(A_v)$.}

\begin{proof} 
Let 
$r=r(B)$. Thus $2^r\le |B|$.
Since $r(e)\le r$,
$d(v)-d(w)<2^{r+1}\le 2|B|$. Rearranging to
$d(v)-2|B| <d(w)$ puts $w$ in the stem of $B$.
\end{proof}

Let \M. be the set of all 
blossoms $B$ formed in the algorithm that are rank-maximal,
i.e., $B$
is not properly contained in any blossom of the same rank.
A given vertex belongs to at most $\log n$ such rank-maximal blossoms.
Thus $\Sigma_{B\in \M.} |B|\le n\log n$.
Our plan for achieving the desired time bound involves
using the proposition to
charge each blossom $B\in \M.$ $O(1)$ time for each vertex 
in  $stem(B)$. The total of all such charges is $O(n\log n)$
and so is within our time bound.


\paragraph*{The data structure}
We describe the entire data structure as well as its motivation.

Each current blossom $B$ has $\le \log n$ 
lists of edges designated as $\pkt(B,r)$ for $r\in [r(B)+1..\log n]$.
$\pkt(B,r)$ consists of edges whose classification parameters $u,r$
have $A_u\con B$ and $r$
matching the packet index. As motivation note all these edges are similar
in the sense of producing the same rank increase
(Proposition \ref{TypePropertiesProp}\xii). This allows the edges of a packet
to be processed as a single group rather than individually.%
\footnote{Our notion of packet is similar in spirit, but not detail, 
to the data structure of the same name in \cite{GGS, GGST}.}

Each  $\pkt(B,r)$ is implemented as a ring, i.e., 
a circularly linked list.
This allows two packets to be concatenated in $O(1)$ time.
A header records  the packet rank $r$ and 
$smallest(B,r)$, the edge of smallest cost in the packet.

The packets of blossom $B$ are stored in a list in arbitrary order,
i.e., they are not sorted on $r$. Also if $\pkt(B,r)$ currently contains
no edges, its header is omitted.
These two rules  limit the space for the algorithm to $O(m)$.
For graphs with $m=\Omega(n\log n)$ an alternate organization is possible:
Each current blossom has an array of $\log n$ pointers to its packets.%
\footnote{Various devices can be added to this approach to achieve linear space
for all $m$.
Such an organization is used in the conference version of this paper
\cite{G90}.
However our current approach is more uniform and simpler.}

Each current blossom $B$ has an additional list
 $loose(B)$. It consists of all edges classified with $A_u=B$
but not yet in a packet of $B$.
This list is a ``waiting area'' for edges to be added to a packet.
It allows us to omit the array of pointers to packets mentioned above.
$loose(B)$ is a linked list.

The value $smallest(B)$ is maintained as the minimum cost edge
in one of $B$'s  lists (i.e., a  $B$-packet or $loose(B)$). 

A Fibonacci heap \H.  stores 
the values $smallest(B)$ for every current blossom $B$.
It is convenient to do lazy deletions in \H.:
An operation $merge(A,B)$, which replaces blossoms $A$ and $B$ by
the combined blossom $C$, marks the entries for $A$ and $B$ in \H.
as deleted, and inserts a new entry for $C$.
To do a \fm. operation
we  perform a Fibonacci heap $find\_min$  in \H.. 
If the minimum corresponds to a deleted blossom,
that entry is deleted (using $delete\_min$) and the procedure is
repeated. Eventually we get
the smallest key for a current blossom. Its edge is
returned as the \fm. value.

An auxiliary array $I[1..n]$ is used
to perform a {\em gather} operation, defined as follows:
We have a collection of $c$ objects, each one having an associated
index in $[1..n]$. We wish to gather together all objects with the
same index. We accomplish this in time $O(c)$, by placing objects with
index $i$ in a list associated with $I[i]$. An auxiliary list of indices
that have a nonempty list allows the lists to be collected when we are done,
in $O(c)$ time. Gathering operations will be done to maintain packets
in $merges$.

Finally each 
blossom $B$ has a number of
bookkeeping items:
Its rank $r(B)$ is recorded.
There is also a representation of the partition of $\cal O$ into blossoms.
A data structure for set-merging \cite{T83} can be used: The 
blossom-merging operation
$merge(A,B)$ executes a set-merging operation  $union(A,B)$
to construct the new blossom; for any vertex $v$,
the set-merging operation $find(v)$
gives $B_v$, the blossom currently containing $v$.

For simplicity we will
omit the obvious details associated with this bookkeeping.
We can also ignore the
time and space. For suppose we use
a simple set-merging algorithm
that does one $find$ in $O(1)$ time, all $unions$ in $O(n\log n)$
time, and uses $O(n)$ space (e.g., \cite{AHU}). Clearly 
the space and the time for $unions$ are within the desired bounds for Edmonds'
algorithm; the time for $finds$ can be associated with  other
operations. Hence we shall ignore this bookkeeping.

\paragraph*{The algorithms}
We present the algorithms for
tree-blossom-merging, verify their correctness, and prove the desired time bound
$O(m+n\log n)$.


The algorithm maintains this invariant: 

\bigskip

{\noindent
(I1)
{The set $S$ of all edges in a
$loose$-list or packet satisfies:

\i every edge of $S$ joins 2 distinct blossoms;

\ii for every 2 blossoms joined by an edge of \E.,
$S$ contains
such a joining edge
of smallest cost.
}
}

\bigskip

\noindent
(I1) guarantees that for every blossom $B$,
$smallest(B)$ is a minimum-cost edge joining $B$ to another blossom.
In proof, \ii guarantees that such a minimum-cost edge belongs to
some $loose$-list or packet of $B$. \i guarantees that this edge
gives the value of $smallest(B)$ (i.e., without \i it is possible
that the minimum-cost edge has both ends in the same blossom and so
is not useful).

As mentioned \fm. is
a Fibonacci heap {\it find\_min} in \H..
This edge gives the next blossom step by the definition of 
$smallest(B)$ and invariant (I1).
The total time for all \fms. is $O(n\log n)$, since there
are $O(n)$ blossoms total and each can be deleted from \H..

$make\_edge(vw)$ is implemented in a lazy fashion as follows:
If $B_v=B_w$ then $e$ is discarded.
Otherwise we 
classify $e=vw$ as type \l., \s., or \d.,
by computing $r(e), r(B_v)$ and $r(B_w)$, as well
as vertex $u$ and rank $r$,  in time $O(1)$.
$e$ is  added to $loose(B_u)$ and $smallest(B_u)$ is updated, with
a possible $decrease\_key$ in \H.. 
The time is $O(1)$ (amortized in the case
of $decrease\_key$). This time is charged to  the creation of $e$.

We now present the algorithm for $merge(X,Y)$.
It forms the new blossom $Z$.
(Note that a given edge $e_0$ that is selected
by the search algorithm for  the next blossom step
will cause one or more such $merge$ operations. The $merge$ algorithm,
and its analysis, does not
refer to $e_0$.)

Let the set $R_0$ consist of all edges 
that must be reclassified because of the $merge$, i.e.,
the 
edges in packets of $X$ or $Y$ of rank $r\le r(Z)$,
and the edges in $loose(X)$ or $loose(Y)$.
The edges of $R_0$ are  pruned to eliminate redundancies, i.e.,
 we ensure that at most one such edge joins $Z$
to any blossom $B\ne Z$. This is done with a gather operation
using $I[1..n]$.
Among all edges joining $Z$ and a given blossom $B\ne Z$, 
only one of smallest cost
is retained. 
Edges with both ends in $Z$ are discarded.
These actions preserve (I1).

Let $R$ denote the set of remaining edges. We  
assign each edge of $R$ to its appropriate packet or
$loose$-list, and form the final packets of $Z$, as follows.

Take any edge $e=vw\in R$.
Compute the new type of $vw$ using $r(e), r(B_v)$ and 
$r(B_w)$. 
Two types of edges get added to $loose$-lists:
An \s.-edge $vw$ with $w\in Z$ gets added to
$loose(B_v)$.
As usual we also update $smallest(B_v)$, possibly doing  $decrease\_key$.
 Similarly
 a \d.-edge $vw$ with $v\in Z$ gets added to
$loose(B_w)$.

The other edges are added to packets of $Z$ as follows.
Use the subarray  $I[r(Z)+1..\log n]$ in a gather operation.
Specifically $I[r]$ gathers all individual edges that are $r$-promoters
for $Z$ and forms them into a list.
It also gathers
$\pkt(X,r)$ and $\pkt(Y,r)$, if they exist.
These two packets are treated as lists, not examining the individual edges,
so $O(1)$ time is spent adding them to the list $I(r)$. 
The final $I(r)$ is taken as the list for
$\pkt(Z,r)$.

We complete \pkt(Z,r) by computing 
$smallest(Z,r)$ from $smallest(X,r)$, $smallest(Y,r)$,
and the costs of all its other edges.
The smallest of all these values gives  $smallest(Z)$.
This value is inserted into \H..
The $loose$ list of $Z$ is empty.

\example 2.
{The algorithm achieves the states given in
Table \ref{Ex1Table} as follows.

{
\parindent=0pt

\bigskip

State 1: $make\_edge(e_1)$ is executed when  
$|B_v|=|B_w|=4$, so the algorithm adds $e_1$ to $loose(B_v)$.
A merge makes $|B_v|=5$ and adds $e_1$ to $\pkt(B_v, 3)$.

\bigskip

State 2:
$make\_edge(e_2)$ is executed and merges increase $|B_w|$ to 16. 
A merge makes $|B_v|=6$, transferring   $e_2$ from
$loose(B_v)$ to $\pkt(B_v,4)$.

\bigskip

State 3: Achieved similar to state 2, with $|B_w|$ increasing
to 32 and then $|B_v|$ increasing to 7.

\bigskip

State 4: Merges increase
$|B_w|$ to 64. A merge makes $|B_v|=16$, so
the edges in the rank 3 and rank 4 packets of $B_v$ are added to $R_0$.
$e_2$ is discarded by the gather operation 
and $e_3$ is added to $\pkt(B_v,6)$.
}

\bigskip

Table \ref{Ex2Table} is similar.
$|B_w|$ starts as 4 and increases to 5 (giving state 1), 
then 6 (state 2),
then 7 (state 3). When $|B_w|$ increases to  8 
edges $e_1$ and $e_2$ are added
to $R_0$, $e_2$ is discarded and $e_1$ is added to
the rank 6 packet (state 4).
}

For correctness of this algorithm observe that
Proposition \ref{TypePropertiesProp}\i shows 
the new packet assigned to an edge of $R$ actually exists.
To show  invariant (I1\emi) for $Z$ consider the edges
in packets of $Z$ that are not examined individually, i.e., edges
$e$ in a packet of  $X$ or $Y$ of rank $r>r(Z)$.
Proposition \ref{TypePropertiesProp}\ii  shows 
a blossom containing both ends of $e$ has rank $\ge r >r(Z)$. Thus
$e$ satisfies (I1\emi).

The time for $merge$ is $O(\log n)$ plus $O(1)$ for each edge
added to $R_0$.
The first term amounts to $O(n\log n)$ for the entire algorithm
(there are $\le n$ $merges$). 

We account for the second term in two 
ways. The first is to charge time to blossoms, via their stems.
To do this recall that an $s$- or $d$-edge $vw$ 
has $w\in stem(B_v)$ (for the current blossom $B_v$ containing $v$;
Proposition \ref{StemLemma}).
We will charge the stems of blossoms $B\in \M.$.
To do this it is important to verify that
that $e$ is the only edge from $B$ to $w$ making the charge.

The second accounting mechanism is a system of credits. 
It pays for
reclassifications that involve $loose$ lists. 
As motivation note that successive merges may move an edge $e=vw$
from
$loose(B_w)$ to 
a packet of $B_w$ to
$loose(B_v)$ to a packet of $B_v$; credits will
pay for these moves.
Define a credit to account for $O(1)$ units of processing time,
specifically the time to process an edge $e\in R_0$.
Credits are
 maintained according to this invariant (I2):
\bigskip

{

(I2\emi) An $l$- or $s$-edge in a $loose$ list has 1 credit.

(I2\emii) A $d$-edge in a packet has 2 credits.

(I2\emiii) A $d$-edge in a $loose$ list  has 3 credits.

}


\begin{lemma}
\label{MergeCreditLemma}
The time spent on all edges of $R_0$ in all merges
is $O(m+n\log n)$.
\end{lemma}

\begin{proof}
A $make\_edge$ operation adds $e$ to the appropriate $loose$ list
with credits corresponding to (I2\emi) or (I2\emiii).
Note this accounts for the case of $e$ starting as an $l$-edge.

Now consider an edge $e=vw$ that belongs to $R_0$ in
an operation 
$merge(X,Y)$ which forms the new blossom $Z$.
An edge $e\notin R$  pays for
the $O(1)$  time spent processing $e$. This is acceptable since
no future merge has  $e\in R_0$.

We can now assume $e\in R$ starts as type $s$ or $d$.
Let $e$ start out 
with classification parameters $A_v,A_w,u,r$.
By symmetry assume $X$ contains an end $v$ or $w$ of $e$.

If $e$ starts out in a packet, i.e.,
$\pkt(X,r)$, then $X\in \M.$. In proof,
the definition of $X$'s packets shows $r(X)<r$.
Since the merge has $e\in R$, $r \le r(Z)$.
Combining gives
$r(X)< r(Z)$, so $X$ is rank-maximal.

Now consider the two possible types for $e$.

\case {$e$ starts as an $s$-edge}
We consider two possibilities.

\subcase{$e$ starts in a packet}
$e$ is the unique edge of $R$ that is directed to $w$.
We charge the reclassification of $e$ to $w\in stem(X)$.
Clearly $X\in \M.$ implies this is the only time the algorithm
charges $w\in stem(X)$.

The 
charge to $w\in stem(X)$ 
is $O(1)$ time to account for  the processing of $e$ in this merge.
In addition if $e$ is added to $loose(B_w)$
there is a
charge of 3 credits which, given to
$e$,  establish (I2\emiii).

\subcase{$e$ starts in $loose(X)$}
(I2\emi) shows it has 1 credit.
If $e$ gets added to a packet
the credit pays for the processing of $e$.
Suppose $e$ gets added to $loose(B_w)$.
Since $e$ changes from $s$ to $d$ we have
$r(A_v)\le r(A_w)\le r(B_w)<r(B_v)$.
Note $A_v=X$, $B_v=Z$.
(The former holds since in general, an edge in a list $loose(B)$ 
is in $R_0$ in the first merge involving $B$.)
So $r(X)<r(Z)$.
Thus $X\in \M.$. So we can charge $w\in Stem(X)$ as in the previous subcase.

\case{$e$ starts as a $d$-edge}

\subcase{$e$ starts in a packet}
$e$ is the only edge of $R$ directed from  $B_v$ to $w$. 
Let $A^+_v$ be the smallest blossom of \M. that contains $A_v$.
We  charge the reclassification of $e$ to $w\in stem(A^+_v)$.
The reclassification has $r(Z)\ge r=r(A_v)$.
So future blossoms $B_w$ have $r(B_w)\ge r(A_v)$, and 
these blossoms will not have a packet of rank $r(A_v)=r(A^+_v)$.
Thus this
case for $w$ and $A^+_v$ will never hold again.
So this is the only charge the algorithm makes to $w\in stem(A^+_v)$.
(Note  $A_v$ need not be rank-maximal, we might even have $A_v=B_v\pcon
A^+_v$.)

The charge
is $O(1)$ time to account for  the processing of $e$ in this merge.
In addition we charge 2 credits 
(1 credit)
if $e$ is added to a packet of $Z$ 
($loose(B_v)$) respectively. 

\subcase{$e$ starts in $loose(X)$}
(I2\emiii) shows $e$ has 3 credits. 
If $e$ gets added to a packet 1 credit pays for the processing and
the other 2 establish (I2\emii). If $e$ gets added to $loose(B_v)$
we can discard 1 credit and still pay for the processing and
(I2\emi).
\end{proof}

\example 3. {In Table \ref{Ex2Table} the transfer of $e_1$ 
from the rank 3 packet to rank 6 
is charged to $w\in stem(A^+_v)$.
Here $A^+_v\in \M.$ is the rank 3 blossom corresponding to states 1 and 2.
If a future merge transfers $e_3$ from the rank 5 packet to a higher one,
the time is charged to   $w\in stem(A^+_v)$ where 
$A^+_v\in \M.$ is the rank 5 blossom of state 3.

In Table \ref{Ex1Table} the transfer of $e_1$ 
from the packet of rank 3 to rank 6 
is charged to $w\in stem(A^+_v)$
where $A^+_v\in \M.$ is the rank 2 blossom corresponding to states 1--3.
This merge increases $r(B_v)$ from 2 to 4.
If a future merge transfers $e_3$ from the rank 5 packet to a higher one,
the time is charged to   $w\in stem(A^+_v)$ where 
$A^+_v\in \M.$ will have rank $\ge 4$.
}

Having analyzed the algorithms for 
\fm., $make\_edge$, and $merge$, 
we can conclude our 
 tree-blossom-merging algorithm achieves the desired time bound:

\begin{theorem}
\label{TreeBlossomMergingTheorem}
The tree-blossom-merging problem can be solved in time $O(m+n\log n)$.
\hfill\qed\end{theorem}


\section{{\boldmath $b$}-matching}
\label{bMatchingSec} 
This section presents a simple generalization of Edmonds' algorithm
to $b$-matching. The major difference from ordinary matching is that 
$b$-matchings
allow 2 varieties of blossoms, which we  call
``light'' (analogous to ordinary matching) and ``heavy''. Our
goal is an algorithm that is as similar to ordinary matching as possible,
by minimizing the use of heavy blossoms. Heavy blossoms
 seem impossible to avoid but our algorithm keeps them ``hidden''.
Section \ref{bBlossomSec} gives the basic properties of $b$-matching blossoms.
Section \ref{bAlgSec}  presents the generalized algorithm and shows
we can find a maximum $b$-matching in time 
$O(b(V)(m+n\log n))$.
Section \ref{bStrongSec} extends the algorithm to achieve the strongly polynomial time bound 
$O(n\log n\ (m+n\log n))$, the same bound as known for bipartite graphs.

A degree constraint function $b$ assigns a nonnegative integer to each
vertex.
We view $b$-matching as being defined on a multigraph.
Every edge has an unlimited number of copies. 
In the context of a given $b$-matching, an edge of the given
graph
has an unlimited
number of unmatched copies; the number of matched copies is specified
by the $b$-matching.
In a {\em partial $b$-matching} every vertex $v$ has degree $\le b(v)$.
In  a {\em (perfect) $b$-matching} every vertex $v$ has degree exactly $ b(v)$.
Note that  ``$b$-matching'' (unmodified) refers
to a perfect $b$-matching, our main concern.

We use these multigraph conventions:
Loops are allowed. A cycle is a connected degree 2 subgraph,  be it a loop,  2 parallel edges,
or an undirected graph cycle.

Contracting a subgraph 
does not add a loop at the contracted
vertex (all internal edges including internal loops disappear).
We will even contract subgraphs that just contain a loop.
We use the following notation for contractions.
Let $\o G$ be a graph derived from $G$ by contracting 
a number of vertex-disjoint subgraphs.
$V$ ($\o V$) denotes the vertex set of $G$ ($\o G$), respectively. 
A vertex of $\o G$ that belongs to $V$ (i.e., it is not in a contracted subgraph) is an {\em atom}.
We identify an   edge of $\o G$
with its corresponding edge in $G$. Thus an edge of $\o G$ is denoted as
$xy$, where $x$ and $y$ are $V$-vertices in distinct $\o V$-vertices, 
and $xy\in E$.
Let $H$ be a subgraph of 
$\o G$. The {\em preimage} of $H$ is a subgraph of $G$ 
consisting of the edges  
of $H$, plus the subgraphs whose contractions are vertices of $H$,
plus the atoms of $H$.
$\o V(H)$ ($V(H)$) denotes the vertex set of $H$ (the preimage of $H$), respectively.
Similarly $\o E(H)$ ($E(H)$) denotes the edge set of $H$ (the preimage of $H$), respectively.

\subsection{Blossoms}
\label{bBlossomSec}
This section presents the basic properties of
$b$-matching blossoms.
We define blossoms in two steps, first
specifying the topology, then specializing to 
``mature'' blossoms which can have positive dual variables.
We give a data structure for blossoms and show how blossoms are
 updated when the matching gets augmented.

\begin{figure}[th]
\centering
\input{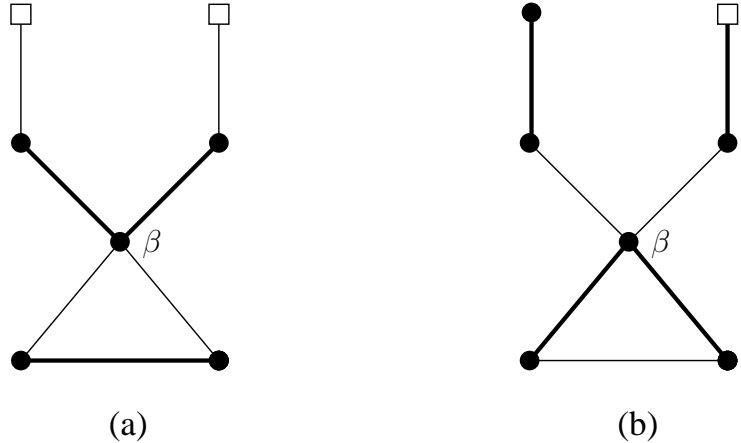}
 \caption{An augmenting trail (a) and the augmented matching (b).}
 \label{InnerOuterFig}
 \end{figure}

Unlike ordinary matching, contracted $b$-matching  
blossoms do not behave exactly like original vertices.
For instance Fig.\ref{InnerOuterFig}(a) shows an augmenting trail --
interchanging matched and unmatched edges along this (nonsimple) trail
enlarges the matching
to  Fig.\ref{InnerOuterFig}(b). 
(As in all figures square vertices are free and
heavy edges are matched. In
 Fig.\ref{InnerOuterFig}(b) one vertex remains free after the augment.) 
The triangle is a $b$-matching blossom (just like ordinary matching).
In
 Fig.\ref{InnerOuterFig}(a)
contracting this blossom gives a graph of 5 vertices that
has no augmenting path. 
Contracted blossoms behave in a more general way than
ordinary vertices. 

When a blossom becomes ``mature'' it behaves just like a vertex -- in fact
a vertex with $b$-value 1 just like ordinary matching! It also behaves like
an ordinary matching blossom in that its $z$-value can be positive
(in contrast an  immature blossom, e.g., the blossom
of Fig.\ref{InnerOuterFig}, cannot have  positive $z$). 

We will define a blossom in terms of its topology --
it is a subgraph
that when 
contracted can behave like a vertex or like
Fig.\ref{InnerOuterFig}. We will then
specialize this notion to the case of mature blossoms.
For completeness we give 2 versions of the topological  definition
(Definitions \ref{EBlossomDefn} and \ref{ABlossomDefn})
the second one being more useful algorithmically.
The first simpler definition is a type of ear decomposition.

Let $G$ be a graph with a partial $b$-matching, i.e., every vertex $v$ is on
$\le b(v)$
matched edges.
A trail is {\em closed} if it starts and ends at the same vertex.
A trail is {\em alternating} if for every 2 consecutive edges exactly one
is matched. The first and last edges of a closed trail are not consecutive.
Say that the {\em $M$-type} of 
an edge is $M$ or $E-M$, according to the set that contains it. 
The following definition is illustrated in Fig.\ref{EarFig}.

\begin{figure}[th]
\centering
\input{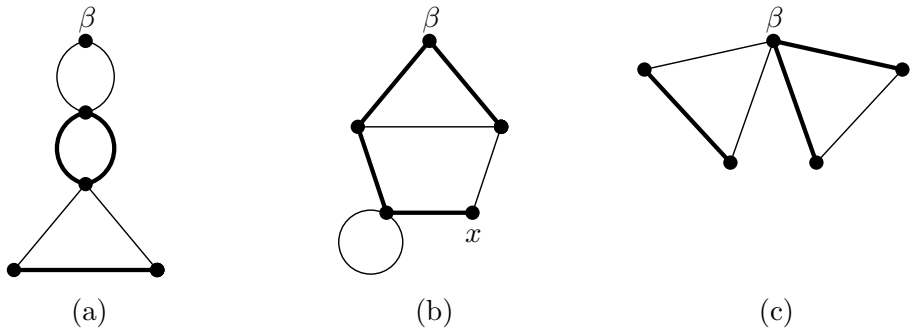}
 \caption{Blossoms with base vertex $\beta$.
(a) is light, (b) is heavy, (c) is heavy or light.
}
 \label{EarFig}
 \end{figure}

\begin{definition}[Ear Blossom]
\label{EBlossomDefn}
A {\em blossom} is a subgraph of $G$ recursively
defined by  two rules:

(a) A closed alternating trail of $G$
is a blossom if it
 starts and ends with edges of the same $M$-type.

(b) 
In a graph derived from $G$ by contracting 
a blossom $\alpha$, 
let $C$ be a
closed alternating trail  that starts and ends at 
$\alpha$ but has no other occurrence of $\alpha$.
The preimage of $C$ in $G$ is
a {blossom}. 
\end{definition}

An example of (a) is a loop.

For uniformity we extend the notation of (b) to (a):
 $C$ denotes the blossom's trail  and
$\alpha$ denotes its first vertex.
The {\em base vertex} of a blossom is
$\alpha$ in (a) and the base vertex of
$\alpha$ in (b). Clearly the base is always in $V$.
The base vertex of blossom $B$ is denoted $\beta(B)$ 
(or $\beta$ if $B$ is clear).
The {\em $M$-type} of $B$
is the $M$-type of the starting and ending edges in (a),
and the $M$-type of $\alpha$ in (b).
It is convenient to call  a blossom of $M$-type $M$
{\em heavy} 
and $M$-type  $E-M$
{\em light}.

\example 1. {In Fig.\ref{EarFig}(a)
the entire graph forms a closed alternating trail that 
starts and ends at $\beta$. It 
is a blossom with base vertex $\beta$,
by part (a) of the definition. 
There is essentially no other way to view
the graph as a blossom, since the two edges at $\beta$ do not alternate.

In  Fig.\ref{EarFig}(b) the unique closed alternating trail starting and ending
at $\beta$ is the triangle.
So it is a minimal blossom. Contracting
it to $\alpha$ gives a graph with a unique closed alternating trail
that starts and ends at $\alpha$.
So the graph is a blossom with base vertex $\beta$.
Again this is essentially the only way to parse this graph as a blossom.

Fig.\ref{EarFig}(c) is a light blossom with base $\beta$
if we start the decomposition with the left triangle.
Starting with the right triangle gives a heavy blossom based at $\beta$.
No other blossom decomposition is possible.}

Only light blossoms occur in ordinary matching, and they
are the main type in $b$-matching.
We note two instances of the definition that 
will not
be treated as blossoms in the algorithm. 
Both instances are for a light blossom $B$.
If
$d(\beta,M)\le b(\beta)-2$ then (a) actually gives an augmenting trail.
Secondly  the definition allows $d(\beta, \gamma(V(B),M) )=b(\beta)$. This
never holds in the algorithm -- 
$\beta$ is either on an  edge
of $\delta(V(B),M)$ or
$d(\beta, M)<b(\beta)$.

Consider a blossom $B$ with base vertex $\beta$.
Similar to ordinary matching,
each vertex 
$v\in V(B)$ has 2 associated  $v\beta$-trails
in $E(B)$,
$P_0(v,\beta)$  and $P_1(v,\beta)$, with even and odd lengths respectively.
Both trails are alternating and
both end with an edge whose $M$-type is that of $B$
(unless the trail has no edges). 
The starting edge for  $P_1(v,\beta)$ 
has the same $M$-type as $B$; it has  the opposite $M$-type for
$P_0(v,\beta)$.
As examples,
$P_1(\beta,\beta)$  is the entire trail in (a).
This trail could be a loop.
$P_0(\beta,\beta)$
is always the trivial trail ($\beta$). It is the only trivial $P_i$ trail.

 The recursive definitions of
the $P_i$ trails follow easily from Definition
\ref{EBlossomDefn}. We omit them since they are a special case of
the $P_i$ 
trails defined below.

We will use
another definition of blossom
that is oriented to the structures discovered in the algorithm.
We always work in graphs $\o G$ where zero or more blossoms have been contracted.
So a vertex of $\o G$ is an atom (defined at the start of the section)
or a blossom.

\begin{figure}[th]
\centering
\input{earB.pstex_t}
 \caption{More blossom examples: (a) is a blossom with base $\beta$.
Adding (b) or (c) to 
Fig. \ref{EarFig}(b) gives a larger blossom.
}
 \label{EarBFig}
 \end{figure}

\begin{definition}[Algorithmic Blossom]
\label{ABlossomDefn}
Let $\o G$ be a graph derived from $G$ by contracting 
a family \A. of zero or more vertex-disjoint blossoms.
Let $C$ be a
closed trail in $\o G$ that starts and ends at 
a vertex $\alpha\in \o V( G)$.
  The preimage of $C$ is
a {\em blossom} $B$
with {\em base vertex} 
$\beta(B)$ if $C$ has the following properties:

\bigskip


If $\alpha$ is an atom then
$C$ starts and ends with edges of the same $M$-type.
$B$ has this   $M$-type and $\beta(B)=\alpha$.

If $\alpha\in \A.$ then 
$B$ has the same $M$-type as $\alpha$ and
$\beta(B)=\beta(\alpha)$.

\bigskip

If $v$ is an atom of $C$ then 
every 2 consecutive edges of $\delta(v, C) $ alternate.

If $v\in \A.\cap C$ then $d(v,C)=2$.
Furthermore 
if $v\ne \alpha$ then
$\delta(\beta(v),C)$ contains an edge of  opposite $M$-type from $v$.
\end{definition}

As before we abbreviate $\beta(B)$ to $\beta$ when possible.
Also $B$ is {\em heavy} ({\em light}) if its M-type is $M$ ($E-M$), respectively. 

\example 2. {The graph of Fig.\ref{EarFig}(a) can be parsed as a blossom
by starting with the triangle (a light blossom), enlarging it 
with the two incident edges (heavy blossom), and enlarging that
with its two incident edges (light blossom). An advantage over Definition
\ref{EBlossomDefn} is that each of these blossoms is a cycle rather than a closed trail. The algorithm will use this property.

Fig.\ref{EarBFig}(a) is a blossom. It can be decomposed
starting with the 5-cycle or starting with the triangle. If we replace
edge $e$ by edge $f$ in the matching, the triangle remains a blossom
but the overall graph does not. 

Suppose the graph of Fig.\ref{EarFig}(b) is enlarged by adding
the triangle of Fig.\ref{EarBFig}(b) at the vertex $x$. 
The graph is a blossom. A decomposition can start by contracting
the triangle. Alternatively it can  delay the triangle
 contraction until the end.
If we use  Definition \ref{EBlossomDefn}
we must delay the triangle contraction until the end.

Suppose instead that Fig.\ref{EarFig}(b) is enlarged by adding
the loop of Fig.\ref{EarBFig}(c) at $x$.
The graph remains a blossom using Definition \ref{ABlossomDefn},
since we can start by contracting the loop. This is the only possibility --
if we start by contracting  Fig.\ref{EarFig}(b) as in Example 1,
the loop disappears in the final contraction, 
so it is not part of the blossom.
So this graph is not a blossom using
 Definition \ref{EBlossomDefn}.
}

When all $b$-values are 1 the problem is ordinary matching and it is
easy to see Definition \ref{ABlossomDefn} is equivalent to ordinary matching blossoms.
We will show the two definitions of blossom are essentially equivalent.
The main difference is that they need not provide the same edge sets;
Fig.\ref{EarBFig}(c) gives the simplest of examples.
Instead we show they provide blossoms with the same vertex sets $V(B)$.
Strictly speaking the lemma is not needed in our development
since our algorithm only uses algorithmic blossoms.

Say  two blossoms are {\em equivalent} if
they have the same 
M-type,
base vertex and vertex set (the latter meaning $V(B)=V(B')$).
For instance the blossoms of 
Fig.\ref{EarFig}(b) and its enlargement with Fig.\ref{EarBFig}(c) 
are equivalent.

Let $\A.^*$ denote the family of blossoms involved in the recursive 
construction of $B$, i.e., $\A.^*$ consists of \A. plus the $\A.^*$ family
of every blossom $A\in \A.$. 
Define $\mu(B)=|\A.^*-\alpha^*|$,
the total number of steps in decompositions for all the blossoms
of $\A.-\alpha$. The next proof will induct on this quantity.

\begin{lemma}
\label{EarLemma}
The ear blossoms and algorithmic blossoms are equivalent families.
More precisely every 
ear blossom is 
an algorithmic blossom.
Every algorithmic blossom
has an equivalent ear
blossom.
\end{lemma}

\begin{proof}
Clearly we need only establish the second assertion.
So consider an algorithmic blossom $B$.
Let $\A., C$, and $\alpha$ be as in the definition.
We prove that  $B$ has an equivalent
ear blossom using 
by induction on $\mu(B)$.
The base case $\mu(B)=0$ corresponds directly to 
Definition \ref{EBlossomDefn}.

Take any algorithmic blossom $A\in \A.-\alpha$. 
Let $e$ be an
 edge of $\delta(\beta(A),C)$ with opposite M-type from $A$, let $f$
be the other edge of  $\delta(A,C)$, and let
$v$ be the end of $f$ in $V(A)$.
(If there are two possibilities for $e$ choose arbitrarily.)

\case {$v=\alpha(A)$ and $e$ and $f$ have the same M-type}
So $v=\beta(A)$.
In $C$ replace the contracted blossom $A$ by $C(A)$.
This gives an algorithmic blossom $B_1$ equivalent to $B$.
Since $\mu(B_1)<\mu(B)$  induction shows $B_1$ has an equivalent ear blossom.
This is the desired ear blossom equivalent to $B$.

\case {$v=\alpha(A)$ and $e$ and $f$ alternate, or
$\{v\}\pcon \alpha(A)$} 
In $C$ replace $A$ by
$\alpha(A)$. 
In both cases this gives an algorithmic blossom $B_1$.
Since $\mu(B_1)<\mu(B)$ it has
an equivalent
ear blossom $E_1$.
Contract it to $\o{E}_1$. The closed trail
consisting of   $\o{E}_1$ and $C(A)$ is an algorithmic blossom
with $\mu$-value $<\mu(B)$.
(This motivates the definition of $\mu$: Inducting on $|\A.^*|$
can fail here.)
By induction it has an equivalent ear blossom $E_2$.
$E_2$
is the desired ear blossom equivalent to $B$.

\case {$v\notin \alpha(A)$}
Choose an  occurrence of $v$ in $C(A)$. 
There is a corresponding partition of $C(A)$ into 2 
$v\alpha(A)$-trails, one starting with a matched edge, the other unmatched.
Let $P$ be the trail whose starting edge alternates with $f$, and let $Q$
be the other trail.
Replacing the contraction of $A$ by $P$ in
$C$ gives another alternating trail. This gives an algorithmic blossom,
with an equivalent ear blossom $E_1$.
Contract $E_1$
to a vertex $\o E_1$. We will form an algorithmic blossom equivalent to $B$
by adding $Q$ as follows. Let $Q_1$ be the subtrail of $Q$ starting with its first edge and ending with the first edge that enters $E_1$.
$Q_1$, which starts and ends with vertex $\o E_1$, 
is the closed trail of an algorithmic blossom. 
It has  an equivalent ear blossom $E_2$. 
If $V(Q-Q_1)\con V(E_2$ then $E_2$ is the desired ear blossom
equivalent to $B$. Otherwise continue in the same manner, defining $Q_2$
as the subtrail of $Q-Q_1$ starting with the first edge  that leaves
$E_2$ and ending with the first edge that enters it.
Eventually $Q$ is exhausted and we get the desired ear blossom.
\end{proof}

From now on we only use algorithmic blossoms.
Our next step is to show
that the $P_i$ trails defined above
exist in these blossoms. 
Observe that 
the definition of $P_i$ implies that
for a blossom $B$ with edge $f\in \delta(B)$ and $v$ the end of $f$ in
$B$, $P_i(v,\beta(B))$ alternates with $f$ iff 
\begin{equation}
\label{iForPEqn}
i=\begin{cases}
0&\text{$f$ and $B$ have the same M-type}\\
1&\text{$f$ and $B$ have opposite M-types.}
\end{cases}
\end{equation}
Clearly any trail $P_i(v,\beta(B))$ is required to contain
at least one edge unless $i=0$ and $v=\beta(B)$. 
We shall define
$P_0(\beta(B),\beta(B))$ to be the trail of no edges
$(\beta(B))$.
(Indeed it is the only possibility for this trail
in blossoms where all edges incident to $\beta(B)$ have the same
M-type.)  This choice satisfies \eqref{iForPEqn} vacuously.

We shall use the following property of the $P_i$ trails:
If $f$ alternates with $P_i(v,\beta(B))$ as in
\eqref{iForPEqn}, and $e\in \delta(\beta(B))$ is
of opposite M-type from $B$, then
 $f,P_i(v,\beta(B)),e$ is an alternating trail.
This is obvious for all cases, even when 
$P_i(v,\beta(B))$ has no edges.

\begin{lemma}
\label{PcdotLemma}
Trails $P_i(v,\beta)$, $i=0,1$ always exist (in an algorithmic
blossom).
\end{lemma}

\begin{proof}
The overall argument is by induction on $|V(B)|$.
Consider two possibilities for $v$.


\case {$v$ is an atom}
We will first specify
$\o P$, the image of the desired trail
$P_i(v,\beta)$ in $\o G$. Then we will enlarge $\o P$ to
$P_i(v,\beta)$ by specifying how it traverses the various
blossoms of \A..

If $v=\alpha$ there is nothing to prove for $i=0$ ($P_0(v,\beta)=(\beta)$).
If $i=1$ then $\o P$ is the entire trail $C$.

If $v\ne \alpha$ choose an arbitrary occurrence of
$v$ in $C$. The edges of $C$ can be partitioned into
2 $v\alpha$-trails in $\o G$,
one starting with a matched edge, the other unmatched.
Choose $\o P$ from among these trails so its starting edge
has the $M$-type of the desired trail $P_i(v,\beta)$.

To convert $\o P$ to the desired trail,
for every
contracted blossom $\o A$ in $\o P$ we will  enlarge $\o P$ by adding
a trail $Q$ that traverses  $V(A)$ correctly.
To do this let
$e$ and $f$ be the edges of $\o P$ incident to $\o A$,
 with $e\in \delta(\beta(A))$ of opposite M-type from $A$. (If there are 2 possibilities
for $e$ choose arbitrarily. If $\o A = \alpha$ then 
only $f$ exists, since $v\notin \alpha$.)
Let $u$ be the end of $f$ in $V(A)$. Let $Q$ be
the trail $P_j(u,\beta(A))$ 
of blossom $A$, with $j$ chosen so the trail alternates with $f$ at $u$,
specifically \eqref{iForPEqn} holds. 
(For example if $u=\beta(A)$ 
and $f$ has $M$-type opposite from $A$ then
$Q=P_1(\beta(A),\beta(A))$.)  
By definition $Q$  alternates 
with $e$ at its other end $\beta(A)$. So $Q$ (or its reverse)
is the desired trail traversing $V(A)$.
Enlarge $\o P$ with every such $Q$. Each
$\o A$ occurs only once in $B$, so no edge is added to 
$\o P$ more than once, i.e., the final enlarged $\o P$ is a trail.

\case {$v$ is a vertex of blossom $A\in \A.$} 
If $A=\alpha$ then $P_i(v,\beta)$ for
blossom $B$ is that trail as defined for  blossom $A$.

\def\xmod #1{\hbox{ (mod #1)}}

If $A\ne\alpha$ we  construct the desired trail as the concatenation
of two trails,
\begin{equation}
\label{PiPjPkEqn}
P_i(v,\beta)
=P_j(v,\beta(A)) P_k(\beta(A),\beta).
\end{equation}
Recall that a $P_i$ trail is alternating with length
congruent to $i \xmod 2$. 
Apply the argument of the previous case ($v$ atomic) 
to vertex $\beta(A)$, using  the edge of
$\delta(\beta(A),C)$ of opposite M-type from $A$ as the starting edge.
We get an alternating $\beta(A)\beta$ trail
of length congruent to $k \xmod 2$
for
\[k=\begin{cases}
0&\mbox{$A$ and $B$ have the same M-type}\\
1&\mbox{otherwise.}
\end{cases}
\]
Use this trail as
$P_k(\beta(A),\beta)$ in \eqref{PiPjPkEqn}.
Define 
\begin{equation}
\label{jkiEqn}
j= (k+i)\xmod 2.
\end{equation}
The
trail
$P_j(v,\beta(A))$ exists by induction.
Using it in
\eqref{PiPjPkEqn} gives  an alternating $v\beta$-trail.
(Note the special case $v=\beta(A), j=0, P_j(v,\beta(A))= (v)$.) 
The trail's  length is congruent to $j+k \equiv i \xmod 2$. 
Hence it qualifies as
$P_i(v,\beta)$.
\end{proof}

Paths $P_i$ are not  simple in general. However it is  easy to see the
above  proof  implies  any  trail $P_i(v,\beta)$  passes  through  any
$A\in  \A.^*$  at  most  once,  and if  so  it  traverses  some  trail
$P_j(v,\beta(A))$, $v\in V(A)$ (possibly in the reverse direction).

The blossoms $B$ in our algorithm have a bit more structure than
Definition \ref{ABlossomDefn}.  (Most of the following properties are the same
as
ordinary matching.)  As already mentioned a light blossom will always have
$d(\beta, \gamma(V(B),M))<b(\beta)$.  
(In contrast a general blossom may have $d(v, \gamma(V(B),M))=b(v)$ for
every $v\in V(B)$.)
Furthermore $C=C(B)$ will always be a
cycle.  Thus $P_i$-trails have a path-like structure, more precisely:
For any $v\in V(B)$, $P_i(v,\beta)\cap C$ is a path in $\o G$ (i.e.,
no repeated vertex) with one exception: When $\alpha$ is atomic (so
$\alpha=\beta$), $P_1(\beta,\beta)\cap C$ repeats vertex $\beta$. 
Repeated vertices present a  difficulty for expanding  a blossom --
an expanded blossom must be replaced in the search forest \os.  by a
path, not a trail.  The special structure of blossoms allows this to
be done, as shown in the Section \ref{bAlgSec}.

\subsection*{Augmenting trails}

\begin{figure}[th]
\centering
\input{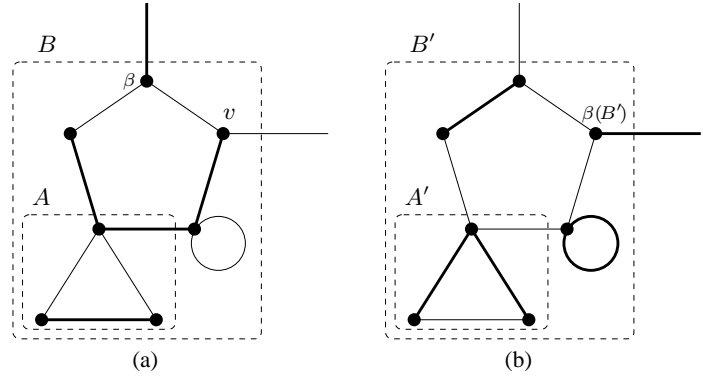}
 \caption{Rematching a trail.}
 \label{AugFig}
 \end{figure}

The algorithm enlarges a $b$-matching using a
generalization of the augmenting paths  of ordinary
matching. Throughout this subsection let $M$ be a partial $b$-matching.
An {\em augmenting trail} for $M$
is an alternating trail $A$ such that
$M\oplus A$ is a $b$-matching with one more edge than $M$.
To {\em augment $M$} we replace it by $M\oplus A$.
Our algorithm finds augmenting trails in blossoms as follows.

\def\av.{\mathy{\varepsilon}}

A vertex $v\in V(G)$  
 is {\em free}  if $d(v,M) \le b(v)-1$.  
Consider a multiset 
$\{v,v'\}$ of two free vertices, where $v=v'$ only if $d(v,M)\le b(v)-2$.
We will specify an augmenting trail with ends
$v$ and $v'$. Start by creating
an artificial  vertex \av..
Add an artificial matched edge from \av. to each of $v,v'$.
(So $v=v'$ implies the 2 matched edges are parallel.)
An {\em augmenting blossom}  for $v,v'$ is a blossom $AB$
with base vertex $\av.$;
furthermore each of $v,v'$ is either an atom
or the base of a light blossom in $\A.(AB)$.

The corresponding augmenting trail $A$ is
$P_0(v,\av.)$ with its last edge $(\av.,v')$ deleted.
Clearly $A$ is an alternating $vv'$-trail.
The two possibilities for each end $v,v'$ both ensure
the 
 first and last edges of $A$ are unmatched, i.e.,
$M\oplus A$ is a valid partial $b$-matching with
greater cardinality than $M$.
An {\em augment step} starts with an augmenting blossom and replaces
rematches the trail $P_0(v,\av.)$ as above.

This definition allows an end $v$ with
$d(v,M)\le b(v)-2$ to be the base of a light blossom $B$.
In this case 
$B$ itself gives an augmenting blossom and there is no need
for a larger blossom $AB$.
Our algorithm will
always use $B$, not a larger blossom.

In Fig.\ref{AugFig} an augment rematches the trail $P_0(v,\beta)$
through blossom $B$.
Blossom $A$ changes from light to heavy while $B$ remains light. 
The next lemma shows that in general blossoms
are preserved by the rematching of an augment step.

\begin{lemma}
\label{RematchBlossomLemma}
Consider  a graph with partial $b$-matching $M$.
Let $B$ be a blossom (not necessarily maximal) with base
vertex $\beta$.
Suppose a trail $P_i(v,\beta)$ through $B$ ($v\in V(B)$) is rematched.
The new matching $M'=M\oplus P_i(v,\beta)$
has a blossom $B'$ with the same subgraph as $B$ and base
vertex $\beta(B')=v$. 
The M-type of $B'$ is that of
the first edge of $P_i(v,\beta)$ in $M'$, unless $M'=M$.
\end{lemma}

\remarks {$M'=M$ only for the trail $P_0(\beta,\beta)$.
Obviously $B'$ is identical to $B$ in this case. 

Also note that 
the lemma implies
$B$ and $B'$ have the same M-type for $i=0$ 
(since $P_0(v,\beta)$ begins with an edge of opposite M-type
from $B$).
Similarly $B$ and $B'$ have the opposite M-types for $i=1$.}
 
\begin{proof}
By induction assume the lemma holds for blossoms smaller than $B$.


We will use the following notation.
$C(B)$ contains its end vertex  $\alpha$ and a vertex $\zeta$
that is either the atom $v$ or a blossom containing  
$v$. 
We will show the new blossom $B'$ has $\alpha(B')=\zeta$.
Thus $\alpha$ is an interior vertex of $C(B')$,
 unless $\zeta=\alpha$.
The other vertices of $C(B)$ remain interior vertices in $C(B')$.

For the main argument consider a vertex
$x\in V(\o G)$ on $C(B)$. Assume $x$ is on the portion of $C(B)$ 
that gets rematched else there is nothing to prove.
We will show that in $M'$, $x$ satisfies 
all the relevant conditions in Definition \ref{ABlossomDefn}
for $B'$ as well as the relevant conditions of the lemma.

First suppose $x$ is an atom.
Consider an occurrence of $x$ in $C(B)$. 
Two edges $e,f$ are associated with this occurrence --
$e$ and $f$ are either consecutive in $C(B)$ or the two ends
of $C(B)$
when $x=\beta$. The following two cases refer to the role 
$x$ plays in this occurrence.

\case{$x=\beta$}

\subcase{$x =v$} 
If $P_0(\beta,\beta)$ gets rematched then nothing changes in $B$.
If $P_1(\beta,\beta)$ gets rematched
then $B'$ has the M-type of the rematched $e$ and $f$.
In both cases $x=\beta(B')$ and all claims of the lemma hold.

\subcase{$x\ne v$} 
$e$ and $f$ have the same M-type in $M$.
Since $\beta=x\ne v$, $P_i(v,\beta)$ contains exactly 
one of $e,f$. So these edges alternate in $M'$.

\case{$x\ne \beta$}
$e$ and $f$ alternate in $M$.
If $x\ne v$ then both edges are in  
 $P_i(v,\beta)$ 
and so
they alternate in $M'$.

The remaining case is $x=v$. Exactly one
edge of the pair $e, f$ is on  $P_i(v,\beta)$, say
$e$. So $e$  gets rematched but $f$ does not. 
Thus $e$ and $f$ have the same M-type in $M'$, 
$x$ is the base of the blossom $B'$, and its M-type
is that of the rematched $e$.

\bigskip

Now assume $x$ is a contracted blossom. 
The inductive assumption shows $x$ is a blossom in $M'$, call it $x'$.

\case{$x\ne \zeta$} 
Let $P_i(v,\beta)$ traverse 
$x$ along the trail $P_j(u,\beta(x))$.
So there is an edge
$f\in C(B)\cap P_i(v,\beta)$
ending at $u$, where $j$ and $f$ satisfy \eqref{iForPEqn}.

First suppose $P_j(u,\beta(x))$ has at least one edge of $x$.
Its first edge $e$ has opposite M-type
from $f$. The inductive assertion shows $x'$ has
M-type that of the rematched $e$, which is opposite that
of  the rematched $f$. Since $f$ is incident to $u=\beta(x')$
Definition \ref{ABlossomDefn} is satisfied for  $x'$.

In the remaining case
no edge of subgraph $x$ is rematched, i.e., $f$ is incident to
$u=\beta(x')$ and 
has the same M-type as $x$
(in this case \eqref{iForPEqn} gives the trail
$P_0( \beta(x'),\beta(x') ) $).
The rematched $f$ has opposite M-type from $x=x'$. As before
Definition \ref{ABlossomDefn} is satisfied for  $x'$.

\case{$x=\zeta$} This implies $x'=\alpha(B')$.
Let $e$ and $f$ be the two edges of $C(B)$ incident to $x$.
In $B'$
there is no constraint on the M-types of $e$ and $f$.
$B'$ has the same M-type as $x'$. 

Suppose  $x$ contains an edge of  $P_i(v,\beta)$.
The inductive assertion for $x$
shows 
$B'$ has the M-type of the rematched first edge of $P_i(v,\beta)$.

Suppose  $x$ does not contain an edge of  $P_i(v,\beta)$.
Then $v=\beta(x)$ and  $P_i(v,\beta)$ uses the trail 
$P_0(v,v)$ through $x$.
If $x\ne \alpha$ then before rematching
the first edge of  $P_i(v,\beta)$
has opposite M-type from $x$. So after rematching
this edge has the same M-type as $x=x'$, as claimed in the lemma.
If $x=\alpha$ then no edge of $B$ is rematched, and again the lemma holds. 
\end{proof}



\subsection*{Mature blossoms}
We turn to the completeness property for blossoms.
Like ordinary matching 
the linear programming $z$-values will 
be positive only on
blossoms. However
complementary slackness requires that a blossom
$B$ with $z(B)>0$
has precisely one incident matched edge,
\begin{equation}
\label{CSforBMatchingEqn}
b(V(B))= 2|\gamma(V(B),M)|+1.
\end{equation}
(This principle is reviewed in Appendix \ref{bfAppendix}.)
For $B$ a light blossom this matched edge will be incident to $\beta(B)$.
(This is required by the definition of blossom
if $B$ is not maximal. More generally as previously mentioned 
our search algorithm always works with light blossoms
that are either free or have this matched edge.)
A heavy blossom $B$ may have exactly one incident
matched edge, but this is irrelevant in our algorithm.
(Heavy blossoms created in a search are
immediately absorbed into a light blossom
(lines 1--2 of Fig. \ref{BMAlg} and line 5 of Fig. \ref{ExpandAlg}).
Heavy blossoms created in an augment 
as in Fig.\ref{AugFig} are ``hidden'', as shown below.)
This motivates the following definition.

For any vertex $v$  with  $b(v)\ge 1$ define  the function
$b_v$ by decreasing $b(v)$ by 1,
keeping all other values the same. 
A blossom based at $\beta$ is {\em mature} if 
$\gamma(V(B),M)$ is a $b_\beta$-matching.

Our search algorithm will extend a blossom to make it mature before 
any dual adjustment makes its $z$-value positive (see Lemma \ref{DualsMatureLemma}).

Now consider the transition
from one search to the next.
First  consider an augment step. 
Blossoms with positive dual values must remain mature.
This is guaranteed by the next lemma, illustrated by 
$B$ in Fig.\ref{AugFig} assuming it is mature.

\begin{lemma}
\label{LMBlossomsLemma}
Let $B$ be a blossom $B$, not necessarily maximal.
If $B$ starts out 
light and mature 
and gets rematched in an augmentation, it remains light and mature.
\end{lemma}

\begin{proof}
Let $A$ be  the augmenting trail and consider the trail
$A'=A+(\av.,v)+(\av.,v')$.
$A'$ contains exactly two edges incident to $B$,
at least one of which is incident to
$\beta(B)$ and matched. 
(If $B$ contains $v$ then $v=\beta(B)$, since $B$ is mature,
so $(\av.,v)$ is the claimed edge. Similarly for $v'$.)
Let $f$ be the other edge.
Since $B$ is mature $f$
is unmatched. 
Thus Lemma \ref{RematchBlossomLemma} shows the rematched
blossom is light and mature. 
\end{proof}

Now consider the contracted graph $\o G$
immediately
after the matching is augmented.
Some maximal blossoms may be immature
(discovery of the augmenting path
prevents these blossoms from growing to maturity).
Such
immature blossoms $B$
should be discarded.
(The contracted blossom $B$ is incident to $>1$  matched edge and complicates
the growth of a search tree, as discussed in Fig.\ref{AlgExFig}(a) below.) 
To discard $B$, we replace its contraction
in the current graph $\o G$
by 
the
atoms and contracted blossoms of $C(B)$ and their incident edges. 
So after augmenting the matching
and before proceeding to the next search, the algorithm
does the following {\em  discard} step:

\b

Repeatedly discard a maximal blossom  unless it is light and mature.

\b

\noindent
At the end of the discard step every maximal blossom is light and mature
(and still contracted).
There can still be blossoms that are immature
and/or heavy, like
$A$ and $A'$ in Fig.\ref{AugFig}, but they are not maximal and so 
they are ``hidden'',
essentially irrelevant to the algorithm.

\subsection*{Data structure for blossoms}
\label{BlossomDataStructureSection}
Blossoms are represented using 
a data structure similar to ordinary matching, with two new additions.
First we review the previous data structure:
Each maximal blossom
$B^*$ has a tree $T(B^*)$.
Every blossom $B\con B^*$ has an interior node in $T(B^*)$; each
vertex of $V(B)$ has a leaf.
The children of an interior node $B$  
are the nodes for the vertices of $C(B)$
ordered as in that trail; there is only one node for
$\alpha(B)$, even though it occurs both first and last in the trail.
The children are doubly linked.
Each link records the corresponding edge of $C(B)$, represented by
its ends in $V(G)$.
The tree $T(B^*)$ has parent pointers.

We add two new values to facilitate construction of the $P_i$ trails:
Each blossom $B$ records its base vertex $\beta(B)$. Also any vertex
$\beta\ne \beta(B^*)$
that is the base vertex of one or more blossoms $B$
records an edge denoted $\eta(B)$.
$\eta(B)$ is the edge incident to the base of $B$
required by Definition \ref{ABlossomDefn}. 
Specifically $\eta(B)$ is the edge of $\delta(\beta,C(A))$ 
of opposite M-type from $B$,
for $A$  the first ancestor of $\beta$ 
where $\beta(A)\ne \beta$.
(The notation $\eta(B)$ comes from our discussion of  $f$-factors, 
Definition \ref{EtaDefn}.)

As with ordinary matching this data structure has size $O(|V(B^*)|)$.
The proof is similar: The
leaves of  $T(B^*)$ form the set $V(B^*)$.
Any interior node has at least 2 children unless it
is the parent of a leaf (i.e., it represents a loop blossom).
So there are $\le 2|V(B^*)|$ interior nodes.
Note that for  general blossoms $B$ where atoms may occur multiple times
in closed trails $C(B)$ this quantity is  
$|V(B^*)|=O(b(V))$.
For our algorithm where atoms occur just once, $|V(B^*)|=O(n)$.

The M-type of a blossom $B$ can be determined in $O(1)$ time:
Any $B$ with $\beta(B)\ne\beta(B^*)$ has M-type
the opposite that of $\eta(B)$.
More generally any $B$
has M-type that of
 the two edges of
$\delta(\beta(B),B_0)$, for $B_0$ 
the minimal blossom
containing $\beta(B)$. These edges are found as the two edges incident
to the node for the parent of leaf   $\beta(B)$
in $T(B^*)$.

The $P_i$ trails are computed by the algorithm for ordinary matching
extended to allow two types of trails 
($P_0,P_1$) and two types of blossoms (heavy and light).
We will describe the extension, relying on the original description
(Section \ref{BlossomSec})
and its notation  as much as possible.
The extended routine to compute  $P_i(x,\beta(B))$ 
uses the original 4 arguments plus 
the value $i\in \{0,1\}$.
As before let the blossoms/atoms on the trail 
$P_i(x,\beta(B))\cap C(B)$ 
be $B_j, 0\le j\le k$. We will describe 
the case $r=0$ and $ B_0\ne B_k$.
The other cases are similar.

As before 
we follow links starting at $B_0$ to get the desired edges of
$C(B)$. 
%
%
The edge leaving $B_0$ is determined as follows:
If $B_0$ is an atom, choose it to have opposite M-type from
$B$ iff $i=0$. (Note the M-type of $B$ is available using
its base, the argument $\beta(B)$.)
If $B_0$ is a blossom choose edge $\eta(B_0)$.
The last blossom/atom $B_k$ is determined 
as either 
the blossom with base vertex $\beta(B)$
or the atom $\beta(B)$.

If $B_j$ is a blossom it engenders a recursive call
(there is no such call for an atom). 
To specify it
let the link traversal enter $B_j$ on the link for edge $e$ and leave it on $f$
($e=\emptyset$ for $j=0$, $f=\emptyset$ for $j=k$).
Define edges $e',f'$ so $\{e',f'\}=\{e,f\}$ and 
$f'=\eta(B_j)$.
(For $j=0$, only define
$f'=\eta(B_j)$. For $j=k$ only define
$e'=e$.)
The recursive call for $B_j$ 
finds the trail $P_{i'}(x',\beta(B_j))$ with reversal bit $r'$, 
where 
$i',x',$ and $r'$ are determined as follows:
\[\begin{array}{l}
i'=\begin{cases}
i&\hskip 28pt \text{$j=0$, $B_0$ and $B$ have the same M-type}\\
1-i&\hskip 28pt \text{$j=0$, $B_0$ and $B$ have opposite M-types}\\
0&\hskip 28pt \text{$e'$ and $B_j$ have the same M-type}\\
1&\hskip 28pt \text{$e'$ and $B_j$ have opposite M-types}
\end{cases}\\
x'=\begin{cases}
x&j=0\\
e'\cap V(B_j)&\text{otherwise}
\end{cases}\\
r'=\begin{cases}
0&\hskip 46pt j\in \{0,k\} \text{ or } f'=f\\
1&\hskip 46pt \text{otherwise}
\end{cases}
\end{array}
\]

As before the time for this procedure is $O(|V(B)|)$, which we have noted is
$O(n)$ 
in our algorithm.
As before this bound 
can be improved to the output size $O(|P_i(x,\beta(B))|)$
by careful path-tracing.
Specifically we 
augment the data structure so each edge $\eta(B)$
has a pointer to the corresponding link in $T(B^*)$. 
If $x$ is the base vertex $\beta(B)$ of one or more blossoms $B$
the path-tracing 
for $x$ starts at the link for $\eta(B)$.

\subsection{$b$-matching algorithm}
\label{bAlgSec} 

The most important differences of the $b$-matching algorithm
from ordinary matching are that
the search forest need not alternate at outer blossoms, 
a blossom step may create 2 blossoms (to hide a heavy blossom), 
and the Expand Step is more 
involved (to keep the search forest acyclic). 
We begin with terminology for the algorithm.

$\o G$ denotes the graph with all blossoms contracted; 
\oE. denotes its edge set.
\S. denotes the search structure and \os. is that structure in $\o G$.
Recall that in $b$-matching
a vertex of $V$ is not
a blossom, so each vertex of $\o G$ is  an atom or a 
contracted blossom but not both.
 The notation $B_x$ denotes the vertex of  $\o G$
containing $x\in V$; if $x$ is atomic then $B_x=x$.
If $d(x,M)<b(x)$ then 
$B_x$ is {\em free}. (If $B_x$ is a blossom we 
shall see that $d(x,M)<b(x)$ implies $x$ is the base vertex of $B_x$.) 
The roots of the forest \os. are the free atoms and free blossoms.

Let $v$ be a node of \os..
$v$ is {\em inner} if it is joined to its parent by an unmatched edge.
Otherwise 
(i.e.,
$v$ is joined
to its parent by a matched edge, or 
$v$ is a search tree root)
$v$  is {\em outer}.
We refrain from classifying vertices contained in a blossom.
(A vertex in an outer blossom can function as both inner and outer,
because of its $P_i$-trails.)

As before an edge is {\em tight}  if it satisfies the LP
complementary slackness conditions with equality
(see Appendix \ref{bfAppendix}). Again as before
we shall see that every matched edge is tight.
The following notion identifies the edges that can be used to
modify the search forest when they are tight.
An edge $e=xy\in \oE. -\os.$ is {\em eligible for $B_x$}
if any of the following conditions holds:

{\hi

$x$ is an outer atom and $e\notin M$;

$B_x$ is an outer blossom;

$B_x$ is  an inner node and $e\in M$.

}

The algorithm uses the following conventions:

{\narrower

{\parindent=17pt \narrower 

{\parindent=0pt

\def\bigskip{}

\bigskip

$M$ denotes the  current matching on $G$. 

\bigskip

For any edge $e$, $e'$ denotes a new unmatched copy of $e$.
$e'$ always exists in $b$-matching.

\bigskip

For related nodes  $x,y$ in \os. (i.e., one of $x,y$ descends from
the other)
$\os.(x, y)$ denotes the \os.-path from $x$ to $y$.

\bigskip

}

}}

\begin{algorithm}

\DontPrintSemicolon
\SetKwFunction{Expand}{Expand}

make every free atom or blossom an (outer) root of \os.\;

\KwSty{loop}

\Indp

\If{$\exists$ tight edge $e=xy\in \oE.$ eligible for $B_x$ with $y\notin \S.$} 
{ \tcc*[h]{grow step}\;
 add $e, B_y$ to \S.\;
}
\ElseIf{$\exists$ tight edge $e=xy\in \oE.$, eligible for both $B_x$ and $B_y$}
{\tcc{blossom step}
\If {$B_x$ and $B_y$ are in different search trees}
{\tcc{$e$ plus the \os.-paths to $B_x$ and  $B_y$ give
an augmenting blossom $B$}
augment $M$ using $B$, and end the search\;}
\lnl{B0}$\alpha\gets$ the nca of $B_x$ and $B_y$ in \os.\;
$C\gets$ the cycle $\os.(\alpha, B_x), e, \os. (B_y, \alpha)$ \;
\If(\tcc*[h]{$\alpha$ is atomic})
{$\alpha$ is an inner node of \os.} 
{\lnl{B1}contract  $C$ to a heavy blossom \tcc*[h]{$C$ is the new $B_x$}\;
$f \gets$  the unmatched edge of \S. incident to $\alpha$\;
$\alpha \gets$ the outer node of \os. on $f$\;
$C\gets $ the closed cycle $f,B_x,f'$}
\If(\tcc*[h]{$\alpha$ is a search tree root})
{$\alpha$ is atomic and $d(\alpha,M)\le b(\alpha)-2$}
{augment $M$ using  blossom $C$, and end the search\;}
\lnl{B2}contract  $C$ to an outer blossom \tcc*[h]{$C$ is the new $B_x$}\;
}

\SetKw{true}{true}\SetKw{false}{false}\SetKw{nt}{not}
\SetKw{myif}{if}\SetKw{mythen}{then}\SetKw{myelse}{else}

\ElseIf{$\exists$ inner blossom $B$ with $z(B)=0$
\tcc*[h]{$B$ is mature}}
{\tcc{expand step}
define edges $e\in\delta(B, \os.-M)$, $f\in\delta(B,M)$
\tcc*[h]{$f$ needn't be in \os.}\;
let $P_0$ be the trail $e,P_0(v,\beta(B)),f$, where $v=e\cap V(B)$,
$\beta(B)=f\cap V(B)$\;
remove $B$ from \S. \tcc*[h]{\os. is now invalid}\;
\Expand$(B,e,f)$ \tcc{enlarge \os. by a path
through $B$ to $f$}
}
\lElse {adjust duals}

\Indm

\caption{Pseudocode for a $b$-matching search.}
\label{BMAlg}
\end{algorithm}

\begin{procedure}
\DontPrintSemicolon
\KwSty{Procedure} \FuncSty{Expand}($B,e,f$)\;

\tcc{
$B$: a blossom formed in a previous search, or an atom.\;
\goin \hskip27pt $B$ is a mature blossom in the initial invocation\;
\goin \hskip27pt  but may later be immature\;
\goin $e$: edge entering $B$, already in  \os.\;
\goin $f$: edge leaving $B$\;
\goin $\beta$: base vertex of $B$  if $B$ is a blossom\;
\BlankLine
\goin \Expand enlarges \os. by adding a path of edges of $B$
 that joins $e$ to $f$\; 
}

\If{$B$ a blossom with $z(B)>0$ or $B$ an atom
\tcc*[h]{$e$ \& $f$ alternate}}
{\lnl{B3} make $B$ a node of \os.\;
}
\ElseIf{$B$ light and $e\in \delta(\beta,M)$ \tcc*[h]{e.g., $B$ light 
and $P_0$ contains $P_1(\beta,\beta)$}}
{\lnl{B4}  make $B$ an outer node of \os.\;
}
\ElseIf{$e,f\in \delta(\beta,E-M)$}
{\tcc*[h]{$B$ is heavy and $P_0$ contains trail $P_1(\beta,\beta)$}\;
let $e=u\beta$\;
let $C$ be the  length 2 cycle  $B_{u},e,B,e',B_{u}$\;
\If{$d(u,M)\le b(u)-2$ \tcc*[h]{$u$ is a search tree root}}
{augment $M$ along  $C$ and end the search\;}
let $B'$ be the blossom defined by $C(B')=C$\;
{\lnl{B5}replace $e$ in \os. by outer node $B'$\;
}}
\Else{
\lnl{L6}let $P=(e_1,B_1,e_2,B_2,\ldots, e_k,B_k,e_{k+1})$
be the trail $E(P_0)\cap (C(B)\cup \{e,f\})$\;
\tcc{$e_1=e,\ e_{k+1}=f,\ B_i \text{ a contracted blossom or 
an atom}$\;
\goin{$P$ is a path}\;
}
\For {$i=1$ \KwTo $k$}{\lIf{$i>1$}{add $e_i$ to \os.};
\Expand($B_i,e_i,e_{i+1}$)}
\For {every blossom $B'$ of $C(B)-P$}
{repeatedly discard a maximal blossom in $V(B')$ unless it is light and 
mature}
}

\BlankLine
\caption{Expand($B,e,f$) for $b$-matching blossoms $B$.} %
\label{ExpandAlg}
\end{procedure}

The algorithm is presented in Fig.\ref{BMAlg}--\ref{ExpandAlg}.
The next  three subsections
clarify how it works as follows.
First we give some simple remarks.
Then we state the invariants of the algorithm (which are proved in the
analysis). Lastly we give examples of the execution.
These three subsections 
should be regarded 
as commentary -- the
formal proof of correctness and time bound
is presented in Section \ref{bMAnalSec}.


\paragraph*{Remarks}
The grow step adds only one edge, unlike ordinary matching.
One reason is that an inner vertex may have an arbitrary number
of children: The possibility of $>1$ child comes from an inner atom
on many matched edges. The possibility of no children arises when
a new inner vertex has all its matched edges leading to vertices that
are already outer.

A second reason comes from the definition of eligibility. It allows
an outer vertex to have a child using a matched edge. So an outer
vertex may have an arbitrary number of children, using matched or unmatched edges. This also shows that the search forest need not alternate the same way as
ordinary matching.

In the blossom step
the test $e\in \oE.$
is equivalent to the condition $B_x\ne B_y$ or $B_x=B_y$ is atomic.
The second alternative
allows a blossom whose circuit is a loop.
This can occur for an atom that is either inner with
a matched loop or outer with a tight loop.
(Loop blossoms do not occur for blossoms $B_x$ since our  contraction operation
discards loops.) 

The contraction of line \ref{B2} creates an outer blossom that is light. 
When a
heavy blossom is  created in line \ref{B1} it gets absorbed
in the light blossom of line \ref{B2}. We shall see this is the only
way a heavy blossom occurs in the search algorithm.

In the expand step note that edge $f$ may or may  not be in 
\os. (as mentioned above for the grow step).
This motivates the structure of {\tt Expand}$(B,e,f)$, which
assumes on entry that $e$ is an edge of \os. but makes no assumption 
on $f$.
Also the trail $P_0$ is used for succinctness.
An efficient implementation is described in
the last subsection of Section \ref{bMAnalSec}.

If the duals are adjusted (last line of
Fig.\ref{BMAlg})
our assumption that the graph has a (perfect) 
$b$-matching guarantees the new duals allow further progress
(i.e., a grow, blossom, or expand step can be done; this is proved 
in Lemma \ref{FailedLemma} and the discussion following it).

A tight edge $xy$ with $B_x\in \os.$ is ignored by the algorithm
in 2 cases: $B_x$ an outer atom with $xy\in M$, and
$B_x$  inner
with $xy\notin M$.

In procedure {\tt Expand} the last case (starting at line \ref{L6}) corresponds
roughly to the expand step for ordinary matching. The purpose of the
preceding cases
is to eliminate repeated vertices in  $P_0$, which of course cannot be added
to the search forest \S..{}

\paragraph*{Invariants}
The first several invariants describe the topology of \os..
As before say  that  \os. {\em alternates}
at an \os.-node $v$ if any edge to a child of $v$ has opposite $M$-type
from the edge to the parent of $v$; if $v$ is a root then any edge to  a child
is unmatched. We can treat a root $v$ as in the general case by using
an artificial vertex
$AV$ as its parent (as done for augmenting trails): 
$AV$ has an artificial matched edge
to each atomic root as well as  the base vertex of each blossom root. 

\bigskip

{\parindent=0pt

(I1) \hskip20pt $\os.$ alternates at any node that is not an
outer blossom.

\bigskip

(I2) \hskip20pt Let $B$ be a maximal
blossom with base vertex $\beta$. 
$B$ is light.

\b

\hskip42pt If $B$ is inner or not in \S. then it is mature, and $\beta$ is on a matched edge incident to $B$.

\b

\hskip42pt
If $B$ is  outer then 
every vertex $x\in V(B)$
has $d(x,M)=b(x)$ unless  $B$ is a root of \os., $x=\beta$, and
$d(x,M)=b(x)-1$. If $B$ is a nonroot of \os. then 
$\beta$ is on the matched edge leading to the parent of $B$.

\bigskip

(I3)\hskip20pt
For every blossom $B$ (maximal or not) $C(B)$ is a cycle.

\bigskip

}

In detail (I1) means the following:
Any child of an inner \os.-node is outer.
Any child of an outer atom is inner (this includes the case of an atomic
 search
tree root).
A child of an outer blossom may be outer or inner.
Note that the first of these properties implies
the parent of an inner node is outer.
Equivalently, \os. alternates at both ends of an unmatched edge in \os..

Note that (I2) holds for a blossom that is maximal at any point in time.

The remaining invariants deal with the dual variables.
(The duals for $b$-matching are reviewed in Appendix \ref{bfAppendix}.)

\b

{\parindent=0pt

(I4)\hskip20pt An edge is tight if 
it is matched,
or it is an edge of \os., or it is an edge of a contracted blossom. 

\b

(I5)\hskip20pt A blossom with $z(B)>0$ is light and mature.

}

\b

Note that in (I5)  blossom $B$ need not be maximal.

\begin{figure}[th]
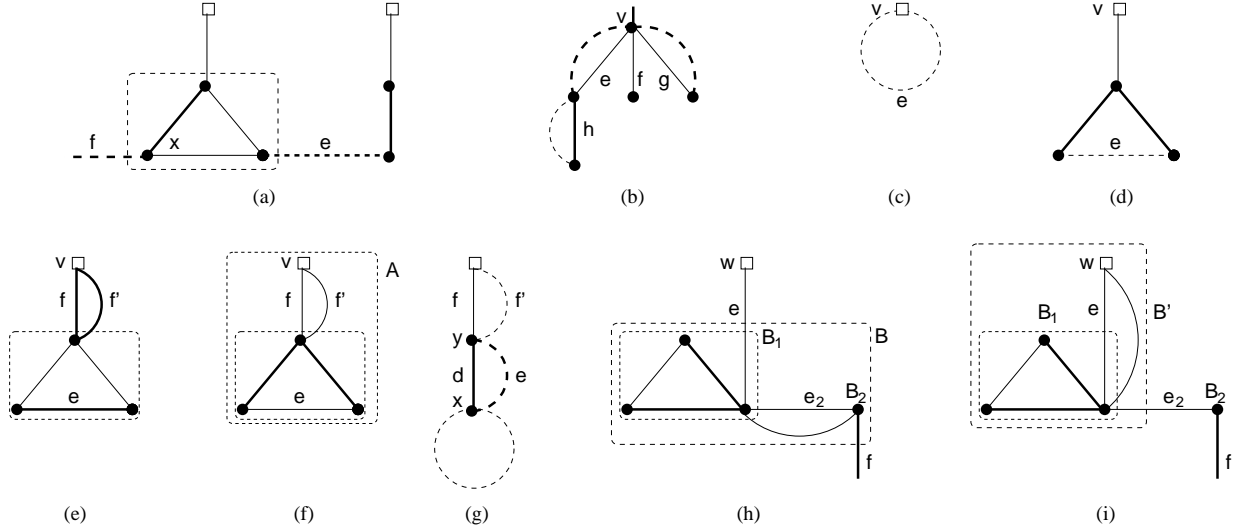

\centering
\input{AlgEx.pstex_t}

\bigskip

\input{AlgEx3.pstex_t}
 \caption{Algorithm examples.}
 \label{AlgExFig}
 \end{figure}

\paragraph*{Examples}
{\em Grow Step}: 
In Fig.\ref{AlgExFig}(a) the triangular blossom 
$B_x$ is inner.
Its
incident matched edge $e$ leads to a node that is already
outer.
So unlike ordinary matching
$B_x$ has  no corresponding
grow step. $B_x$ also illustrates the possibility of
an expand step
where the inner blossom is a leaf (unlike ordinary matching).

(I2) requires
the inner blossom $B_x$ to be mature. Suppose
it was
immature. A matched edge like $f$ incident to $B_x$ might exist.
A grow step for $f$ would be invalid, since \S. does not
contain an alternating trail from $f$ 
to a free vertex.
Requiring inner blossoms to be mature
avoids this complication.

\b
\noindent
{\em Blossom Step}: 
In Fig.\ref{AlgExFig}(b)  $v$ is an outer vertex.
An atomic $v$ is the root of the
alternating subtree of edges $e,f,g,h$. If $v$ is a blossom
a blossom step can be done for the matched copy of $e$, as
well as
the matched copy of $g$.
The unmatched copy of $h$ is necessarily tight so a blossom step
can be done for it. These blossom steps illustrate 
how outer blossoms can be  enlarged to become mature.
(Edge $e$ of Fig.\ref{AlgExFig}(g) illustrates another case.)

In Fig.\ref{AlgExFig}(c) the unmatched loop $e$ may not be tight.
If it becomes tight it forms an augmenting blossom
if the free vertex $v$ lacks at least 2 matched edges.
The algorithm can repeatedly match
copies of $e$ as long as this condition holds.
If eventually $v$ lacks 1 matched edge another copy of $e$
forms a light blossom.

Fig.\ref{AlgExFig}(d) shows  a search tree when
a blossom step for $e$ is discovered. 
The triangle is made a heavy blossom.
If $v$ lacks at least two edges  the matching is augmented as in
Fig.\ref{AlgExFig}(e). The discard step  then abandons the triangle blossom.
(Edges $f,f'$ show the blossom is immature.)
If $v$ lacks only one  edge the light blossom $A$ of 
Fig.\ref{AlgExFig}(f) is formed.
This illustrates that the algorithm never creates 
heavy blossoms that are maximal.

In Fig.\ref{AlgExFig}(g) a blossom step creates the loop
blossom $B_x$. Edge $e$ is a matched copy of the search tree edge
$d$.
A blossom step may be done for $e$, $B_x$ and $y$.
($y$ is necessarily an atom because of $d,e$.)
The unmatched edges $f,f'$ give an
augmenting trail or
 blossom as in
Fig.\ref{AlgExFig}(e)--(f).

\b
\noindent {\em Expand Step}:
In Fig.\ref{AlgExFig}(h)  blossom $B$ is inner and mature. (As an example of
how this occurs, 
blossom $B$ and  atom $B_2$ may be, respectively,
blossom $A$ of Fig.\ref{AlgExFig}(f)
and its base vertex $v$. In
 Fig.\ref{AlgExFig}(f)
blossom $A$ 
gets positive $z$ when duals are adjusted. An augment is done using an 
unmatched edge
of $\delta(v)\cap \delta(A)$, so $A$ does not change. 
$v$ is no longer free.
$A$ then becomes inner in a grow step from the free vertex $w$
of 
  Fig.\ref{AlgExFig}(h).) A dual adjustment makes $z(B)=0$ and $B$
is expanded. Line \ref{L6} generates
two recursive calls, with arguments $e,B_1,e_2$ and
$e_2,B_2,f$. 
The first recursive call
either
augments the matching as in
Fig.\ref{AlgExFig}(e) or forms a blossom as in
Fig.\ref{AlgExFig}(f). In the second case 
Fig.\ref{AlgExFig}(i) gives
the result of the entire expand step.

Finally suppose in Fig.\ref{AlgExFig}(h)  blossom $B$ is not maximal:
It is included in a blossom $B_3$ where
$\alpha(B_3)=B$, $C(B_3)$ is
a length 2 closed trail containing
$B$ and an atom $x$, with 2 copies of edge
$vx$, say
$h$ and $i$ where $h\in M$, $i\notin M$.
The initial call to Expand issues one recursive call
with arguments $e,B,f$.
The rest of the expand step is as before. 
It results in
Fig.\ref{AlgExFig}(i) with 
$x$ no longer in a blossom.
A
grow step for $x$ is now possible using the matched
edge $h$. 
Such grow steps -- involving vertices contained in a blossom that
has been expanded -- are not possible in ordinary matching.
They motivate
 the structure
of  the grow step
in Fig.\ref{BMAlg}.

\subsection{Analysis} 
\label{bMAnalSec}
The analysis is presented in three subsections.
First we prove the invariants. Then we prove the algorithm is correct,
adding some details about initialization and termination.
Lastly we prove the desired time bound, adding some implementation details 
for
dual variables.

\subsection*{Proof of invariants}
We   show that
all the invariants are preserved by every step of the algorithm.
The first four lemmas treat
grow steps, blossom steps,
augment steps, and expand steps, respectively.
(The lemmas
for the
blossom step and 
the expand step assume 
any augment is done correctly; 
the augment step lemma treats the detailed
augment and discard steps of Section \ref{bBlossomSec}.)
Then we check  dual adjustment.
Many details  are straightforward 
so we discuss just the most interesting and representative cases.


\begin{lemma}
The grow step preserves all the invariants.
\end{lemma}

\begin{proof}
We will verify the case $e\in M$ and $B_y$ a blossom.

$B_y$ starts out not in \S., so  (I2) shows it is mature and $\beta$ is on
its unique
incident matched edge. Thus (I2) holds when $B_y$ becomes an outer blossom.
(I4) shows $e$ is tight, so adding it to \os. preserves (I4).
(I3) and (I5) are unchanged.
\end{proof}

\begin{lemma}
The blossom step preserves all the invariants.
\end{lemma}

\begin{proof}
We will verify the case of
a blossom step that starts by making
$\alpha$ in line \ref{B0} an outer node.
Line \ref{B2} makes $C$ an outer blossom. 
(The other cases are similar to this one.)

(I1) clearly holds trivially, as do
(I4) and (I5). We check (I2), (I3),
and the fact that $C$ is a valid outer blossom, as follows:

\b
{\parindent=0pt

{\em (I3)}: $C$ is the fundamental cycle of $e$ in the forest \os..

\b

{\em (I2)}: We check (I2) for the outer blossom $C$.

{\hi 

\case{$\alpha$ is an atom} 
This makes $\alpha$ the base vertex $\beta$ of the blossom $C$.
To show $C$ is light observe that at least one of the two edges
of $\delta(\alpha,C)$ goes to a child of $\alpha$. (I1) shows 
the edges at $\alpha$ alternate, so that edge is unmatched.
Thus $C$ is light.

Since $\alpha$ is outer, so is $C$.
The test preceding line \ref{B2} ensures
$d(x,M)\ge b(x)-1$ for $x=\alpha=\beta$;
$d(x,M)=b(x)$ 
for the remaining vertices of $V(C)$ since
no atom or blossom of $C-\alpha$ is free.
If $\alpha$ is a root of \os. clearly it has
$d(x,M)=b(x)-1$.
If $\alpha$ is not a root it 
gives the desired matched edge to the parent of $C$.

\case{$\alpha$ is a blossom} (I2) shows $\alpha$ is light, so $C$ is light.
The other properties follow the atomic case.
}

\b

{\em $C$ satisfies Definition \ref{ABlossomDefn} of a  blossom}:  
Note that $\alpha$ and $C$ of the definition
are the same as $\alpha$ and $C$ of the algorithm.
The conditions to verify are for the
3 types of vertices $v$ of cycle $C$:

{
\case {$v$ is an atom and $v=\alpha$} An
edge $e\in \delta(\alpha,C)$ either goes to 
a child of $\alpha$ or $e$ is the blossom edge joining
$B_x$ and $B_y$. In both cases $e$ is eligible for the
outer atom $\alpha$, so $e\notin M$. 
Thus the first and last edges of $C$ have the same M-type.

\case {$v$ is an atom, $v \ne \alpha$}
If $v\notin \{B_x,B_y\}$ then (I1) shows 
the 2 edges  of $\delta(v,C)$ alternate.
If $v\in \{B_x,B_y\}$ then 
one edge of $\delta(v,C)$ goes to its parent and the other
is eligible for $v$.
The two cases of eligiblity for an atom show the edges of
$\delta(v,C)$ alternate.

\case{$v$ is a contracted blossom}
$C$ is a cycle so $d(v,C)=2$.
The rest of the verification concerns the case $v\ne \alpha$.
Let $\beta$ be the base vertex of $v$.
(I2) shows $v$ is light, so we must show $\beta$ is on a matched edge of $C$.
If $v$ is outer (I2) shows 
the matched edge $f$ going to $v$'s parent 
is incident to $\beta$. Furthermore $f\in C$.

{\hi
If $v$ is inner first assume $v\notin \{B_x,B_y\}$. 
(I2) implies $\beta$ is on a matched edge $f$ going to $v$'s
unique child, and $f\in C$.
Now suppose $v\in \{B_x,B_y\}$.
Edge  $e$ of the blossom step is eligible for $v$, so by definition
$e\in M$.
(I2) shows $e$ is  
incident to $\beta$, so
$e$ is the desired matched edge for blossom $v$.  
}}}
\end{proof}

In the next two lemmas some invariants are 
explicitly violated during the execution of a step. However we only require
that all the invariants are satisfied at the end of the step,
assuming they hold at the start of the step.

\begin{lemma}
The augment and discard steps preserve all the invariants.
\end{lemma}

\begin{proof}
Consider the rematching of the augment step. It
preserves (I5) by Lemma \ref{LMBlossomsLemma}.
(I4) holds since the rematching is done along the trail
$P_0(v,\epsilon)$
of the augmenting blossom.
(I3) is unchanged. (I1) becomes irrelevant after the rematching,
since $\os.$ is empty for the next search.

The rematching can violate (I2), since
a maximal blossom that is
incident to 2 matched edges on the augmenting trail
becomes heavy. But
this violation is only temporary since the discard step eliminates any 
maximal blossom that is not light and mature. Thus
(I2) holds at the start of the next search.
The discard preserves (I4), since (I5) shows no blossom with positive
$z$ is eliminated, and thus the set of tight edges does not change.
\end{proof}

\def\de{\dot{e}} \def\df{\dot{f}}

We turn to the expand step of Fig.\ref{BMAlg}. 
Let {\tt Expand}$(B,\de,\df)$ denote the call made in the expand step.
Recall that \os. is defined as a forest rooted at the free atoms and
blossoms. 
When the expand step removes $B$ from  \S., 
this definition and (I1)
will be violated if 
\os. has a nonempty subtree descending from $\df$. 
(Unlike ordinary matching this subtree needn't exist, e.g., Fig.\ref{AlgExFig}(a).)

\def\I1p.{(I1$'$)}

To remedy this we modify (I1) to 
a condition \I1p. defined
as follows.
Define \oP. to be the set of edges of the trail $P_0$
(defined in the expand step) that have been added to \os. so far.
Say that a trail in \os. has {\em permissible alternation} if
it  alternates at any node that is not an outer blossom.

\b

{\parindent=0pt

\I1p.\hskip20pt
If on entry to {\tt Expand}$(B,e,f)$
\oP. is a trail 
that has permissible alternation and has
$e$ as its last edge,
then on exit  $\oP.+f$
has permissible alternation and has
$f$ as its last edge.
}

\b

\noindent
Note that in general 
the exit trail \oP. is an extension of the entry trail.
The exit condition does not mean that $f$ belongs to \oP..

Call {\tt Expand}$(B,e,f)$  a {\em base case execution} if
it executes
 line \ref{B3}, \ref{B4}, or \ref{B5} without recursing in line \ref{L6}.

\bclaim {\em If \I1p. holds for every base case execution of
{\tt Expand} then 
the expand step ends with
\oP. having permissible alternation and joining
$\de$ to the end of $\df$ in $V(B)$.} 


\b

\remark{Note that
if $\df$ was an edge of \os. at the start of the expand step,
the claim implies
$\df$ and its descendants are once again connected to a root
of \os..}


\bproof
Suppose {\tt Expand}$(B,e,f)$ is a base case execution
and {\tt Expand}$(B,e',f')$ is the next base case execution.
Observe that when {\tt Expand}$(B,e',f')$ is entered
$f$ is the last edge added to \os.
and $e'=f$.
This observation follows by 
a simple induction using the structure of the recursive calls
when line \ref{L6} is executed.
The observation implies
that if the call to
{\tt Expand}$(B,e,f)$ satisfies the exit condition of \I1p. then
the entry condition of \I1p. holds for
{\tt Expand}$(B,e',f')$.

The initial call {\tt Expand}$(B,\de,\df)$ 
starts with $\oP.=\de$. So stringing together all the base case executions
shows that when 
{\tt Expand}$(B,\de,\df)$ exits,
$\oP.+\df$
has permissible alternation and has
$\df$ as its last edge. This gives the claim. 
\ecproof

\begin{lemma}
The expand step preserves all the invariants.
\end{lemma}

\begin{proof}
We examine
the four possibilities for an execution of {\tt Expand},
lines \ref{B3} -- \ref{L6}. We will show
the three base cases satisfy \I1p. and
also preserve invariants (I2)--(I5).
In the recursive case line \ref{L6} we will 
verify that \oP. is a path, i.e., no repeated vertices. This will complete the
verification of (I1). (By definition \os. is a forest and so it 
must be acyclic.) 



\case{line \ref{B3} is executed}
First we show that $e$ and
$f$ alternate (as in the comment).
The test guarding line \ref{B3} ensures $B$ is
atomic or mature, by (I5).  Line \ref{B3} is not executed in the
initial call 
{\tt Expand}$(B,\de,\df)$.
So the current invocation was made 
from line \ref{L6}, as {\tt Expand}$(B_i,e_i,e_{i+1})$ 
for some $i\le k$.
The rest of the argument
switches to this parent invocation. 
We must show that
$e_i$ and $e_{i+1}$ alternate when $B_i$ is atomic or mature.

When
$i\ne 1,k$, $e_i$ and $e_{i+1}$ are the edges of $\delta(B_i, C(B))$.
The edges alternate for atomic $B_i$ by the definition of blossom.
For mature $B_i$ we use that definition and the definition of maturity.
When $i$ is 1 or $k$
(i.e.,
$e_i=e$ or $e_{i+1}=f$, or both)
a similar inspection, using the definitions of $P$
and $P_0$,
 shows $e_i$ and $e_{i+1}$ alternate.

The alternation of $e$ and $f$ implies \I1p..
For (I2) a blossom $B$ is light by (I5).
The properties of (I2) for inner and outer $B$ follow easily.
(I3)--(I5) are unchanged.

\case {line \ref{B4} is executed}
$e\in M$ makes $B$ outer. So \I1p.  holds trivially 
for $f$ and $B$.
The other invariants hold trivially.

\case {line \ref{B5} is executed}
It is easy to see $B'$ is a valid blossom.
To establish (I2) first observe that 
$B'$ is light:
If $u$ is atomic then $e\notin M$ makes 
$B'$ light. If $B_u$ is  a
contracted blossom it is
 light by
(I2). So again $B'$ is light.

Next observe $B'$ is outer:
$e$ is unmatched
and in \oP. and \os. when (this invocation of) {\tt Expand} starts.
So 
$B_u$ is outer by the permissible alternation of $e$. This makes $B'$ outer. 
 The
other properties of (I2) as well as (I3)--(I5) follow easily.
\I1p. holds as in the previous case.

\case{line \ref{L6} is executed}
$P$ is a path since (I3) implies the
only $P_i$-trail that repeats a node of $C(B)$ is
$P_1(\beta,\beta)$. Such blossoms $B$ are processed in line \ref{B4} or
line \ref{B5} (see the comments).
So  the recursive calls for line \ref{L6} keep \oP. and \os. acyclic.

The discard step of this case ensures (I2).
The remaining invariants hold by induction.
\end{proof}

We turn to the dual adjustment step.
The following lemma and its corollary describe \os. 
when duals are adjusted.

\begin{lemma}
\label{DualsMatureLemma}
When duals are adjusted,
\os. alternates at every node and every maximal blossom is light and mature.
\end{lemma}

\begin{proof}
\os. alternates at every node when
every blossom is mature. This follows from (I1) plus the observation that
(I2) prohibits a mature outer blossom being joined to a child
by a matched edge. 
(I2) also shows that every maximal blossom is light.
Thus to prove the lemma we need only show that
every maximal blossom  is mature when duals are adjusted.
(I2) shows this for inner and non-\S. blossoms.
So we need only consider outer blossoms.

Let $B$ be an immature outer blossom.
We will show that a grow or blossom step can be done for
$B$. This proves the lemma, since duals are adjusted only when
no such step can be done.

Let $\beta$ be the base vertex of $B$. Let
$f\in \delta(\beta,M)$  be the  edge leading to
the parent of $B$ in \os.; $f$ is undefined if $B$ is free.
(I2) implies either $B$ is free and 
every $x\in V(B)$ has $d(x,M)=b_{\beta}(x)$ 
or $f$ exists and every $x\in V(B)$ has $d(x,M)=b(x)$. 
In both cases, $B$ immature implies there is
an edge $e=xy\in \delta(B,M-f)$ with $x\in V(B)$.
Clearly $e$ is tight.

\case{$B_y \notin \os.$} Since $B_x$ is an outer blossom,
$e$ is eligible for $B_x$. Thus a grow step can be done
to add $B_y$ to \S..

\case{$B_y \in \os.$} Suppose $B_y$ is inner.
Then $e\notin \os.$. ($e\in \os.-f$ makes $B_y$ a child
of $B_x$, so $B_y$ is outer.)
As before $e$ is eligible for $B_x$, and $e\in M$ shows it is eligible for
$B_y$. So
a blossom step can be done for
$e$.

Suppose $B_y$ is outer. 
Any matched edge is  tight, so the unmatched copy $e'$ of $e$ is tight.
$e'\notin M$ is eligible for both outer vertices $B_x,B_y$. 
So a blossom step can be done for $e'$. 
\end{proof}

In contrast to ordinary matching, 
duals may be  adjusted with matched edges $xy$
not in \os. but
incident to nodes of \os.:
$x$ and $y$ can be atoms with $x$ inner and $y$ outer.
This possibility is governed by the following lemma.

\begin{corollary}
\label{DualsMTightLemma}
When duals are adjusted,
any matched edge incident to a node of \os. joins an inner 
node  to an outer node.
\end{corollary}

\begin{proof}
Take $e=xy \in M$ with $B_x$ a node of \os.. 

\case {$B_x$ is inner}
Since no grow step can be done,  $y \in \S.$.
If $e\in \os.$ it goes to a child of $B_x$, which is outer as desired.
If $e\notin \os.$, since no  blossom step can be done
$B_y$ is an outer atom, as desired.

\case{$B_x$ is outer}
An unmatched copy $e'$ of $e$ is tight and not in $\os.$.
Since no grow step can be done for $e'$,  $y \in \S.$.
Since no  blossom step can be done for $e'$,  $B_y$ is  inner, as claimed.
\end{proof}

Let us check that the invariants are preserved when
duals are adjusted. (I1)--(I3) are unaffected by the adjustment.
For (I4)--(I5) recall the dual adjustment step
(Fig.\ref{DualbMatch} of Appendix \ref{bfAppendix}).

\b

{\parindent=0pt

(I4): 
Edges of \os. remain tight when duals are adjusted, 
by the alternation of the lemma.
Matched edges incident to \os. remain tight,  by  
the corollary.
Edges in contracted blossoms
remain tight by definition of the dual adjustment. Thus
(I4) is preserved.

\b

(I5): A dual adjustment only increases the duals of maximal blossoms 
(that are  outer).
So the lemma implies (I5) for the blossoms whose 
$z$ value increases from 0 to positive.

}

\b

We have shown the dual adjustment preserves all the invariants.
Furthermore
the dual variables continue to be
valid,
 i.e., they are feasible for the $b$-matching linear program
(reviewed in Appendix \ref{bfAppendix}).
We have verified this for all cases except 
unmatched edges not in a blossom subgraph.
This case follows
by exactly the same argument as ordinary matching.

\subsection*{Termination, correctness,  and initialization}
As in ordinary matching each step of the algorithm makes progress
-- it either finds an augmenting path or
modifies
the graph in a way that 
the current search
will not  undo.
(Each step creates a new node of \os..
Obviously the number of new outer atoms, outer blossoms, or inner atoms
is limited. The number of new inner blossoms is also limited --
a given blossom from a previous search can become an inner blossom only once.)
Thus the algorithm does not loop, it eventually halts.

Next we show
the algorithm halts with a perfect matching. 
We prove this using the fact that the
maximum size of a $b$-matching is

\begin{equation}
\label{MaxBMatchEqn}
\min_{I\con V}\ b(I)+\sum _C \f{b(C)/2}
\end{equation}

\noindent where $C$ ranges over all nontrivial
connected components of $G-I$ (a component is trivial if
it consists of one vertex $x$ but no loop, i.e.,  $xx\notin E$)
\cite[Theorem 31.1]{S}. 
In fact it is straightforward to see that the above quantity upper-bounds the size of a $b$-matching (a matched edge either contains a vertex of $I$
or has both vertices in a component $C$). 
Our derivation gives an alternate proof that 
the bound is tight.

Consider a search that {\em fails}, i.e., 
no grow, blossom,  or expand step
is possible, and duals cannot be adjusted to remedy this.

\begin{lemma}
\label{FailedLemma}
In a failed search,
any edge $e\in E$ incident to an \S.-vertex  
is either spanned by a blossom
or incident to an inner atom.
\end{lemma}

\begin{proof}
A dual adjustment decreases the $z$-value of any
inner blossom, and lowering it to 0 allows an expand step to be done.
So there are no inner blossoms.

We can assume $e\notin M\cup \os.$ by using an unmatched copy.
Let $e=uv$. If
the lemma does not hold we can assume
at least one of $B_u,B_v$ is outer, say $B_u$. 
Furthermore either $B_v\notin \os.$ or 
$B_v$
is outer with either $B_u\ne B_v$ or $B_u=B_v$  atomic (since $e$ is not spanned by a blossom).
Since no grow or blossom step can be done,  
$e$
is not tight, in each of these cases. A dual adjustment decreases the $y$-value of any
vertex in an outer node.
So duals can be adjusted to make $e$ tight. This
makes a grow or blossom step possible, contradiction.
\end{proof}

Consider a failed search.
Let $I$ be the set of inner atoms.
The lemma shows that
deleting $I$ gives  a collection of connected components, 
each of which is 
either 
(a) an outer blossom, (b) an outer atom that has no loop,
or (c) a set of non-\S. vertices. 
(For (b) note that the existence of a loop ensures
the vertex is in a blossom, by the lemma.)
 Corollary \ref{DualsMTightLemma} shows no matched edge joins
two vertices of $I$.
Observe 
that \[|M|=b(I)+ \sum_C \f{b(C)/2}\]
where the sum ranges over the components of type (a) or (c).
(For (a) Lemma \ref{DualsMatureLemma} shows 
the blossom is mature.
For (c) 
note that
the
non-\S.-vertices are perfectly matched, i.e., each $v\notin \S.$
is on $b(v)$ matched edges leading to other non-\S.-vertices.)
This shows the upper bound 
\eqref{MaxBMatchEqn}
is tight.
It also shows that our algorithm, executed on an arbitrary input graph,
halts with a maximum cardinality $b$-matching. 

\def\H#1{\widehat{#1}} 

When a perfect $b$-matching exists our algorithm finds such a matching
of  maximum weight. 
This follows simply from the LP formulation of maximum $b$-matching
of Appendix \ref{bfAppendix}.
The argument is the same as ordinary matching, so we just summarize it
as follows.

The primal LP is satisfied by
any (perfect) $b$-matching. 
The dual LP requires every edge to be dual-feasible,
i.e.,
\[\H{yz}(e)  = y(e)  + z\set {B} {e \con B}\ge w(e).\]
The dual adjustment step enforces this.
Complementary slackness requires  tightness in the above inequality 
for every $e\in M$.
Complementary slackness also requires every blossom
with positive $z$ to be mature.
These two complementary slackness conditions are guaranteed by (I4) and
(I5) respectively. We conclude the final $b$-matching has maximum weight, and
our algorithm is correct.
In fact the algorithm provides an alternate proof that the LP
of Appendix \ref{bfAppendix} 
is a correct
formulation of maximum $b$-matching.

\b

As in ordinary matching, the algorithm can be advantageously initialized
to use any information at hand.
The initialization must specify
a partial $b$-matching $M$, 
dual functions $y,z$, and 
a collection of blossoms $\B.$. The requirements are that 
the duals must be  
feasible on every edge,
tight on every edge of $M$ or a blossom subgraph, 
and invariants (I2), (I3) and (I5) must
hold. 
The simplest choice is
a function $y$ on vertices where
$y(e)\ge w(e)$ for every edge $e$, $z\equiv 0$, $\B.=\emptyset$,
and $M$ a partial $b$-matching consisting of tight edges.
(Here and elsewhere $z\equiv 0$ means $z$ is the function that is 
0 everywhere.)
This initialization is used  in 
Section \ref{bStrongSec}.
A handy special case is when every vertex is assigned the same initial
$y$-value. This gives the invariant that every free vertex
has the same $y$-value, which is the minimum
$y$-value.
Using this initialization
 when the input graph does not have a perfect $b$-matching,
our
algorithm finds
a {\em maximum cardinality maximum weight $b$-matching}, i.e.,
a partial $b$-matching that
has the greatest number of edges possible,
and subject to that constraint, has the greatest weight possible.
This is shown in Appendix \ref{bfAppendix}, along with other maximum weight
variants. 
Other 
choices  for initialization  
allow
various forms of  sensitivity analysis to be accomplished in
$O(1)$ searches after finding a maximum $b$-matching
(as in ordinary matching).

\subsection*{Dual variables and efficiency analysis}
The numerical computations of the
algorithm are organized around a parameter
$\Delta$  maintained as
the total of all dual adjustment quantities $\delta$
in the current search.
($\delta$ is computed by
the dual adjustment algorithm of 
Fig.\ref{DualbMatch}, Appendix \ref{bfAppendix}.)
We use $\Delta$ as an offset to
compute dual variables as they change, and in a data structure to
adjust duals and determine the next step to execute. The details are as follows.

The data structure records values $Y(v),Z(B)$ that are used to find
the current value of any dual $y(v), z(B)$ ($v\in V, B$ a blossom)
in $O(1)$ time.  For
example let $v$ be a vertex
currently in an outer node of \os., 
with $\Delta_0$ the smallest value of $\Delta$ for which this is true,
and
$y_0(v)$ the value of $y(v)$ at that point.
($\Delta_0$ may be less than the value of $\Delta$ when the current blossom $B_v$ was
formed.)
Then
\begin{eqnarray*}
Y(v)&=&y_0(v)+\Delta_0,\\
y(v)&=&Y(v)-\Delta,
\end{eqnarray*}
since $y(v)$ decreases by $\delta$ in every dual adjustment
where it is outer.
Similarly a blossom $B$ that is currently an outer node of \os. has
\begin{eqnarray*}
Z(B)&=&z_0(B)-2\Delta_0,\\
z(B)&=&Z(B)+2\Delta,
\end{eqnarray*}
for $\Delta_0$ the value of $\Delta$ when $B$ became outer
and $z_0(B)$ the value of $z(B)$ at that point.
($z_0(B)$ is the value of $z(B)$ at the start of the search if $B$ was formed prior to that,
or 0 if $B$ was formed in the current search.)
Modulo  changes of sign, similar
 equations are used to compute
$y(v)$ for $v$ currently an inner atom and
$z(B)$ for $B$ 
currently a maximal inner blossom.
Other
$y$ and $z$ values 
are computed as described in Appendix \ref{GrowExpandSection}.

To adjust duals and  determine the next step of the algorithm 
to be executed
we use
a Fibonacci heap \F.. \F. contains a node for each 
grow, blossom, and expand step that is a candidate for
the next step to execute.
The key of each such node is the (future) value of $\Delta$ 
when the step can be done. 
Specifically when a blossom $B$ becomes inner in a grow or expand step,
a node for expanding $B$  is inserted in \F. 
with key equal to the current value of $\Delta$ plus $z(B)/2$.
(This node may get deleted before the expansion, in a blossom step.)
Theorem \ref{TreeBlossomMergingTheorem}
provides the node of \F. for the
next candidate blossom step.
Gabow \cite{G85b} gives an algorithm
 for future grow steps that uses total time $O(m\alpha(m,n))$.
For completeness 
Appendix \ref{GrowExpandSection}
gives a simpler algorithm for grow steps. It  uses total time
$O(m+n\log n)$ and so suffices for our purposes. The algorithm is also valid
for a pointer machine.

\b

The algorithms of Fig.\ref{BMAlg}--\ref{ExpandAlg} use linear
time. We use the data structure for blossoms of Section
\ref{BlossomDataStructureSection}.  Note that in the expand step of
Fig.\ref{BMAlg} $P_0$ should not be computed:%
\footnote{The pseudocode uses $P_0$ to facilitate specification of $P$.
Explicitly computing $P_0$ could lead to quadratic time when there is
a nesting of blossoms that get expanded, i.e., we have $B=B_0
\supset B_1\supset B_2\supset\ldots$ where each $B_i$ becomes inner when
$B_{i-1}$ gets expanded.}
The path $P$ in 
line \ref{L6}
of Fig.\ref{ExpandAlg} is easily computed 
without $P_0$ by
following appropriate links in the representation of $C(B)$.
The details are as described for  the procedure that computes $P_i$ trails
(given at the end of Section \ref{bMatchingSec}) 
except no recursive calls are made.

Blossom steps are implemented using the tree-blossom-merging algorithm
of Section \ref{TBMAlgSec}. 
This algorithm is unchanged from ordinary matching, assuming the supporting
forest is maintained correctly. 
(A new case is that matched edges
can cause blossom steps, and such edges may be incident to
inner vertices. Such blossom steps are handled seamlessly by the
tree-blossom merging algorithm. Alternatively they can be easily
handled outside
that algorithm, since matched edges are always tight.)

Now consider the supporting forest.
We modify Definition \ref{STreeDefn} 
of the supporting forest, which
uses the paths $P(x,\beta)$ of ordinary matching for inner blossoms $B$.
The corresponding trail for $b$-matching is $P_0(x,\beta)$.
To keep the supporting forest acyclic
we modify this trail as follows. Let 
$P_0^-(x,\beta)$ be 
$P_0(x,\beta)$ with every
maximal subtrail of the form 
$P_1(\beta(A),\beta(A))$ replaced by the vertex $\beta(A)$.
We modify  Definition \ref{STreeDefn} so that inner blossoms $B$
use  $P_0^-(x,\beta)$ as $T_B$
rather than $P(x,\beta)$. 

Note that an inner blossom $B$ still has $\beta(B)$ a vertex of the supporting
forest. This allows a blossom step for the matched edge incident to $\beta(B)$ to be handled correctly.

The algorithm for maintaining $T$ is 
essentially unchanged for grow and blossom steps. 
For expand steps suppose
{\tt Expand} is executed for a blossom $A$
which as above is represented only by the vertex $\beta(A)$.
We cannot have $z(A)>0$.
(That would make $A$ mature, by
(I5). But maturity implies a subtrail
$P_i(\beta(A),\beta(A))$ of $P_0(x,\beta)$ has $i=0$, contradiction.)
So line \ref{B4} or \ref{B5} 
is executed, making
$A$ a new outer vertex.
$\beta(A)$ is already in the supporting tree, and
the remaining vertices of $A$ are added to $T$ using
$add\_leaf$
operations.

We conclude that our algorithm finds a maximum $b$-matching in total time
$O(b(V)(m+n\log n))$.

\subsection{Strongly polynomial algorithm}
\label{bStrongSec}
The algorithm easily extends to a strongly polynomial version.
We follow  previous approaches 
that use  bipartite matching (i.e., network flow) to reduce the
size of the problem solved on the given nonbipartite graph
\cite{Anstee,Ger,S}.
The high-level algorithm is as follows:

\bigskip

Set $b'=2\f{b/2}$. Let $M$ be a
maximum cardinality 
maximum weight $b'$-matching 
with corresponding optimal dual function $y$. Using
$M,y$ (and $z\equiv 0,\ \B.=\emptyset$) as the initial solution, execute
the $b$-matching algorithm of Section \ref{bAlgSec}.

\bigskip

This is a straightforward combination of  previous algorithms
\cite{Anstee,Ger,S}.
For completeness we give the analysis. 

Correctness of this
algorithm -- that an optimum $M,y$ actually exists
-- may not be immediately clear. We  establish  correctness
below
as part of the efficiency analysis.

Since we assume $G$ has a perfect $b$-matching it has a 
partial $b'$-matching with $\ge 
\frac{b(V)}{2}-n$ edges.
So our $b$-matching algorithm performs $\le n$ augmentations.
Thus the time for the entire algorithm is $O(n(m+n\log n))$ plus the time
to find $M,y$.
We will see the latter strictly dominates the time.

We
find $M,y$ as follows.
Extend the given graph $G$ to $G^+$ 
by adding a vertex $s$ with 
$b'(s)=b'(V)$,
edges $vs$, $v\in V$ of weight 0 and edge $ss$
of weight $Wb'(s)$ for
\[W=\max \set{1, |w(e)|} {e\in E(G)}.\]
It is easy to see there is a 1-1 correspondence between 
partial $b'$-matchings of $G$ and (perfect)
$b'$-matchings of $G^+$, wherein a cardinality $c$ partial matching 
corresponds to a perfect matching with $c$ loops $ss$.
Furthermore a
maximum cardinality maximum weight $b'$-matching of $G$
corresponds to a maximum $b'$-matching of $G^+$.
To verify this last assertion 
it suffices to show
any $b'$-matching  of $G^+$
with $c$ loops $ss$, say $M_c$, weighs more than 
any such matching with $d<c$ loops, say $M_{d}$.
This 
follows 
since the relation
$b'(s)/2\ge c\ge d+1$ gives
\[\begin{array}{r@{\hspace {4pt}}l}
w(M_c)\ge (Wb'(s)-W)c &\ge \big(Wb'(s)d+Wb'(s)\big)-Wb'(s)/2\\
&= Wb'(s)d+ Wb'(s)/2> (Wb'(s)+W)d\ge w(M_d).
\end{array}
\]

We find a maximum $b'$-matching of $G^+$ by reducing to 
a bipartite graph $BG$. $BG$ has
vertex set
$\set {v_1,v_2}{v\in V(G^+)}$, edge set $\set {u_1v_2,u_2v_1}{uv\in E(G^+)}$,
and edge weights 
and degree constraints given respectively by
\[w(u_1v_2)=w(u_2v_1)=w(uv) \hbox{ and }
b'(v_1)=b'(v_2)={b'(v)/2}.\]
(Note  $s$ has even $b'$ value. Also a loop $uu$ of $G^+$ gives one
edge $u_1u_2$ in
$BG$.)
Let $x$ be a maximum 
$b'$-matching 
on $BG$
with optimum dual function $y$ (we show  $x$ and $y$ exist below).
Define a $b'$-matching $M$ on $G^+$
by taking $x\{u_1v_2, u_2v_1\}$ copies of each edge $uv\in E(G^+)$
(by our summing convention this means 
a loop  $uu$ has
$x(u_1u_2)$ copies).
Define a dual function $y$ by 
\begin{equation}
\label{yDefnTransportationEqn}
y(v)=y\{v_1,v_2\}/2.
\end{equation}
We will show that restricting $M$ and $y$ to $G$ gives the 
optimum values
desired for the main algorithm.

We first prove that
the $b'$-matching
$x$ exists, and
$M$ is a maximum $b'$-matching on $G^+$. 
These properties follow from the facts that 

(a) any 
$b'$-matching $x$ on $BG$ gives a
$b'$-matching $M$ on $G^+$ 
of the same  weight;

(b) any $b'$-matching $M$ on $G^+$ gives a 
$b'$-matching $x$ on $BG$ of the same  weight.

\noindent 
Recall that $G^+$ has a $b'$-matching. So (b) implies $x$ exists.
Also (a) is obvious from the above construction of
$M$ on $G^+$. So we need only prove (b).

We prove 
(b) using 
the Euler tour technique:
Let $M$ be a $b'$-matching  on $G^+$.
Since $b'$ is even on $G^+$, the edges of $M$ 
form a collection of closed trails.  Traverse each trail, and for each edge
$uv$ traversed from $u$ to $v$ match edge $u_1v_2$.
This applies to loops $uu$ too. For
each vertex $v$, the $BG$-matching has exactly $b'(v)/2$ edges incident
to each of  $v_1,v_2$.
Clearly we have the desired matching on $BG$.

Applying complementary slackness to the definition of 
the dual function for bipartite 
$b$-matching 
\cite[Ch.21]{S}
we get that
a $b'$-matching $x$ on $BG$ and a dual function $y$ are both optimum 
iff
\begin{equation}
\label{SOptEqn}
\text{$y(e)\ge w(e)$ for all edges $e$ of $BG$, with equality
when $x(e)>0$.}
\end{equation}
(Recall our summing convention means that if $e=uv$ then
$y(e)=y(u)+y(v)$.) Thus for every edge $e=uv$ of $G^+$,
\[y(e)=(y(u_1)+y(u_2)+ y(v_1)+y(v_2))/2
\ge (w(u_1v_2) +w(u_2v_1))/2=w(e).\]
Furthermore  equality holds for $e\in M$.
This follows because 
the matching $x$ on $BG$ has a mirror image
$x'$ defined by $x'(a_1b_2)=x(b_1a_2)$.
Thus
 $x(u_1v_2)>0$ implies
$y(u_1v_2)= w(e)$ as well as $y(v_1u_2)= w(e)$.
This calculation remains valid when $e$ is a loop $uu$ (i.e., $u=v$).
So the  functions $y,0$ are optimum duals for $b'$-matching on $G^+$.
Restricting $M$ and $y$ to $G$, and
taking $z\equiv 0,\ \B.=\emptyset$, gives permissible initial values for
the $b$-matching algorithm of  Section \ref{bAlgSec}.
(Recall
the discussion of initialization
at the end of Section \ref{bAlgSec}.)
We have proved the main algorithm is correct.

We find $x,y$ on $BG$ using an algorithm for minimum cost network flow.
Specifically the problem on $BG$ is a transportation problem,
where $x$ is an optimum integral solution and $y$ is an optimum dual function
\cite[Ch.21]{S}. The optimality conditions \eqref{SOptEqn}
are precisely those for the transportation problem
(assuming the trivial sign flip to convert our maximum weight problem to
a minimum cost problem).
Orlin 
solves the transportation problem (more generally the transhipment problem)
 in time
$O(n\log n(m+n\log n))$
\cite{Orlin}. It gives both $x$ and $y$. Using this we
obtain our strongly polynomial bound:

\begin{theorem}
A maximum $b$-matching can be found in time
$O(\min\{b(V), n\log n\}(m+n\log n))$.
\hfill\qed\end{theorem}

\section{{\boldmath $f$}-factors}
\label{fFactorSec} 

\begin{figure}[th]
\centering
\input{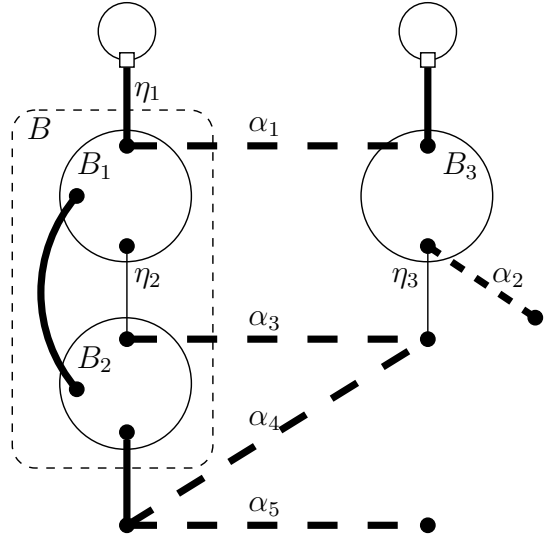}
 \caption{$f$-factor search structure.
As usual dashed edges are not part of the structure;
in this figure they are $\alpha_i$, all matched.}
 \label{fBlossomFig}
 \end{figure}

The fundamental difference between $f$-factors and $b$-matching
is illustrated in
Fig.\ref{fBlossomFig}, which  shows a search structure for $f$-factors. 
Recall that $f$-factors are defined on arbitrary multigraphs --
unlike $b$-matching, edges have  a limited number of parallel copies. 
The parallel copies in 
$b$-matching often allow a blossom to be enlarged using an incident
matched edge and its parallel unmatched edge
(e.g., in Fig.\ref{fBlossomFig} blossom $B$ and 
$\alpha_3$; also $B_3$ and $\alpha_2$).
 The parallel unmatched edge needn't exist for $f$-factors (in fact the search 
structure of Fig.\ref{fBlossomFig} might be maximal).  
This in turn leads to another major difference: the linear programming
$z$ dual variables are assigned
to blossom/incident-edge-set pairs rather than just blossoms.

The  organization of the $f$-factor section is the same as $b$-matching:
Section \ref{fBlossomSec} gives the basic properties of blossoms.
Section \ref{fAlgSec}  presents our algorithm that
finds a maximum $f$-factor in time 
$O(f(V)(m+n\log n))$.
Section \ref{fStrongSec} extends the algorithm to achieve the strongly polynomial time bound 
$O(m\log n\ (m+n\log n))$, the same bound as known for bipartite graphs.

We use the same terminology as $b$-matchings whenever possible 
--
we  introduce each such  duplicate term and 
point back to its complete definition in
Section \ref{bMatchingSec}. For instance
the degree constraint function $f$,
a {\em partial $f$-factor},
and all terminology regarding multigraphs and contractions
are the same as before.

\subsection{Blossoms}
\label{fBlossomSec}
Similar to $b$-matching
we define immature and mature blossoms,
give a data structure, and  show how blossoms are
 updated when the matching gets augmented.

Blossoms are defined as before by Definition \ref{ABlossomDefn}.
We add  the notion of  the ``base edge'' of a blossom. It is the
``exitting edge'' of a blossom.
(It  is used implicitly for $b$-matchings.)

\begin{definition}
\label{EtaDefn}
The {\em base edge} of a blossom $A$, denoted $\eta(A)$,
is either an edge of $\delta(\beta(A)) \cap \delta(V(A))$
with opposite $M$-type from $A$, or $\emptyset$.
It  satisfies these properties:

Blossoms with the same base vertex have the same base edge.

If $\beta(A)$ is not the base of a maximal blossom, 
i.e., $A\in \A.(B)-\alpha(B)$ for
some blossom $B$, then $\eta(A)$ is an edge of $C(B)$.  
\end{definition}

\noindent 
It is easy to see this notion is well-defined. In particular
in the last case
the definition of blossom $B$ shows the edge of opposite M-type
from $A$ exists.
Note that  $\eta(A)=\emptyset$ only if
$\beta(A)$ is the base of a maximal blossom.
Using $\emptyset$ as a base edge is handy notation below, e.g.,
\eqref{IBDefnEqn}.

As before we abbreviate $\beta(B)$ 
to $\beta$ when possible, and similarly for $\eta(B)$.
{\em Heavy} and {\em light} blossoms are defined as before.
In
Fig.\ref{fBlossomFig} the 2 free loop blossoms have
$\emptyset$  base edges. $B$ and $B_1$ are light blossoms and
have base edge $\eta_1$.
In Fig.\ref{EarFig}(a) the triangle blossom
has $\eta$ arbitrarily chosen as one of 2 matched edges
incident to its base; similarly
the heavy blossom has $\eta$
as one of the 2 unmatched edges incident to its base.

The family 
of blossoms $\A.^*$ constructing
$B$ is defined as before. 
The trails $P_i(v,\beta)$ for $f$-factor   blossoms are 
the same as before.
As before
any 
 trail $P_i(v,\beta(B))$ passes through any $A\in \A.^*(B)$ at most once,
and if so it traverses a trail $P_j(v,\beta(A))$, $v\in V(A)$
(possibly in the reverse direction). 

Furthermore for such blossoms $A$, $\eta(A)$ is an edge in 
$P_i(v,\beta(B))$ unless $\beta(A)=\beta(B)$.
To prove this we can assume $A$ is a maximal blossom with base
$\beta(A)\ne\beta(B)$.
Let $D$ be the blossom with $A\in \A.(D)$ ($D\in \A.^*(B)\cup \{B\}$).
$P_i(v,\beta(B))$  traverses the blossom trail $C(D)$
on a subtrail, denoted $\o P$ in the proof of Lemma \ref{PcdotLemma}.
For blossom $A$ on $\o P$ with $A\ne \alpha(D)$,
$\o P$
either contains both edges of
$\delta(A,C(D))$ or, when $A$ is the first vertex of
$\o P$, the edge of $\delta(\beta(A),C(D))$ of opposite M-type from $A$.
Both alternatives have $\eta(A)$ in $\o P$.

\subsection*{Mature blossoms}

As in $b$-matching complementary slackness dictates the sets
that may have positive dual variables, and we use this to define
which blossoms are mature. Dual variables are associated with
blossom/incident-edge-set pairs (see the review in Appendix \ref{bfAppendix})
but the blossoms must satisfy the following ``completeness'' property.
A blossom $B$ with base vertex $\beta$ and base edge $\eta$ is {\em mature}
if 

\bigskip

{\hi

every $x\in V(B)-\beta$ has $d(x,M)=f(x)$, and furthermore,

either $d(\beta,M)=f(\beta)$ and $\eta$ is an edge, or
$d(\beta,M)=f(\beta)-1$ and $\eta=\emptyset$.
}

\bigskip

We shall see that in contrast to $b$-matching the algorithm never creates
immature blossoms. Thus the $f$-factor algorithm does not use a discard step.

\subsection*{Augmenting trails}
Augmenting trails, augmenting blossoms, and the augment step are defined
exactly as in $b$-matching.
Any blossom $B$ on the augmenting trail, maximal or not,
remains a blossom 
after rematching. 
Lemma \ref{RematchBlossomLemma} shows this except for exhibiting
the base edges of blossoms. We define base edges
for rematched blossoms as follows.


Let $AT$ denote the augmenting trail $P_0(v,\av.)-(v',\av.)$.
To be consistent with 
Lemma \ref{RematchBlossomLemma}  we use primes to denote
blossoms after rematching, e.g., $B'$ is the rematched blossom $B$. 

\begin{lemma}
For any mature blossom $B$ with $\delta(B,AT)\ne \emptyset$,
\label{NewBaseEdgeLemma}
$\eta(B')$ is the unique edge satisfying
\[\delta(B,AT) -\eta(B)=\{\eta(B')\}.
\]
\end{lemma}

\remark{The lemma applies to all blossoms of our algorithm
since they are mature. The lemma also shows that after augmenting
every blossom remains mature -- even a free blossom that occurs at an end of
$AT$.}

\begin{proof}
For some $x\in V(B)$ let 
$xx'$ be the edge of $\delta(B,AT)-\eta(B)$.
To show this edge is uniquely defined
first note the definition of $AT$ implies
$d(B,AT)$ is 1 or 2, so consider two cases:
 If $d(B,AT)=2$ then $\eta(B) \in E(AT)$,
by the definition of augmenting blossom.
If $d(B,AT)=1$ then $B$ contains a free vertex
$v$ or $v'$ so $\eta(B) =\emptyset$, by the definition of maturity.
In both cases  $\delta(B,AT)-\eta(B)$ has exactly 1 edge.

Next note that $AT=P_0(v,\av.)-(v',\av.)$ passes through $B$ 
on a trail $P_j(x,\beta(B))$
(possibly in the reverse direction). 
Suppose $P_j(x,\beta(B))$ starts with an edge $e$.
Lemma \ref{RematchBlossomLemma} 
 shows 
$\beta(B')=x$ and
the M-type of $B'$ is that of the rematched $e$. This is
the opposite of the M-type of the rematched $xx'$. So we can take
$\eta(B')=xx'$.

The remaining possibility is that $P_j(x,\beta(B))$ has no edges.  
So $x=\beta(B)$ and $B'=B$. There are two possibilities.

{\parindent=0pt

\case
{$B$ is not free} $ xx'$ alternates with
$\eta(B)$. 
So the rematched $xx'$ has the original M-type
of $\eta(B)$.
The M-type of $B$ does not
change so we can again take $\eta(B')=xx'$.

\case
{$B$ is free} $B$ is a light blossom (even if
it is not maximal, by the definition of augmenting blossom).
This makes $xx'$  unmatched before rematching ($j=0$).
So the augment makes $xx'$  matched and we can take
$\eta(B')=xx'$.

}
\end{proof}

For any mature blossom $B$ define
\begin{equation}
\label{IBDefnEqn}
I(B)= \delta(B,M)\oplus \eta(B).
\end{equation}
The algorithm will assign positive values to  dual variables
of blossom/incident-edge-set pairs of the form $B,I(B)$ (recall
 Appendix \ref{bfAppendix}). 
As an example note this is consistent with ordinary matching and
$b$-matching:
Duals are associated with blossoms only because any blossom has
$I(B)=\emptyset$.
The latter follows since 
$\eta(B)$ is either the unique matched edge incident to $B$, or
$\emptyset$ when this edge does not exist.

The following lemma will be used to show that
an augment step maintains validity of the dual variables
(see Lemma \ref{fAugmentPreservedLemma}).

\begin{lemma}
\label{IUnchangedLemma}
An augment step does not change $I(B)$ for any mature 
blossom $B$ (maximal or not), i.e, $I(B)=I(B')$.
\end{lemma}

\begin{proof}
Let $M'$ be the augmented matching $M'=M\oplus AT$.
Thus
\[\delta(B,M')=\delta(B,M)\oplus \delta(B,AT).\]
Lemma \ref{NewBaseEdgeLemma}
shows
\[\{\eta(B')\}=\delta(B,AT) \oplus \eta(B).\]
By definition $I(B')=
\delta(B,M')\oplus \eta(B')$. Substituting the 
displayed equations transforms this  to
$\delta(B,M)\oplus \eta(B)=I(B)$.
\end{proof}

\subsection*{Data structure for blossoms}
The data structure is the one used in  $b$-matching with one minor extension:
The base edge $\eta(B)$ is stored even for maximal blossoms,
if it exists. This edge is not needed in 
$b$-matching, but it is
required for $f$-factors.
For instance it 
defines $I(B)$-sets
(via (\ref{IBDefnEqn})) which in turn defines the dual variables.

\subsection{$f$-factor algorithm}
\label{fAlgSec} 
Compared to $b$-matching an $f$-factor algorithm has more restrictions
on grow and blossom steps, since edges have a limited number of copies.
This necessitates assigning
$z$ dual variables to blossom/incident-edge-set pairs, and introduces the
case of matched edges that are not tight.
It also simplifies the algorithm by making matched and unmatched edges more symmetric.
Similarly heavy blossoms are required in both
the linear programming formulation
and the algorithm. 

We present both the search 
algorithm and the dual adjustment step -- 
the latter differs enough 
from ordinary matching and $b$-matching to merit detailed discussion. 

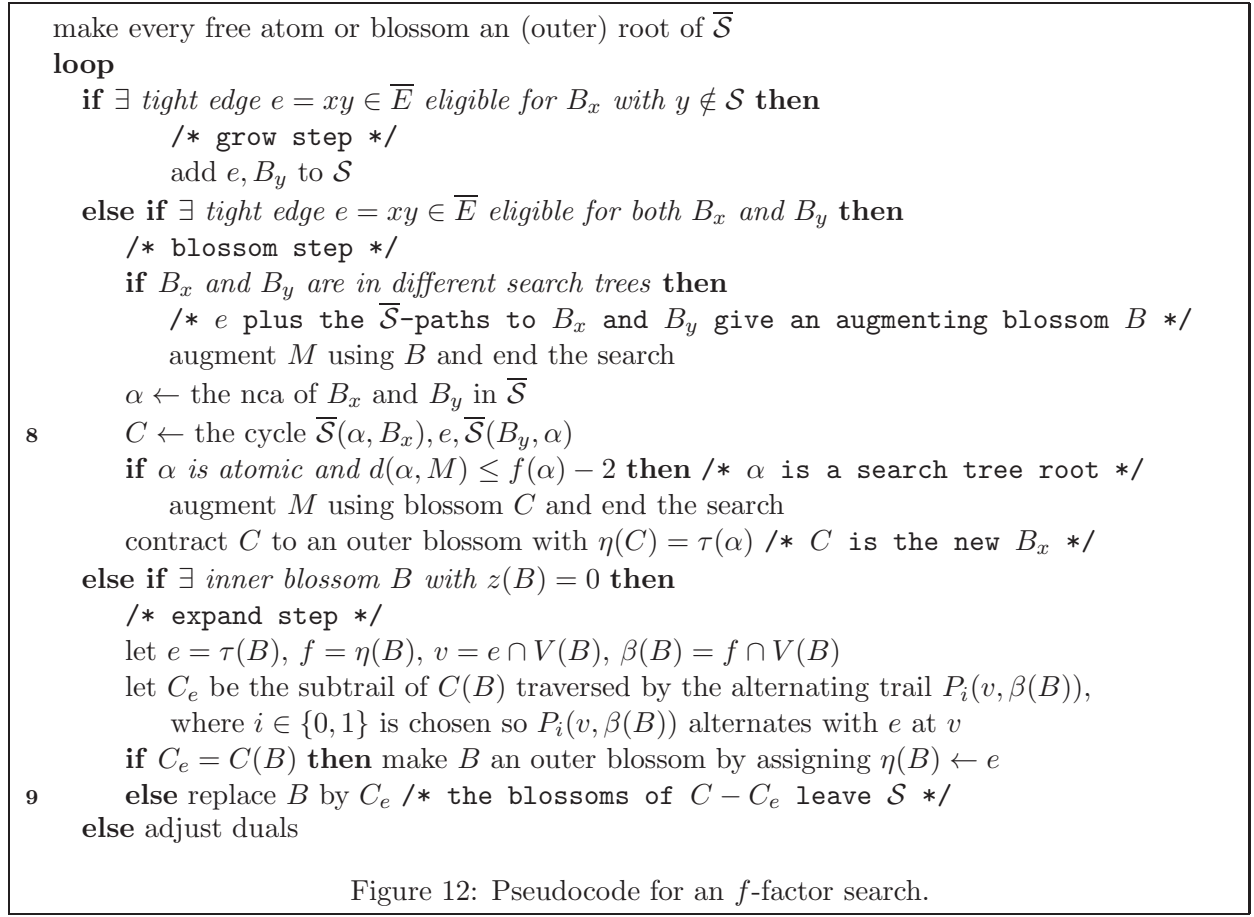
\def\goin{\hspace{17pt}}

\begin{algorithm}[h]

\DontPrintSemicolon
\SetKwFunction{Expand}{Expand}

make every free atom or blossom an (outer) root of \os.\;

\KwSty{loop}

\Indp

\If{$\exists$  tight edge $e=xy\in \oE.$ eligible for $B_x$ with  $y\notin \S.$}
{\goin\tcc{grow step}
 \goin add $e, B_y$ to \S.\;
}
\ElseIf{$\exists$ tight edge $e=xy\in \oE.$
 eligible for both  $B_x$ and $B_y$}
{\tcc{blossom step}
\If {$B_x$ and $B_y$ are in different search trees}
{\tcc{$e$ plus the \os.-paths to $B_x$ and  $B_y$ give 
an augmenting blossom $B$}
augment $M$ using $B$ and end the search\;}
$\alpha\gets$ the nca of $B_x$ and $B_y$ in \os.\;
\lnl{CConstructionLine}$C\gets$ the cycle $\os.(\alpha, B_x), e, \os. (B_y, \alpha)$ \;

\If(\tcc*[h]{$\alpha$ is a search tree root})
{$\alpha$ is atomic and $d(\alpha,M)\le f(\alpha)-2$}
{augment $M$ using blossom $C$ and end the search\;}
contract  $C$ to an outer blossom with $\eta(C)=\tau(\alpha)$  \tcc*[h]{$C$ is the new $B_x$}\;
}

\ElseIf{$\exists$ inner blossom $B$ with $z(B)=0$}
{\tcc{expand step}
let $e=\tau(B)$, $f=\eta(B)$, $v=e\cap V(B)$,  $\beta(B)=f\cap V(B)$\;
let $C_e$ be the subtrail of $C(B)$ traversed by the
alternating trail $P_i(v,\beta(B))$,\; 
\goin where $i\in \{0,1\}$  is chosen so 
$P_i(v,\beta(B))$ 
alternates with $e$ at $v$\;
\lIf{$C_e=C(B)$}{make $B$ an outer blossom by assigning 
$\eta(B)\gets e$}\;
\lnl{ReplaceBCLine}\lElse{replace $B$ by $C_e$ \tcc*[h]{the blossoms of $C-C_e$ leave \S.}\;}
}

\lElse {adjust duals}

\Indm

\caption{Pseudocode for an $f$-factor search.}
\label{fMAlg}
\end{algorithm}

The 
search algorithm is  presented in Fig.\ref{fMAlg}.
The definitions of the contracted graph $\o G$
its edge set
\oE.,
the search structure  \S.,
its contraction \os., the
$B_x$ sets denoting blossoms or atoms, and free nodes are the same as
 $b$-matching.
To define inner/outer classification let $v$ be a node of \os..
If $v$ is not a search tree root  let $\tau(v)$ be the edge to its parent.

\b

Node $v$ of \os. is {\em outer} if any of the following conditions holds:

{\hi

$v$ is a search tree root;

$v$ is an atom with $\tau(v)\in M$; 

$v$ is a blossom with $\tau(v)=\eta(v)$.

}

Otherwise $v$ is {\em inner}, i.e., $\tau(v)$ exists but
either of the following holds:

{\hi

$v$ is an atom with $\tau(v)\notin M$;

$v$ is a blossom with $\tau(v)\ne \eta(v)$.

}

\b

In contrast with $b$-matching,
an outer blossom can have $\tau(v)\notin M$ 
(i.e., when it is heavy) and 
 an inner blossom can have $\tau(v)\in M$.
These possibilities are illustrated by $B_2$ and $B_3$ respectively in 
Fig.\ref{fBlossomFig}.

Eligibility is defined as follows.  The motivation is to ensure that
paths in \os.  have the same structure as trails in blossoms:

\b
An edge $e=xy\in \oE.-\os.$ is {\em eligible for $B_x$}
if any of the following conditions holds:

{\hi

$x$ is an outer atom and $e\notin M$;

$x$ is an inner atom and $e\in M$;

$B_x$ is an outer blossom;

$B_x$ is an inner blossom and $e=\eta(B_x)$.

}

\bigskip

In the algorithm statement the current matching $M$ 
and the paths 
$\os.(x, y)$ 
are as before.

\begin{algorithm}[h]
\DontPrintSemicolon

\def\Or{\KwSty{or}\ }
$\delta_1\gets\min \set{   |\H{yz}(e)-w(e)| }{e=xy\in E 
\mbox{ eligible for $B_x$ with  $y\notin \S.$}}$\;
$\delta_2=\min \{ |\H{yz}(e)-w(e)|/2 \,:\,
e=xy\in \oE.
\mbox{ eligible for both  $B_x$ and $B_y$} \}$\;
$\delta_3=\min \set{z(B)/2}{B \mbox{ an inner blossom of }\os.}$\;
$\delta=\min \{\delta_1, \delta_2, \delta_3\}$\;

\lFor{every vertex $v\in \S.$}\\
\Indp\lIf{$B_v$ is inner}{$y(v)\gets y(v)+\delta$}
\lElse{$y(v)\gets y(v)-\delta$}\;
\Indm\lFor{every blossom $B$ in \os.}\\
\Indp\lIf{$B$ is inner}{$z(B)\gets z(B) -2\delta$}
\lElse{$z(B)\gets z(B) +2\delta$}\;

\caption{Dual adjustment step for $f$-factors.}
\label{DualfFactor}
\end{algorithm}

We turn to the dual
adjustment step, Fig.\ref{DualfFactor}.
We first recall terminology explained in detail in Appendix \ref{bfAppendix}.
Similar to $b$-matching the function
$\H{yz}:E\to \mathbb {R}$ is defined by
\begin{equation}
\label{fHyzTextEqn}
\H{yz}(e) = y(e)  + z\set {B} {e \in \gamma(B)\cup I(B)}.
\end{equation}
Say edge $e$ is {\em dominated, tight,} or {\em underrated}
depending on whether
$\H{yz}(e)$ is  $\ge w(e)$, $= w(e)$, or $\le w(e)$, respectively;
{\em strictly dominated} and {\em strictly underrated} 
designate the possibilities  $>w(e)$ and $< w(e)$ respectively.
The complementary slackness conditions for optimality require
$e$ to be dominated if it is unmatched, as in $b$-matching.
The new requirement is that $e$ must be underrated if it is matched.

As usual our algorithm maintains duals to satisfy these requirements.
As in $b$-matching there may be strictly dominated unmatched edges;
symmetrically there may be 
strictly underrated matched edges.
(This is expected since our algorithm has minimum cost network flow
as a special case.)
The 
absolute values in the definitions of $\delta_1$ and $\delta_2$ reflect 
these  possibilities, as 
$\H{yz}(e)-w(e)$ may have arbtrary sign.
The use of $\H{yz}(e)$ rather than $y(e)$ (as in ordinary matching and
$b$-matching, Figs. \ref{DualEdmonds}--\ref{DualbMatch})
reflects the possibility that eligible edges can be in $I(B)$ sets
and so have positive $z$ contributions in $\H{yz}(e)$.

The rest of this section
follows the same organization as before,
giving clarifying remarks,
invariants, examples, and then
the formal 
proof of correctness.

\paragraph*{Remarks}
Many remarks for $b$-matching still apply, the exceptions being the simplifications
in the  blossom  and expand steps.

In the blossom  step
consider the cycle $C$, which is constructed in line
\ref{CConstructionLine} and then processed as either
an augmenting blossom or an outer blossom.
When the augment step is executed,
$C$ is not considered an ordinary blossom (since it is not mature).
When $C$ is processed as an outer blossom,  the definition $\eta(C)
=\tau(\alpha)$ assumes $\tau(\alpha)=\emptyset $ when $\alpha$ is 
the root of \os..

The expand step is simpler than $b$-matching since all blossoms of
$C(B)$ are mature. As in ordinary matching
a new blossom in \os. may have $z$-value 0. 

The algorithm maintains
$\eta$ values in blossom, augment and expand steps.
In line \ref{ReplaceBCLine} $\eta$-values of blossoms in
$C_e$ are unchanged, so new blossoms of \os. may be inner or outer.

As before when duals 
are modified, our assumption that the graph has an
$f$-factor guarantees the new duals allow further progress
(see the remark at the end of the proof of Lemma
)

\paragraph*{Invariants}
The definition of \os. {\em alternating} at node $v$ is unchanged.

\bigskip

{\parindent=0pt

(I1) \hskip20pt $\os.$ alternates at any atomic node. Any root blossom is light.

\bigskip

(I2)\hskip20pt Every blossom $B$ (maximal or not) is mature.

\bigskip

\hskip42pt If $B$ is inner then it is either a leaf of \os.
or its base edge leads to its unique child.

\b

(I3)\hskip20pt
For every blossom $B$ (maximal or not) $C(B)$ is a cycle.

\bigskip

(I4)\hskip20pt An edge is tight if
it is an edge of \os. or an edge of a contracted blossom.
Any nontight edge is 
dominated if it is unmatched and underrated if 
matched.

}

\begin{figure}[th]
\centering
\input{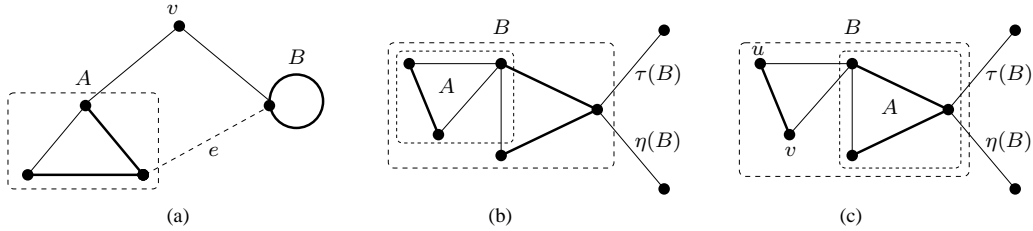}

 \caption{$f$-factor algorithm examples.}
 \label{fAlgExFig}
 \end{figure}

\paragraph*{Examples}
{\em Strictly Underrated Edges}: 
Fig.\ref{fBlossomFig} illustrates how matched edges can become underrated.
Suppose a dual adjustment is done.
(This requires $\alpha_3$ to be strictly underrated.
Alternatively we can assume $\alpha_3$ does not exist.)
Since $\alpha_5$ is incident to an outer atom a dual adjustment decreases
$\H{yz}(\alpha_5)$ by $\delta$.
Since $\alpha_2$ is incident to an inner blossom
and belongs to its $I$ set,
a dual adjustment increases
$\H{yz}(\alpha_2)$ by $\delta-2\delta =-\delta$.
These adjustments
exemplify the two main cases where strictly underrated matched edges
are created. (Note also that  $\alpha_2$ or $\alpha_5$ may
 be the base edge of a blossom not in \os..
So even $\eta$ edges need not be tight.)
The underrated edge $\alpha_3$ may become tight
in the dual adjustment since $\H{yz}(\alpha_3)$ increases by 
$(-\delta+2\delta) +\delta=2\delta$.

\b

\noindent {\em Eligibility}: 
Consider Fig.\ref{fBlossomFig}.
No grow step can be done for the ineligible edge $\alpha_2$.
No augment step can be done for $\alpha_1$
or $\alpha_4$ since they are
ineligible at one end. 

\b \noindent
{\em Grow Step}: 
In Fig.\ref{fBlossomFig}
the grow step that added the inner blossom
$B_3$ used a matched edge. This is not possible
in $b$-matching.

\b
\noindent
{\em Blossom Step}:
In Fig.\ref{fAlgExFig}(a) a blossom step can be done for $e$
if  $e=\eta(A)=\eta(B)$ and $e$ is tight. 
Such a blossom step -- for two inner vertices joined by an unmatched edge --
cannot be done in $b$-matching. If $e\ne \eta(A)$ 
then a blossom step
cannot be done for $e$. In this case $e$ 
may even complete an augmenting trail (when $e$ is tight and
$d(v,M)\le f(v)-2$) yet the algorithm does not augment.
The same holds if $e\ne \eta(B)$, assuming $B$ remains inner.  
This situation is further discussed in the Dual Adjustment Step example below.

\b
\noindent
{\em Augment Step}:
If an augment step is done for $\alpha_3$ in Fig.\ref{fBlossomFig},
the edge joining blossom $B_3$ to the free root blossom
becomes the (unmatched) base edge of both blossoms.

\b

\noindent {\em Expand Step}: 
In Fig.\ref{fAlgExFig}(b)
an expand step is done for blossom $B$.
$B$ is made an outer node, with base edge $\tau(B)$.
So unlike $b$-matching an expand step may 
preserve $B$ as a maximal contracted blossom.
 If a similar expand step is done for $B$
in Fig.\ref{fAlgExFig}(c) 
$A$ is made an inner blossom in \S. and 
$u$ and $v$ 
become atoms  not in \S..
Blossom $A$ may not get expanded in this search
even if $z(A)=0$. If it does get expanded then $A$ 
becomes an outer blossom.

\b

\noindent {\em Dual Adjustment Step}: 
Returning to Fig.\ref{fAlgExFig}(a)
suppose $e$ is 
not the base edge of either blossom $A,B$
and $B$ remains inner. 
A dual adjustment increases $\H{yz}(e)$ by $\delta+\delta=2\delta$.
$e$ becomes strictly dominated, 
and the topologically valid blossom or augment
step for $e$ has been destroyed. 
Subsequent dual adjustments take $e$ even further from being tight. 
However this cannot continue forever:
Eventually one of the inner blossoms
gets expanded, and  dual adjustments now increase  
$\H{yz}(e)$ by $-\delta+\delta=0$.
When the other blossom gets expanded $e$ becomes 
an unmatched edge joining two outer blossoms/vertices.
Now dual adjustments decrease the slack 
($\H{yz}(e)$ decreases by $2\delta$)
and eventually a blossom step can be done for $e$.

\subsection{Analysis} 
\label{fMAnalSec}
We follow the same organization as before:
The first subsection 
proves the invariants. The second subsection proves correctness and
adds details on initialization and termination. The last subsection
proves the desired time bound.


\subsection*{Proof of invariants}
As before
we  show all the invariants are preserved by every step of the algorithm.
Again our discussion of  grow, blossom, augment and expand steps just treats
the most interesting cases.
Start by noting that when a search begins any free blossom is light,
by (I1) from the previous search. So (I1) is preserved 
when \os. is
initialized.

\begin{lemma}
The grow step preserves all the invariants.
\end{lemma}

\begin{proof}
If $B_x$ is inner then the definition of eligibility implies
$e=\eta(B_x)$ and $B_y$ will be the unique child of $B_x$. So (I2) holds
for $B_x$. (I4) continues to hold.
\end{proof}

\begin{lemma}
The blossom step preserves all the invariants.
\end{lemma}

\begin{proof}
Using the algorithm's notation,
$e=xy$ is the edge triggering the blossom step and
$C$ is the new outer blossom.
We will verify that $C$ satisfies
Definition \ref{ABlossomDefn} as well as the definition of outer.
Furthermore since (I2) requires every blossom to be mature,
we will verify the base edges satisfy Definition \ref{EtaDefn}.
We first verify the conditions on $\alpha$.
There are  4 possibilities for $\alpha$.

If $\alpha$ is an outer atom then
(I1) shows all edges to its children are unmatched.
Also if $e$ is incident to $\alpha$ then eligibility implies
$e$ is unmatched. 
So blossom $C$ is light. This preserves (I1) if
$\alpha$ is a search tree root.
If $\alpha$ is not a root then $\tau(\alpha)$ is matched,
so it is a valid choice for $\eta(C)$.
Contracting $C$ makes
$\eta(C)=\tau(C)$ so $C$ is outer.

Similarly if  $\alpha$ is an inner atom
then $C$ is heavy, 
$\tau(\alpha)$ is unmatched, 
and $\eta(C)=\tau(C)$ makes
$C$ a valid outer blossom.

If $\alpha$ is an outer blossom then so is $C$, by definition.

Finally $\alpha$ cannot be an inner blossom $B$:
$\eta(B)$ is the only edge of $\delta(B)$
that can lead to a child of $B$
or be $e$ (by (I2) and the definition of eligibility).
So $B$  cannot be the nca of $B_x$ and $B_y$.

Next 
 consider a node $v\ne \alpha$  that is a blossom
in $C$. 
Definition
\ref{EtaDefn} requires $\eta(v)$ to be an edge of $C$.
If $v$ is outer
this holds since $\eta(v)=\tau(v)$. If $v$ is inner
then as before $\tau(v)$ and $\eta(v)$ 
must be the two edges of $\delta(v,C)$.
\end{proof}

\begin{lemma}
The expand step preserves all the invariants.
\end{lemma}

\begin{proof}
If $C_e=C(B)$ then $i=1$ in $P_i(v,\beta(B))$.
Thus $e$ and $f$ are both incident to $\beta(B)$ and both have the same M-type.
So changing $\eta(B)$ to $e$ preserves Definition \ref{EtaDefn}.
Since $z(B)=0$ the change also preserves $\H{yz}$ values.

If  $C_e\ne C(B)$ then replacing
$B$ by $C_e$ 
implicitly classifies the new maximal blossoms as inner or outer,
so the definition of \os. is maintained.
\end{proof}

\begin{lemma}
\label{fAugmentPreservedLemma}
The augment step preserves all the invariants.
\end{lemma}

\begin{proof}
Recall that an  augment step is given  a valid augmenting blossom --
in particular a nonatomic end $v$ or $v'$ is a light blossom by (I1).
Lemma \ref{IUnchangedLemma} shows no $I(B)$ set changes
in the augment.
Thus \eqref{fHyzTextEqn} shows
every
$\H{yz}(e)$ value remains the same
 and (I4) is preserved.
\end{proof}

\begin{lemma}
\label{fDualsPreservedLemma}
A dual adjustment preserves (I4) unless $\delta=\infty$. 
\end{lemma}

\begin{proof}
Any dual adjustment step has $\delta>0$. Let $e=uv$.
If $B_u=B_v$ is a blossom then clearly 
$\H{yz}(e)$ does not change. Thus suppose $e\in \oE.$. 
Assume $e$ is not a loop -- loops are covered in the second case below,  $e\notin \os.$.

The quantities in 
$\H{yz}(e) = y(e)  + z\set {B} {e \in \gamma(B)\cup I(B)}$
that may change in a dual adjustment
are limited to
 $y(x)$ and $z(B_x)$ for $x\in \{u,v\}$.
Clearly these quantities do not change if $B_x \notin \os.$.
From now on assume  $x\in \{u,v\}$ with $B_x \in \os.$. 
Then $y(x)$ changes by $\pm \delta$ and,
if $B_x$ is a blossom,
$z(B_x)$ changes by $\mp 2\delta$, but this 
contributes to $\H{yz}(e)$ iff $e\in I(B_x)$.
Define
$\Delta(B_x)$ to be the total change in
$\H{yz}(e)$ at the $x$ end. It is easy to see that 
$\Delta(B_x)=\pm \delta$, more precisely,
\begin{equation}
\label{DeltaBxEqn}
\Delta(B_x)=\begin{cases}
-\delta&\text{if $B_x$ is an outer atom, or
an outer blossom with $e\notin I(B_x)$,}\\ 
&\text{or
an inner blossom with $e\in I(B_x)$}\\
+\delta&\text{if $B_x$ is an inner atom, or
an outer blossom with $e\in I(B_x)$,}\\ 
&\text{or
an inner blossom with $e\notin I(B_x)$.
}
\end{cases}
\end{equation}

Define a sign $\sigma$ by
\[\sigma=\begin{cases}
+1&e\notin M\\
-1&e\in M.\\
\end{cases}
\]

\case{$e\in \os.$}
We claim
\begin{equation}
\label{DdeMteqn}
\Delta(B_x)=
\begin{cases}
 +\sigma \delta &e=\tau(B_x)\\
- \sigma \delta &e\ne\tau(B_x).
\end{cases}
\end{equation}
The claim implies  $\Delta(B_u)+\Delta(B_v)=0$ 
since one of  $B_u,B_v$ is the child
of the other. Hence
$\H{yz}(e)$ does not change and (I4) holds.

To prove the claim consider the value of $\Delta(B_x)$ in two symmetric cases.
 
\subcase{$\Delta(B_x)=-\delta$} 
Suppose $e\ne \tau(B_x)$, i.e., $e$ goes to a child of $B_x$. 
In all three cases of
\eqref{DeltaBxEqn} $e$ is unmatched.
Thus $\Delta(B_x)=-\sigma\delta$ as claimed.

Suppose $e=\tau(B_x)$, i.e., $e$ goes to the parent of $B_x$. 
In all three cases of
\eqref{DeltaBxEqn} $e$ is matched.
Thus $\Delta(B_x)=\sigma\delta$ as claimed.

\subcase{$\Delta(B_x)=\delta$} 
Suppose $e\ne \tau(B_x)$. In all three cases of
\eqref{DeltaBxEqn} $e$ is matched.
Thus $\Delta(B_x)=-\sigma\delta$ as claimed.

Suppose $e=\tau(B_x)$.
In all three cases of
\eqref{DeltaBxEqn} $e$ is unmatched.
Thus $\Delta(B_x)=\sigma\delta$ as claimed.

\b

\case {$e\notin \os.$}
Let $B_x$ be a node of \os..
We claim
\begin{equation}
\label{DdeMeqn}
\Delta(B_x)=
\begin{cases}
-\sigma\delta &e \text{ eligible for $B_x$}\\
 +\sigma\delta&e  \text{ ineligible for $B_x$.}
\end{cases}
\end{equation}

To prove the claim consider the value of $\Delta(B_x)$.
 
\subcase{$\Delta(B_x)=-\delta$} 
Suppose $e$ is eligible. In all three cases of
\eqref{DeltaBxEqn} $e$ is unmatched.
Thus $\Delta(B_x)=-\sigma\delta$ as claimed.

Suppose $e$ is ineligible. This is impossible
in the middle case, i.e., $B_x$ outer with $e\notin I(B_x)$.
In the other two cases of
\eqref{DeltaBxEqn} $e$ is matched.
Thus $\Delta(B_x)=\sigma\delta$ as claimed.

\subcase{$\Delta(B_x)=\delta$} 
Suppose $e$ is eligible. In all three cases of
\eqref{DeltaBxEqn} $e$ is matched.
Thus $\Delta(B_x)=-\sigma\delta$ as claimed.

Suppose $e$ is ineligible. This is impossible
in the middle case.
In the other two cases of
\eqref{DeltaBxEqn} $e$ is unmatched.
Thus $\Delta(B_x)=\sigma\delta$ as claimed.

\b

Now we show (I4). For every edge $e$ define
\[
slack(e)=\sigma(\H{yz}-w(e)).
\]
(I4) requires every edge to have nonnegative slack.
A dual adjustment changes 
$slack(e)$ by $\sigma\Delta(B_x)$.
\eqref{DdeMeqn} shows $slack(e)$ decreases iff $e$ is eligible for $B_x$. 

At least one of $B_u,B_v$ is a node of \os., so assume $B_u\in \os.$.
If $B_v$ is also in \os. we can assume
$\Delta(B_u)=\Delta(B_v)$, since otherwise 
$\H{yz}(e)$ does not change and the lemma holds.
So in the two cases below,
when $B_v$ is a node of \os.
either 
$e$ is  eligible for  both $B_u$ and $B_v$ 
or ineligible for  both.

This characterization also holds if $e$ is a loop, i.e., $u=v=x$.
In this case 
\eqref{DdeMeqn} still applies, so 
$\Delta(B_u)=\Delta(B_v)$. Also the total change to 
$\H{yz}(e)$ is
$\Delta(B_u)+\Delta(B_v)$ as before, since
$x$ is an atom.


\subcase{$e$ ineligible for $B_u$}
The dual adjustment increases the slack. 
Clearly (I4) continues to hold.

\subcase{$e$ eligible for $B_u$}
When the dual adjustment starts any edge
has nonnegative slack, i.e., $slack(e)
=|\H{yz}(e)-w(e)|$.
If $B_v\notin \os.$ then initially we have
 $|\H{yz}(e)-w(e)|\ge \delta_1\ge\delta$. 
Since $slack(e)$ decreases by $\delta$,
$slack(e)\ge 0$ after the dual adjustment and (I4) holds.
Similarly if $B_v\in \os.$ 
initially 
 $slack(e)=|\H{yz}(e)-w(e)|\ge 2\delta_2\ge 2\delta$.
$slack(e)$ decreases by $2\delta$ so after the dual adjustment
 $slack(e)\ge 0$ and (I4) holds.

\b

\remark{In this last subcase note that when $\delta=\delta_1$
the corresponding minimizing edge becomes tight. Thus a grow step can be done in
the next iteration. Similarly when $\delta=\delta_2$
a blossom step has become possible. Taking $\delta_3$ into account
we see that  for any $\delta<\infty$,
the dual adjustment step makes at least one grow, blossom, or expand step
possible, just like ordinary matching and $b$-matching.}
\end{proof}

We conclude that all the invariants are preserved throughout the algorithm.

\subsection*{Termination, correctness, and initialization}
The algorithm does not loop, by exactly the same argument as  $b$-matching.
Next we show the algorithm halts with an $f$-factor, i.e., no free vertices.
We shall prove this using the fact that the
maximum size of a partial $f$-factor is
\begin{equation}
\label{fFactorSizeEqn}
\min\ \bigg\{f(I)+ |\gamma(O)| + \sum _C \f{\frac{f(C)+|E[C,O]|}{2}}\bigg\}
\end{equation}
\noindent where 
the set is formed by letting $I$ and $O$  
range over all pairs of disjoint vertex sets,
and in the summation
$C$ ranges over all
connected components of $G-I-O$ 
\cite[Theorem 32.1]{S}.
Our derivation gives an alternate proof of this min-max relation.

We first observe that (\ref{fFactorSizeEqn})
upper-bounds the size of any partial $f$-factor.
In proof note that 
any edge $e$ of $G$ satisfies exactly one of these conditions:

\i $e$ is incident to an $I$ vertex;

\ii $e$ joins 2 $O$ vertices;

\iii $e$ joins a vertex in some component $C$ to another
vertex of $C$ or an $O$-vertex.

\noindent
We call
$e$  type \xi, \xii, or \iii accordingly.
We shall see these three types correspond respectively to the three terms of
\eqref{fFactorSizeEqn}.
Note that a loop $e$ may have any of the three types.

Clearly the number of matched edges of type \i and \ii 
is bounded by the first two terms of
(\ref{fFactorSizeEqn}) respectively. For type \iii
consider
any component $C$.
Counting edge ends shows
the number of matched edges of type \xiii, 
$|E[C,C\cup O] \cap M|$, satisfies
\begin{equation}
\label{GE2Eqn}
2|E[C,C\cup O] \cap M| =
\sum_{x\in C}d(x,M) - |E[C,I] \cap M| +|E[C,O] \cap M|.
\end{equation}
Obviously this implies
\begin{equation}
\label{GEEqn}
2|E[C,C\cup O] \cap M| \le f(C) + |E[C,O]|.
\end{equation}
The third term of 
(\ref{fFactorSizeEqn})
follows.
So (\ref{fFactorSizeEqn}) is a valid upper bound.
Now we prove this bound is tight.

Consider a search that {\em fails}, i.e., 
no grow, blossom, or expand step can be done and
the dual adjustment step gives $\delta=\infty$.
Since $\delta_1=\delta_2=\infty$,
no vertex of \os. has an eligible edge.
Since $\delta_3=\infty$ there are no inner blossoms.

In a failed search
let $I$ be the set of inner atoms and $O$ the set of outer atoms.
Deleting $I\cup O$ 
gives  a collection of connected components $C$.
There are exactly $f(I)$ matched edges of type \xi.
This follows since 
an inner atom is not free, and
a matched edge joining two inner atoms
is eligible (for both ends).
There are exactly $|\gamma(O)|$ matched edges of type \xii, 
since an unmatched edge joining two outer atoms is eligible.
For the type \iii edges
we will prove that any component $C$ has a value $\Delta\in \{0,1\}$ with
\begin{equation}
\label{GE3Eqn}
\sum_{x\in C}d(x,M) - |E[C,I] \cap M| +|E[C,O] \cap M|
=f(C)+|E[C,O]|-\Delta.
\end{equation}
With (\ref{GE2Eqn})
this shows
the left-hand side of 
(\ref{GEEqn}) is within $1$ of the right. So 
$|E[C,C\cup O] \cap M| =\f{f(C)+ |E[C,O]|\over 2}$. This shows
the number of type \iii matched edges equals the third term of
\eqref{fFactorSizeEqn}. Thus \eqref{fFactorSizeEqn} is a
tight upper bound on the size of a partial $f$-factor.

To prove (\ref{GE3Eqn}) consider 2 types of components
$C$:

\case {$C\con V-\S.$}
Since no vertex of $C$ is free the first term  on the left of
(\ref{GE3Eqn})
 is $f(C)$. Take any $e\in \delta(C)$.
$e$ goes to a node $v$ of \os. but $e$ is not eligible.
So $v$ is an inner or outer atom (not a blossom).
$e\in M$ iff $e\in E[C,O]$.
Thus the second term on the left of
(\ref{GE3Eqn}) is 0 and the third term is $|E[C,O]|$.
 So $\Delta=0$ as desired.

\case {$C$ contains an \os.-node}
We first show that $C$ is a collection of blossoms forming a subtree of \os.,
with no other edges (i.e., $\gamma(C,\o E)\con \os.$). 
Any \os.-node $B_x$ of $C$ is an outer blossom ($x\notin I\cup O$).
Consider an edge $e=xy\in \o E$
with $B_x$ a blossom of $C$ and
$B_y$ a node of $C$.
$e$
 is not eligible for $B_x$, so $B_y$ is an \os.-node.
Hence $B_y$ is an outer blossom.
$e$ is not eligible for at least one of $B_x,B_y$,
so
$e$ is an edge of \os., as claimed.

Let $B_r$ be the root of subtree $C$.
Let $e=rs$ be the edge of \os. from $B_r$ to its parent $B_s$,
if such parent exists; $e=\emptyset$ if $B_r$ is a free blossom.
We claim
\begin{equation}
\label{dCMEEqn}
\delta(C,M)-e=E[C,O]-e.
\end{equation}
Take any edge $xy\in \delta(C)-e$, $B_x\in C$.
$B_y=y\in I\cup O$ since $C$ is a connected component of
$G-I-O$.
So the claim
\eqref{dCMEEqn} is equivalent to $xy\in M$ iff $y\in O$.
This follows from two cases:
If $xy$ is an edge of \os. then $y$ is a child of $B_x$.
Thus 
$xy\in M$ iff $y$ is outer.
If $xy$ is not an edge of \os. then it is not eligible
for $y$. Again $xy\in M$ iff $y$ is outer.

Now we show (\ref{GE3Eqn}) holds with $\Delta=1$ in each of
3 possibilities for $e$. 


If $e=\emptyset$ then
$B_r$ contains the unique free vertex of $C$.
Using \eqref{dCMEEqn},
the 3 terms on the left of (\ref{GE3Eqn}) 
are $f(C)-1$, 0 and $|E[C,O]|$.

If $e$ is an edge 
then  $B_s=s$ is atomic.
If $s$ is inner then $e \in M$. 
Using \eqref{dCMEEqn}
the 3 terms on the left of (\ref{GE3Eqn}) 
are $f(C), 1$ and   $|E[C,O]|$.
If $s$ is outer 
then 
$e\notin M$. Using \eqref{dCMEEqn}
the terms are $f(C), 0$ and   $|E[C,O]|-1$.

We conclude the upper bound (\ref{fFactorSizeEqn}) is tight.

\bigskip

We have also shown that our algorithm, executed on an arbitrary input graph,
halts with a partial $f$-factor of
maximum cardinality. In proof
the  analysis of a failed search shows if the algorithm halts because
$\delta=\infty$, the current matching has size
\eqref{fFactorSizeEqn}, so its cardinality is
maximum.


Now we verify that our algorithm is correct, i.e., assuming
an $f$-factor exists the algorithm finds a maximum weight $f$-factor.
We have just verified the algorithm's final matching is an $f$-factor.
It remains to
verify the LP conditions for optimality (Appendix \ref{bfAppendix}).
(I4) gives the complementary slackness conditions for matched and unmatched
edges. We need only discuss the primal inequalities for blossoms
and the corresponding complementary slackness conditions.

The blossom inequalities state that every pair
$B,I$ ($B\con V$, $I\con \delta(B)$) satisfies
\begin{equation}
\label{fLPasMatchingInequality}
|(\gamma(B)\cup I)\cap M| \le \f{f(B)+|I|\over 2}.
\end{equation}
It is easy to see  this holds
for any $f$-factor:
The degree constraints imply
$2|\gamma(B)\cap M| + |I\cap M|\le f(B)$
so arithmetic gives 
$2(|\gamma(B)\cap M| + |I\cap M|)\le 
f(B)+|I\cap M|\le f(B)+|I|$
and integrality gives \eqref{fLPasMatchingInequality}.

Complementary slackness
requires tightness in \eqref{fLPasMatchingInequality}
for every blossom
$B$ and its set $I(B)$. We will show
\begin{equation}
\label{fLPasMatchingEquality}
2(|\gamma(B)\cap M| + |I(B)\cap M|)= f(B)+|I(B)|-1.
\end{equation}
It is easy to see that arithmetic and integrality imply this makes
\eqref{fLPasMatchingInequality}
tight. A
light blossom $B$ has $\eta(B)\in M-I(B)$, so counting degrees gives
$f(B)= 2|\gamma(B)\cap M| + |I(B)\cap M| +1$.
Since $I(B)\cap M=I(B)$ arithmetic gives \eqref{fLPasMatchingEquality}.
A heavy  blossom
 $B$ has $\eta(B)\in I(B)-M$, so counting degrees gives
$f(B)= 2|\gamma(B)\cap M| + |I(B)\cap M|$.
Since $ I(B)\cap M=I(B)-\eta(B)$
arithmetic gives \eqref{fLPasMatchingEquality}.
We conclude the algorithm is correct.

\b

The algorithm can be initialized
with any partial $f$-factor $F$, 
collection of blossoms $\B.$, 
and dual functions $y,z$
that satisfies the invariants: 

\b

Every blossom is mature. Every free blossom is light.

Every blossom  $B$ has $C(B)$ a cycle.

Every edge of a blossom subgraph is tight. Every
nontight edge is
dominated 
(underrated)
if it is unmatched (matched), respectively.

\b

\noindent
In addition $z$ can be nonzero only on pairs $(B,I(B)), B\in \B.$. 

As before
the simplest choice is any partial $f$-factor, no blossoms, $z\equiv 0$, 
and a function $y$ on vertices with
$y(e)\ge w(e)$ ($y(e)\le w(e)$) 
for every edge $e$ that is unmatched (matched), respectively.
This and other initializations are used in Section 
\ref{fStrongSec}.
As with $b$-matching, appropriate initialization shows 
that for arbitrary input graphs our algorithm
finds
a {\em maximum cardinality maximum weight $f$-factor}, i.e.,
a partial $f$-factor that
has the greatest number of edges possible,
and subject to that constraint, has the greatest weight possible.
See Appendix \ref{bfAppendix}.

\subsection*{Efficiency analysis}
The time to find a maximum $f$-factor is $O(f(V)(m+n\log n))$.
The analysis is essentially identical to $b$-matching. 
The biggest difference is 
intepretation of 
the parameter $m$. In the simplest case
every copy of a fixed edge $xy$ has the same weight.
Then as in $b$-matching, 
$m$ denotes the number of nonparallel edges in the given multigraph $G$.
If $G$ has parallel edges with possibly different
weights, the same interpretation of $m$ holds if we assume
the copies of $xy$ are given together with their multiplicities and
weights, and sorted by decreasing weight. 
This follows since a given search 
refers to at most 2 copies of any fixed edge $xy$, 
and a new edge $xy$ is chosen
with the greatest weight possible. If $G$ is not given in
this required form we assume a preprocessing step does the sort.

As in $b$-matching
the algorithms of 
Fig. \ref{fMAlg}--\ref{DualfFactor}
use linear time. In
the expand step 
$C_e$ is easily computed from the blossom data structure,
which represents $C(B)$ by links joining the children of 
the root of $T(B)$. 
For tree-blossom-merging
the supporting tree $T$ 
has minor changes from $b$-matching.
In the definition, an $f$-factor inner blossom is traversed
by a trail $P_i(x,\beta)$, $i\in \{0,1\}$
($i=0$ for $b$-matching). 
Analogous to $b$-matching
we define $P_i^-(x,\beta)$ to be  $P_i(x,\beta)$ with every
maximal subtrail of the form
$P_1(\beta(A),\beta(A))$ replaced by the vertex $\beta(A)$.
In the new  Definition \ref{STreeDefn} an inner blossom $B$
is represented by 
$P_i^-(x,\beta)$
in $T_B$.
(This trail may actually be the single vertex $\beta(B)$,
as illustrated in Fig.\ref{fAlgExFig}(b)--(c).
In the latter note that $P_i(x,\beta)$ does not pass through
$u$ or $v$.)

The supporting tree is maintained similar to $b$-matching. 
A minor difference is that line \ref{ReplaceBCLine} can add
a blossom $A$
represented by $\beta(A)$ to \os..
If this makes $A$ outer the remaining vertices of 
$A-\beta(A)$ are added to the supporting tree. Otherwise $A$ is inner and
the supporting tree is unchanged. 
$A$ may become outer in 
a subsequent  expand step  (when $z(A)=0$) 
or in a blossom step. In both cases the vertices of
 $A-\beta(A)$  are added as above.

The tree-blossom-merging algorithm is used to track blossom steps
for both unmatched and matched edges. No modifications are needed.
To justify this observe that
\eqref{DdeMeqn} shows
once $xy$ becomes eligible at both ends, every dual adjustment
decreases its slack by $2\delta$.
Thus the numerical key used in the blossom-tree merging algorithm
is correct, i.e., it gives
the value of total dual adjustment when $xy$ becomes tight.
 
This argument depends on the simple fact that once
an edge becomes eligible it remains so. This property was also used
in  ordinary matching and $b$-matching, 
but we give a formal proof for $f$-factors here.
Say that an edge $uv$ is {\em eligible at $u$}
if it is eligible for $B_u$. The new term avoids referring to the time-varying
$B_u$. This term and the next lemma come in handy in Appendix \ref{GrowExpandSection}.

\begin{lemma}
\label{AlwaysEligible}
Once an edge $e=uv$ becomes eligible at $u$ it remains so, until
it leaves $\o E - \os.$.
\end{lemma}

\begin{proof}
Once $B_u$ becomes an outer node, it may get absorbed in larger blossoms
but they  are always outer. So $e$ remains eligible 
at $u$
as long as it belongs to
$\delta(B_u)-\os.$.

Suppose $B_u$ transitions from being not in \S. to an inner node. 
The unique eligible edge is $uv=\eta(B_u)$.
If $B_u$ gets expanded vertex $u$ remains in \S..
Let $B'_u$ be the new maximal blossom containing  $u$, $B'_u\in \os.$. 
We have the following possibilities.

$B'_u$ an outer blossom: Clearly $uv$ is eligible for $B'_u$.

$B'_u$ an inner blossom: $uv$ is the base edge of $B'_u$, 
so it remains eligible.

$B'_u$ an atom: The $P_i$ trail used to expand $B_u$ is alternating.
So either $u$ is outer and $uv\notin M$ or $u$ is inner and $uv\in M$.
In both cases $uv$ is eligible at $u$. 

Finally note there is nothing to verify for new nodes of $\os.-B'_u$
created in the expansion.
\end{proof}

\begin{theorem}
\label{fFactorTheorem}
A maximum $f$-factor can be found in time $O(f(V)(m+n\log n))$.
\hfill\qed\end{theorem}

\subsection{Related algorithms} 
\label{fStrongSec}
We start by generalizing  $f$-factors to degree-bounded subgraphs.
For functions $\ell,h:V\to \mathbb{Z_+}$ 
a subgraph $H$ of $G$ is 
an {\em $(\ell,h)$-subgraph} if its
degree function $d_H$ satisfies 
$\ell\le d_H\le h$.

We convert such subgraphs into $f$-factors as follows.
Starting with the given graph $G$ form graph $G_s$
by adding a vertex $s$, with  edges $vs$ of multiplicity
$h(v)-\ell(v)$, $v\in V$, and the loop $ss$ of multiplicity
$\f{h(V)/2}$. Every new edge weighs 0.
Define a degree requirement
function $f$ by 
\[f(v)= \begin{cases}
h(v)& v\in V\\
h(V)& v=s.
\end{cases}
\]

The $(\ell,h)$-subgraphs  $H$ of $G$
correspond to the $f$-factors $F$ of $G_s$,
and corresponding subgraphs have the same weight.
In proof 
starting with an $H$, construct $F$ by
adding $h(v)-d_H(v)$ copies of $vs$ for every vertex $v\in V$,
and $|E(H)|$ copies of $ss$. This gives $s$ degree exactly
$(h(V)- 2|E(H)|) +2|E(H)|= h(V)=f(s)$. Obviously
every $v\in V$ has degree $f(v)$, and $w(H)=w(F)$.
Similarly starting with an $F$, let $H=F-s$.
Clearly
 $H$ is an $(\ell,h)$-subgraph and $w(H)=w(F)$.

\begin{corollary}
A maximum or minimum weight $(\ell,h)$-subgraph 
can be found in time
$O(h(V)(m+n\log n))$. For
a 
minimum weight $(\ell,h)$-subgraph 
the bound improves to
$O(\ell(V)(m+n\log n))$ if 
the weight function is nonnegative or if $h\equiv d_G$ (i.e., we seek a minimum
weight $\ell$-edge cover \cite [Ch.34] {S}).
\end{corollary}

\begin{proof}
To achieve the first time bound execute
the $f$-factor algorithm on $G_s$, using the given weight function
$w$ for maximization and $-w$ for minimization.
Since $f(V+s)=O(h(V))$ and
$G_s$ has $O(m)$ distinct edges,
Theorem \ref{fFactorTheorem} gives the desired bound.

Next consider  minimum weight  $(\ell,h)$-subgraphs with nonnegative $w$.
Use  the $f$-factor algorithm on $G_s$ with weight function $-w$,
and initial dual functions $y\equiv 0$ and $z\equiv 0$ with no blossoms.
These  duals  are feasible for $-w$.
To define the initial matching let  $\delta= \ell(V) \mod 2$.
Match every copy of every edge $vs$, $v\in V$,
and $(\ell(V)-\delta)/2$ copies of $ss$.
To show this matching is valid first note
the degree of $s$ in the matching is
$(h(V)-\ell(V)) + (\ell(V)-\delta)=h(V)-\delta\le f(s)$. 
Also every matched edge is tight since $y\equiv 0$.

The number of searches of the $f$-factor algorithm is
$(\ell(V)+\delta)/2=O(\ell(V))$. The time bound for nonnegative $w$
follows.

Finally
suppose $w$ is arbitrary but $h\equiv d_G$.
Let $N=\set {e}{w(e)<0}$ and let $G'$ be the graph $G-N$.
The minimum weight $(\ell,h)$-subgraph consists of $N$
plus a minimum weight $(\ell',h')$-subgraph on $G'$, where
$\ell'\equiv \max\{\ell-d_N,0\}$ and $h'\equiv d_{G'}$. 
Since $G'$ has a nonnegative  weight function 
the previous case shows the time is 
$O(\ell(V)(m+n\log n))$.
\end{proof}

Next we present a strongly polynomial version of the $f$-factor algorithm. 
We use essentially the same reduction to bipartite matching
as $b$-matching.
Assume the multigraph $G$ is specified by a function 
$c:V \times V \to \mathbb{Z}_+$ that gives the number of parallel copies of each edge.
The algorithm below rounds $c$ up to ensure that edges do not disappear.

\bigskip

Define graph $G'$ by setting
$f'=2\f{f/2}$ and $c'=2\c{c/2}$. Let $M'$ be a
maximum cardinality maximum weight $f'$-factor on $G'$
with corresponding optimal dual function $y$. 
For every edge $e$ 
with $c'(e)>c(e)$ copies of $e$ in $M'$,
remove 1 copy of $e$ from $M'$. 
Let $M$ be the resulting partial $f$-factor on $G$.
Using
$M,y$ (and $z\equiv 0,\ \B.=\emptyset$) as the initial solution, execute
the $f$-factor algorithm of Section \ref{fAlgSec} on $G$.

\bigskip

The analysis is similar to $b$-matching.
Since we assume $G$ has an $f$-factor, $G'$  has a 
partial $f'$-factor with $\ge 
\frac{f(V)}{2}-n$ edges.
At most $m$ matched edges are deleted to form $M$. 
So our $f$-factor algorithm performs $\le m+n$ augmentations.
Thus the time for the entire algorithm is $O(m(m+n\log n))$ plus the time
to find $M',y$. As before
the latter strictly dominates the time.

We find $M',y$ using a graph $G^+$ similar to $b$-matching:
Extend graph $G'$ to $G^+$ 
by adding a vertex $s$  with degree constraint
\[
f'(s)=f'(V)\]
and
edges $vs$ ($v\in V$)
and $ss$ with
multiplicities and weights given respectively by
\[c'(vs)=f'(v),\ 
c'(ss)=f'(V),\ 
w(vs)=0, \ 
w(ss)=Wf'(s) \text{ for } 
W=\max \set{1, |w(e)|} {e\in E(G)}.\]
Note that $f'$ and $c'$ remain even-valued functions.
A maximum cardinality maximum weight $f'$-factor of $G'$
corresponds to a maximum $f'$-factor of $G^+$.
The proof is exactly the same as $b$-matching.

As before we find
 a maximum $f'$-factor of $G^+$ by reducing to 
a bipartite graph $BG$. $BG$ has
vertices
$v_1,v_2
\ (v\in V(G^+))$ and edges  $u_1v_2,v_1u_2\ (uv\in E(G^+))$
with
degree constraints,
multiplicities,
and
edge weights given respectively by
\[f'(v_1)=f'(v_2)={f'(v)/2},\ 
 c'(u_1v_2)=c'(v_1u_2)=c'(uv)/2,\ 
w(u_1v_2)=w(v_1u_2)=w(uv).\]
A loop $uu$ of $G^+$ gives 
edge $u_1u_2$ with multiplicity $c'(uu)$ in
$BG$.
Let $x$ be a maximum 
$f'$-factor
on $BG$
with optimum dual function $y$.
Define an $f'$-factor $M'$ on $G^+$
by taking $x\{u_1v_2, u_2v_1\}$ copies of each edge $uv\in E(G^+)$
(by our summing convention this means $x(u_1u_2)$ copies of a loop  $uu$).
$x$ exists and
$M'$ is a maximum $f'$-factor on $G^+$, by exactly the same proof as $b$-matching.
Define a dual function $y$ by 
$y(v)=y\{v_1,v_2\}/2$.
Applying complementary slackness to the definition of 
the dual function for bipartite $f$-factors \cite[Ch.21]{S}
we get that an $f'$-factor $x$ on $BG$ and a dual function $y$ are both optimum 
iff
for every edge $e$ of $BG$, 
\begin{equation}
\label{fSOptEqn}
x(e)=0 \imp y(e)\ge w(e) ;\
x(e)=1 \imp y(e)\le w(e).
\end{equation}

Now consider an edge $e=uv$ of $G^+$ ($e$ may be a loop).
The matching $x$ on $BG$ has a mirror image
$x'$ defined by $x'(a_1b_2)=x(b_1a_2)$.
Suppose some copy of $e$ in $BG$ is unmatched, say
 $x(u_1v_2)=0$. (\ref{fSOptEqn}) implies
$y(u_1v_2)\ge w(uv)$ as well as $y(v_1u_2)\ge w(uv)$. Thus
\[y(e)=\big((y(u_1)+y(u_2))+ (y(v_1)+y(v_2))\big)/2
\ge 2w(uv)/2=w(e).\]
Similarly if some copy of $e$ in $BG$ is matched
then $y(e)\le w(e)$.
So the  functions $y,0$ are optimum duals for an $f'$-factor on 
(the non-bipartite graph) $G^+$.

The matching $M$ defined from $M'$ is clearly valid on $G$ (i.e.,
nonexistent matched edges are deleted). $y$ is also optimum on $G$
(an unmatched edge of $G$ is present in $G^+$ since $c'$ rounds up).
We conclude that
restricting $M$ and $y$ to $G$, along with 
$z\equiv 0,\ \B.=\emptyset$, gives permissible initial values for
our $f$-factor algorithm. 
In conclusion  the main algorithm is correct.

The problem on $BG$ is a capacitated transportation problem,
where $x$ is an optimum integral solution and $y$ is an optimum dual function
\cite[Ch.21]{S}. 
We solve it using  Orlin's algorithm \cite{Orlin}.
It reduces the
capacitated transportation problem to the uncapacitated case.
The reduction modifies the graph, but it is easy to
see that the optimum dual function $y$ on the modified graph
gives an optimum dual on the given graph. (Alternatively an optimum
dual function can be found from $x$ itself 
using a shortest path computation,
in time $O(nm)$ \cite{AMO}.)

Orlin 
solves the capacitated transportation problem (more generally  
capacitated transhipment)
 in time
$O(m\log n(m+n\log n))$
\cite{Orlin}. It gives both $x$ and $y$. Using this we
obtain our strongly polynomial bound:

\begin{theorem}
A maximum $f$-factor can be found in time
$O(\min\{f(V), m\log n\}(m+n\log n))$.
\hfill\qed\end{theorem}

Next
recall that for
any set of vertices $T$ of even cardinality,
a {\em $T$-join} is a subgraph of $G$ 
that has $T$ as its set of
 odd-degree vertices.
For any edge cost function $c$ it is of interest to find a minimum cost $T$-join. We proceed as follows.

Let $N$ be the set of edges of negative cost.
Define $t=|T|+2|N|$. 
Let $G'$ be the graph $G$ enlarged by adding
$t/2$ loops at every vertex, where each loop has cost 0.
Define a degree-constraint function
\[f(v)= \begin{cases}
t-1& v\in T,\\
t& v\notin T.
\end{cases}
\]
A minimum cost $f$-factor is a minimum cost $T$-join
augmented by enough loops to exactly satisfy the degree constraints.
In proof,
let $J$ be  a minimum $T$-join and $F$ a minimum $f$-factor.
$c(F)\ge c(J)$, since $F$ with all loops deleted 
gives a $T$-join of the same cost.

For the  the opposite inequality note that {\em wlog}
$J$ consists of $|T|/2$ paths, each joining two vertices of $T$,
and $\le |N| $ cycles. The latter holds since we can assume
each cycle contains a negative edge.
Thus any vertex has degree $d(v,J)\le |T|+2|N|=t$, with strict inequality
if $v$ is a terminal.
Furthermore $d(v,J)$ and $f(v)$ have the same parity.
 Hence we can add loops at each vertex to
make $J$ an $f$-factor. We conclude  $c(J)\ge c(F)$.

For our algorithm 
define edge weights to be the negatives of edge costs.
So we seek a maximum weight $f$-factor of $G'$.
Initialize the algorithm with
a matching $M$ consisting of 
every negative edge, and enough loops at each vertex
to make $f(v)\ge d(v,M)\ge f(v)-1$.
Furthermore
 $y\equiv 0$
and there are no blossoms. 
(This initialization is valid
since every loop is tight
and for edges of $G$,
every matched edge is underrated and every other edge is dominated.)
Then execute the $f$-factor algorithm.

The $f$-factor algorithm  performs $\le n/2$ searches. 
A search uses time $O(m+n\log n)$ --
although the graph has  many loops, only 2 loops at each vertex
are processed in any given search.
Also note these special cases: When there are no negative edges
there are $|T|/2$ searches. When there are no terminals
there are $\le |N|$ searches.

\begin{theorem}
A minimum cost $T$-join can be found in time $O(n(m+n\log n))$.
If costs are nonnegative the time is $O(|T|(m+n\log n))$.
If there are no terminals the time is $O(\min\{|N|,n\}(m+n\log n))$.
\end{theorem}

We turn to the shortest path problem on
a connected undirected graph $G$ with a conservative cost function $c$, i.e.,
negative edge costs are allowed but any cycle has nonnegative cost.
We are interested in 
the single source shortest path problem, i.e., given a
source vertex $s$,
we wish to find a shortest path from each vertex $v$ to 
$s$. 
We will show how the blossom tree provides a shortest path tree
that specifies  all these paths.
The discussion is organized into four parts as follows.

$\bullet$ We present a ``base algorithm'' that accomplishes
all our goals for a subfamily of cost functions. 
The goals are to show the existence
of a succinct representation of all shortest paths to $s$, 
and to give  an efficient
algorithm to construct the representation.

$\bullet$ We extend
the base algorithm to show existence of the representation
for arbitrary conservative real-valued cost functions.

$\bullet$ 
For readers familiar with the generalized shortest path tree
(the ``gsp structure'') introduced by Gabow and Sankowski \cite{GS13},
we give a simple verification that
our representation is precisely that structure.

$\bullet$ We extend the base algorithm 
to construct the representation efficiently
for arbitrary conservative integral-valued cost functions.
The time is $O(n(m+n\log n))$,
the best-known time bound to find a shortest $sv$-path for two
given vertices $s,v$.

\bigskip

We will
use the $f$-factor algorithm to find a  search structure
\os.
that handles shortest path queries -- given any vertex $v$, 
a shortest $vs$-path $P$ is composed of  $P_i$ trails 
and can be 
found in time proportional to the length of $P$. 
The {\em base algorithm} accomplishes this assuming
every cycle of the given graph has positive cost
(rather than nonnegative cost).
The base algorithm is as follows.

\b

\noindent
{\em Base Algorithm.} Define edge weights as the negatives of edge costs.
Let $G'$ be the graph $G$ with a loop of weight 0 added at every 
vertex except $s$. 
Define a degree-constraint function
\[f(v)= \begin{cases}
0& v=s,\\ 
2& v\ne s.
\end{cases}
\]
Execute the $f$-factor algorithm to find a maximum $f$-factor $M$ and corresponding duals.
Increase $f(s)$ to 1. Perform a search of the
 $f$-factor algorithm, initialized with $M$ and its duals, 
halting when $\delta=\infty$.
($M$ remains the  matching but the duals and blossoms
may change. Let $y$ denote the final
dual function.)

A query algorithm outputs a shortest $vs$-path, for given $v$,  as follows.
The shortest distance from $v$ to $s$ is  $d(v)=y(v)-y(s)$. 
A shortest
$vs$-path  consists of the nonloop edges in these sets:
the path $P$ in \os. from $B_v$ to $B_s$,
plus
for each blossom $B$ in $P$,
the path $P_i(x,\beta(B))$ with $i$ and $x$ chosen as follows:

{\narrower

{\parindent=0pt

\case{$B=B_v$} $x=v$ and $i$ is chosen so
the first edge of $P_i(v,\beta(B_v))$ is matched.

\case{$B\ne B_v$}
Let $f=xy \in \delta(B,P)-\eta(B)$ with $x\in V(B)$.
Choose $i$ according to \eqref{iForPEqn}, i.e., 
$i=0$ iff 
$f$ and $B$ have the same
M-type. 

}}

\b

Now we prove the base algorithm is correct. 
Note that $M$ is the set of all the loops
since every cycle of $G$ has negative weight.
The rest of the proof is in 3 claims.

Fix an arbitrary vertex $v\ne s$. 
Define the graph $G'_v$ to be $G'$ enlarged with a vertex $v'$ that has
 $f(v')=1$ and a weight 0 edge $vv'$.
Set $y(v')=-y(v)$. 
The new graph satisfies all the invariants
of the $f$-factor algorithm, so we can
imagine a hypothetical search of that algorithm. 

\claim 1 {The hypothetical search executes a blossom step
for edge $vv'$ and augments the matching.}

\bproof 
$vv'$ is tight and eligible for $v'$.
The claim follows if $vv'$ is eligible for $B_v$.
So suppose it is ineligible.
This causes the hypothetical search 
to immediately halt with $\delta=\infty$.
But $G'_v$ is connected so it has an $sv$-path, i.e., an $f$-factor,
contradiction.
\ecproof

The next claim actually holds for arbitrary augmenting paths in the
$f$-factor algorithm, generalizing a property of
Edmonds' algorithm. For notational simplicity we  only prove
the special case needed for our shortest path algorithm.

\claim 2 {The  augment step in the hypothetical algorithm
changes the weight of the matching
by $y(s)+y(v')$.}

\bproof
Every edge of the augmenting trail $A$ is tight, by (I4).
Thus the weight of the matching changes by
$\H{yz}(A-M) - \H{yz}(A\cap M)$. 
The $y$  terms make a net contribution of $y(s)+y(v')$,
since $A$ alternates at every interior vertex.
The $z$ terms make no net contribution. In proof
let $AZ$ be the set of all edges
with a $z(B)$ contribution, i.e., $AZ= A\cap (\gamma(B)\cup I(B))$.
We claim $AZ$ is an alternating trail of even length.
Clearly this implies the $z(B)$ terms make a net contribution of 0.
To prove the claim assume
$AZ\cap \gamma(B)$ is nonempty and denote it as $P_i(x,\beta(B))$.
If $B$ is a maximal blossom we are
using the same notation as the algorithm.
Although $B$ need not be maximal 
$i$ and $x$ are still chosen according to
 the algorithm: If $v\in B$ the case $B
=B_v$ holds.
If $v\notin B$ the case $B\ne B_v$ holds
(with the obvious modification that
we require $f\in A$ rather than $f\in P$).

If $B$ is light then 
$AZ=P_0(x,\beta(B))$. This follows since
a light blossom has  $i=0$
(in both cases); also $I(B)=\emptyset$
since $M$ consists of loops. This shows $AZ$ has even length.
If $B$ is heavy then 
$AZ=P_1(x,\beta(B))+\eta(B)$.
This follows since
a heavy blossom has  $i=1$
(in both cases); also $I(B)=\eta(B)$
since $M$ consists of loops. Again this shows $AZ$ has even length.
\ecproof

\claim 3 {The base algorithm computes correct distances and shortest paths.}


\bcproof 
Since $w(M)=0$,
Claim 2 shows the augmented  matching weighs 
$y(s)+y(v')=y(s)-y(v)$.
The definition of $G'_v$ shows this 
$f$-factor weighs the same as a
maximum weight $sv$-path.
Thus $y(v)-y(s)$ is the cost of a shortest $sv$-path.

The algorithm outputs the nonloops in the augmenting path
of the hypothetical search. These edges are exactly the nonloops
of the optimum $f$-factor, since $M$ consists of loops.
\ecproof

Our second goal is to extend this representation to arbitrary
conservative real-valued cost functions.
We accomplish this using a
symbolic execution of the base algorithm, as follows.

Conceptually increase each edge cost $c(e)$ by the same unknown positive quantity
$\epsilon$.
The algorithm will maintain all numeric quantities as expressions of the form 
$r+s\epsilon$, where $r$ and $s$ are known real-valued quantities
and $\epsilon$ is a symbol. 
The  algorithm maintains the invariant that
the same sequence of grow, blossom, expand, and dual adjustment steps
is executed for all sufficiently small positive values of $\epsilon$.
(The execution for these sufficiently small values uses
real-valued quantities as usual, i.e., no symbolic quantities.
We assume that any ties are broken the same way in every execution.)

The invariant implies that the same sequence of grow, blossom, and expand
steps can be done when $\epsilon=0$. 
In proof observe that 
at any point in the algorithm an edge has slack
$s \epsilon$ (for arbitrary $s$) iff setting $\epsilon=0$
makes the slack 0, i.e., the edge is tight.
These tight edges can be processed in any order when $\epsilon=0$, but they
must all be processed before any others become tight.
The algorithm does this, i.e., all edges with slack $s\epsilon$ become tight
and are processed before any others with slack having $r>0$.
Thus the same sequence of steps is executed for $\epsilon=0$.

To illustrate the workings of the
 base algorithm 
consider
 the computation of  $\delta_1$ in a dual adjustment step.
Each quantity $ |\H{yz}(e)-w(e)|$ in the set defining $\delta_1$
is a linear combination of  quantities
$y(e)$, $z(B)$, and $w(e)$, all of the form $r+s\epsilon$.
Hence it too has that form.
To find $\delta_1$, the minimum quantity, the algorithm declares
$r+s\epsilon<r'+s'\epsilon$
exactly when 
$(r,s)$ precedes $(r',s')$ in
lexicographic order.
If
$r=r'$ this is true for any $\epsilon>0$.
This also holds if $r<r'$ and $s\le s'$. In the remaining case
$r<r'$ and $s> s'$,
it holds for sufficiently small $\epsilon$, i.e.,
$0<\epsilon < (r'-r)/(s-s')$.
Clearly $\delta_1$, as well as $\delta$ and the updated expressions for
$y$ and $z$, all have the desired form
$r+s\epsilon$.

When the algorithm halts, setting $\epsilon$ to $0$
gives valid duals $y,z$ with matching $M$ the set of all loops and
corresponding blossoms. This gives the desired representation.

\bigskip

As mentioned the final search structure \S. is a succinct representation of
all shortest paths from a fixed source $s$. 
Our third goal is to verify this is the gsp-structure.

To do this we will
use two additional properties of 
\S..
Every loop $vv$ is tight and belongs to \S., since $vv\in M$ yet $vv\notin P_v$ 
implies $vv$ is in the augmenting path of the hypothetical search for $v$.
Furthermore wlog $vv$ is a blossom.
In proof $v$ is not an atom, since $\tau(v)$ unmatched would make
$v$ inner. If the minimal blossom $B$ containing $v$ has $v$ atomic in
$C(B)$
we can declare $vv$ a loop blossom
with $z(vv)=0$. $\eta(vv)$ can be chosen as either of the unmatched edges
of $\delta(v,C(B))$.

A formal definition of
the ``generalized shortest path structure'' is presented in Gabow and Sankowski
\cite{GS13}.
The structure  is simply the search structure \S.
with the matched loops removed. To give a brief verification,
the overall structure \os. is a tree 
whose nodes are contracted blossoms that collectively contain
all the vertices of $G$. 
(Note that a node may consist of a single vertex of $G$, 
i.e., a contracted loop. When all edge costs are nonnegative every node is
such a vertex, and the gsp-structure is the usual shortest-path tree.)
The rest of \S. is represented as a collection of cycles, corresponding to
Definitions \ref{ABlossomDefn} and \ref{EtaDefn}
of blossom and the blossom tree $T(\B.)$.
The base edges of blossoms are used
as pointers to follow shortest paths, as in our definition of the 
$P_i$ trails. (These  pointers are called
edge $\tau(N)$ for blossom $N$
in \cite{GS13}; the base vertex of $N$ is called $t_N$.)

The gsp-structure also has numeric labels that prove its validity.
To describe these first note that
every edge of $G$ is unmatched and so
satisfies $\H{yz}(uv)\ge w(uv)$, 
with equality for every edge of \S.
(this includes loops). 
Equivalently  \[d(u)+d(v)+c(uv)= y(u)+ y(v)-2y(s) -w(uv) \ge
-z\set{B} {uv \in \gamma(B)\cup \eta(B)} -2y(s)\] with equality on \S..
Define
$z':2^V\to \mathbb {R}$
by $z'(B)=-z(B)$, $z'(V)=-2y(s)$ to get
\[d(u)+d(v)+c(uv)\ge 
z'\set{B} {uv \in \gamma(B)\cup \eta(B)}
\]
with equality for every edge of the representation (including loops).
This is 
the exact relation satisfied by the labels of the gsp-structure, where
$d$ labels each vertex, $z'$ labels each node as well as $V$
with 
$z'(B)\le 0$ for every $B\ne V$.

For the last goal assume the given 
cost function $c$ is integral-valued, as is the case in algorithms.
We will simply execute the base algorithm using an integral
blow-up   of $c$.
Another approach would be
symbolic execution as given above. 
We prefer using integral values, since we have given a
complete efficiency analysis of this case.
(For instance Appendix \ref{bfAppendix}
 shows the numbers computed by the algorithm
are acceptably small.
We have not analyzed the size of the $s$ coefficients in the symbolic execution.)

Define the cost function 
\[c'=4nc+1.\] 
$c'$ has no 0-cost cycles.
For any $v$, a shortest $sv$-path wrt $c'$ is a 
shortest $sv$-path wrt $c$ that in addition has the smallest length possible.
We execute the base algorithm using $c'$.
Clearly this algorithm can answer shortest path queries for $c$.
(Also the base algorithm verifies that  $c$ is conservative --
if not it returns a matching
with negative cost.)
However we wish to find the complete gsp-representation, which
uses the
optimum dual functions $y,z$ 
for its numerical labels. (These labels complete the representation since
they provide a simple check that $c$ is conservative.)
So our algorithm for general cost functions  requires one more step,
 to transform the duals  given by the algorithm for $c'$ to 
those for $c$. 

The extra step uses the following terminology. For any duals $y,z$ and corresponding blossom structure, define the function $Z$ on blossoms $B$ by
\[Z(B)= z\set {A} {A\supseteq B,\, A \text{ a blossom}}.\]
Obviously $Z$ uniquely defines the dual function $z$ via the relations
$z(B)=Z(B)-Z(p(B))$ for $p(B)$ 
the parent of $B$ in the blossom tree, with the convention
$Z(p(B))=0$ for every maximal blossom $B$.

Also call any edge $e$ a {\em  witness} for blossom $B$ if
$e\in C(B)$ and
$Z(B)= w(e)-y(e)$.
Any edge of $e\in C(B)$ is tight, i.e.,
$w(e)=\H{yz}(e) = y(e)  + z\set {A} {e \in \gamma(A)\cup I(A)}$.
So $e$ is a witness for $B$ if $B$ is the minimal blossom containing
both ends of $e$
and $e\notin I(A)$ for any blossom $A\pcon B$.
In general $B$ may not have a witness. But
a blossom at the end of our algorithm does have a witness.
In proof first note this is clear if $C(B)$ is just a loop.
So assume $C(B)$ contains $r$ nodes, $r> 1$. 
The node $\alpha(B)$ does not have its base edge
in $C(B)$ (this base  edge may not even exist). 
So the $r$ nodes collectively have at most $r-1$ base edges in $C(B)$. 
$C(B)$ has $r$ edges that are not loops.
So at least one of these edges, say $e$,
is not the base edge of either of its ends.
Since $e$ is unmatched
this implies $e\notin I(A)$ for any blossom $A$. Since $B$ is the minimal blossom
containing both ends of $e$, $e$ is a witness for $B$.

The algorithm defines the desired duals as follows.
Let $y',z'$ be the given duals for $c'$, and
$y,z$ the desired duals  for $c$.
To construct $y,z$ first do an extra dual adjustment step to make
$y'(s)$ a multiple of $4n$. (In other words 
 use $\delta= y'(s)-4n\f{y'(s)/4n}$ in a dual adjustment step.
Let $y',z',Z'$ now denote these adjusted duals.)
The base algorithm provides the optimum $f$-factor for $v$, call it $P_v$. 
The new algorithm defines
\[
\begin{array}{llll}
y (v)&=&-w(P_v) +y'(s)/4n & \text{for every vertex } v\in V\\
Z(B)&=&w(e)-y(e)&
\text{for every blossom $B$ and
$e$ a witness for $B$} 
\end{array}
\]
Also $z$ is the dual function corresponding to $Z$.

\begin{lemma}
The above functions $y,z$ are valid optimum duals. Specifically they satisfy (I4) for
the unmodified weight function $w=-c$ and
the $f$-factor $M$ and  blossoms found by the algorithm.
\end{lemma}

\begin{proof}
Let $w'$ denote the modified weight function used in the algorithm,
i.e., 
\[w'(e)= 4nw(e)-\mu(e)  \text{ where } \mu(e)=
\begin{cases}
1 &\text{$e$ an edge of $G$}\\
0&\text{$e$ a  loop.}
\end{cases}\]
(Recall $w(e)=0$ for $e$ a loop.)
Let 
$\ell(P_v)$ be the length of the $sv$-path contained in $P_v$.
So
$y'(v)=-w'(P_v)+y'(s)=-4nw(P_v)+\ell(P_v)+y'(s)$.
Thus \[y'(v)= 4ny(v)+\ell(P_v).\]


Take any blossom $B$ and any witness $e=uv\in C(B)$. So 
\begin{equation}
\begin{split}
\label{Z'ZShtstPthEqn}
Z'(B)&=w'(e)-y'(e)= (4nw(e)-\mu(e))-4ny(e)-\ell(P_u)-\ell(P_v)\\
&= 4nZ(B)-r(B)
\end{split}
\end{equation}
where \[0\le r(B) \le 2n-1.\]
Although there may be several choices for $e$,
\eqref{Z'ZShtstPthEqn} shows $Z(B)$ is uniquely defined, since
the interval $[Z'(B), Z'(B)+2n-1]$ contains a unique multiple of
$4n$.

Next observe that $z(B)$ is nonnegative:
\begin{equation*}
\begin{split}
4nz(B)&=4n(Z(B)-Z(p(B)))=Z'(B)+r(B) - Z'(p(B))-r(p(B))\\
&=z'(B)+r(B)-r(p(B))\ge -(2n-1).
\end{split}
\end{equation*}
The last inequality follows since $z'(B)\ge 0$
and $r(p(B))\le 2n-1$ (even if $B$ is maximal).
Clearly $4nz(B)\ge -(2n-1)$ implies $z(B)\ge 0$.

Finally we show (I4) for any edge $e=uv$ (including loops).
Let $\sigma' = w'(e)-\H{y'z'}(e)$
and $\sigma =4n (w(e)-\H{yz}(e))$.
We claim \[|\sigma'-\sigma|\le 4n-2.\]
The claim  gives the desired conclusion (I4).
In proof consider two cases. 
If $e$ is tight {\em wrt} $y',z'$
then $\sigma'=0$. Since $\sigma$ is a multiple of $4n$
the claim shows it must be 0, 
so $e$ is tight {\em wrt} $y,z$.
If $e$ is not tight 
then $\sigma$ is not uniquely determined, but
the claim shows $\sigma$ cannot have opposite sign from $\sigma'$.
Thus 
either $e$ is dominated {\em wrt} both pairs of duals
or underrated {\em wrt} both.

To prove the claim 
let $B_u$ be the minimal blossom containing $u$ and similarly for $B_v$.
Consider two cases.

\case{$e$ is not the base edge of both $B_u$ and $B_v$}
This case implies there is 
a blossom $B$ such that
$\set {A} {e \in \gamma(A)\cup I(A)}
=\set {A} {A\supseteq B,\, A \text{ a blossom}}$.
In proof,
if $e$ is a loop then clearly we can take $B$ as $B_u=B_v$.
If $e$ is a nonloop witness for a blossom $B$ then $B$ is the desired blossom.
If $e =\eta(B_u) \ne \eta(B_v)$ then
$B=B_u$.

Using blossom $B$ we have
\[\sigma' = w'(e)-y'(e)-Z'(B)=
4n(w(e)-y(e)-Z(B)) -\mu(e)-\ell(P_u)-\ell(P_v)+r(B)
=\sigma \pm r\]
where 
 $0\le r\le 2n-1$. 
The claim follows since $2n-1\le 4n-2$.

\case{$e=\eta(B_u)=\eta(B_v)$}
Let $B$ be the minimal blossom containing both $u$ and $v$.
Thus $e=\eta(A)$ for every blossom $A$ with
$u\in A \pcon B$ or $v\in A \pcon B$. Clearly
$\set {A} {e \in \gamma(A)\cup I(A)}
=\set {A} {A \text{ a blossom containing $B_u$ or $B_v$}}$.
Thus
\begin{align*}
\sigma' &= w'(e)-y'(e)-Z'(B_u)-Z'(B_v)+Z'(B)\\
&=4n(w(e)-y(e)-Z(B_u)-Z(B_v)+Z(B)) -1-\ell(P_u)-\ell(P_v)
+r(B_u)+r(B_v)-r(B)\\
&=\sigma\pm r
\end{align*}
where 
 $0\le r\le 4n-2$. 
The claim follows.
\end{proof}

We conclude that the general algorithm is correct.

Regarding efficiency since $f(V)<2n$ the complete execution of the $f$-factor
algorithm runs in time $O(n(m+n\log n))$. This dominates the total time.
For a more precise estimate,
the set $N$ of negative edges 
 for a conservative cost function is acyclic, so 
$|N|<n$. 
Let $W$ be the largest magnitude of a given cost.
Initialize the dual functions by $z\equiv 0$ with no blossoms
and
\[
y(v)=\begin{cases}
W&d(v,N)>0\\
0&d(v,N)=0
\end{cases}
\]
Match every loop $vv$ where $y(v)=0$.
Now
 the $f$-factor algorithm performs $O(|N|)$ searches, using time
$O(|N|(m+n\log n))$ assuming $N\ne \emptyset$. This again dominates the time.

\begin{theorem}
The generalized shortest path structure 
representing all  shortest paths from $s$
is the $f$-factor algorithm search structure \S.. It
 can be constructed in time 
$O(n(m+n\log n))$. More generally if $N\ne \emptyset$ is the set of
negative cost edges
the 
time is
$O(|N|(m+n\log n))$.
\end{theorem}

The first time bound is given in Gabow and Sankowski
 \cite{GS13}. The second bound
shows how the time increases with more negative edges.
For example in a graph with $O(1)$ negative edges the algorithm is as fast
as Dijkstra's algorithm, which allows no negative edges.

A similar dependence on negative edges holds for conservative directed graphs:
The single source shortest path problem can be solved in
time $O(n_N (m+n\log n))$, for $n_N$ the number of
vertices incident to a negative edge.
In contrast
the Bellman-Ford algorithm runs in time $O(nm)$ with no dependence on $N$.
To achieve our  time bound we
model
the digraph $G$ as an undirected graph: The vertex set is
$\set{v_1,v_2} {v\in V(G)-s}+ s_2$; 
the edge set is
$\set{v_1 v_2} {v\in V(G)-s}\cup 
\set{u_2 v_1} {uv\in E(G)}$, 
with $c(v_1v_2)=0, c(u_2v_1)=c(uv)$,
$f(v)=1$ for every vertex.
The initialization sets $y$ to
$W$ (the largest magnitude of a given cost)
if $v$ is on a negative edge else $0$.
The initial matching consists of the edges $v_1v_2$ where
$v$ is not on a negative edge.

\fi
\or

\input nca
\input alpha

\fi

\ifcase 1 
\or

\section*{Acknowledgments}
The author thanks  Bob Tarjan for some fruitful early conversations, 
as well as
Jim Driscoll. Also thanks to  an anonymous referee for a careful 
reading 
and many 
suggestions.

\clearpage
\setcounter{section}{0}
\renewcommand{\thesection}{\Alph{section}}
\renewcommand{\thetheorem}{\Alph{section}.\arabic{theorem}}

\setcounter{equation}{0}
\renewcommand{\theequation}{\Alph{section}.\arabic{equation}}
\section{Dual adjustment step for Edmonds' algorithm}
\label{EdAppendix}
To state the dual adjustment step we first review the
linear program for  perfect matching.
Its  variables are given by
the function $x:E \to \mathbb {R_+}$ which
 indicates whether or not an edge is matched. 
The following linear program
 for maximum matching uses our summing convention, e.g.,
$x(\delta(v))=\sum_{e\in \delta(v)}x(e)$.

\hskip60pt maximize $\sum_{e\in E} w(e)x(e)$ subject to

\[\begin{array}{llll}
x(\delta(v))&=&1&\hbox{for every } v\in V\\
x(\gamma(B))&\le& \f{|B|\over 2}&\hbox{for every } B\con V\\ 
x(e)&\ge& 0&\hbox{for every }  e\in E
\end{array}
\]

The dual LP uses dual functions
$y:V \to \mathbb {R}$,
$z:2^V\to \mathbb {R_+}$.
Define
$\H{yz}:E\to \mathbb {R}$ by
\begin{equation}
\H{yz}(e) = y(e)  + z\set {B} {e \con B}.
\end{equation}

\noindent
(Note for  $e=vw$,  $y(e)$ denotes
$y(v)+y(w)$ and
$z\set {B} {e \con B}$
 denotes 
$\sum_{e \con B} z(B)$.) 

\hskip60pt minimize 
$y(V) +  \sum_{B\con V} \f{|B|\over 2}\, z(B) $
subject to

\[\begin{array}{llll}
\H{yz}(e) &\ge& w(e)&\hbox{for every } e\in E\\
z(B)&\ge& 0&\hbox{for every }  B\con V
\end{array}
\]

\noindent
$e$ is {\em tight} when
equality holds in its constraint, i.e., $\H{yz}(e) = w(e)$.
The algorithm maintains the complementary slackness conditions:

$x(e)>0 \imp e$ is tight.

$z(B)>0 \imp$ $x(\gamma(B))=
\f{|B|\over2}$.

\noindent
In addition every edge in a blossom subgraph is tight (so blossoms can be rematched). It is easy to see the following dual adjustment step maintains these conditions.

\begin{algorithm}
\DontPrintSemicolon

$\delta_1\gets\min \set{y(e)-w(e)}{e=uv \mbox{ with $u$ outer, } v\notin \S.}$\;
$\delta_2=\min \set{(y(e)-w(e))/2}{e=uv \mbox{ with $u,v$ in distinct
outer blossoms}}$\;
$\delta_3=\min \set{(z(B)/2}{B \mbox{ an inner blossom of }\os.}$\;
$\delta=\min \{\delta_1,\delta_2,\delta_3\}$\;

\lFor{every vertex $v\in \S.$}\\
\Indp\lIf{$v$ is inner}{$y(v)\gets y(v)+\delta$}
\lElse{$y(v)\gets y(v)-\delta$}\;
\Indm\lFor{every blossom $B$ in \os.}\\
\Indp\lIf{$B$ is inner}{$z(B)\gets z(B) -2\delta$}
\lElse{$z(B)\gets z(B) +2\delta$}\;

\caption{Dual adjustment step in Edmonds' algorithm.}
\label{DualEdmonds}
\end{algorithm}

\section{Details for $b$-matching and $f$-factor algorithms}
\label{bfAppendix}
The LPs for $b$-matching are the obvious generalizations
of ordinary matching:

\hskip60pt maximize $\sum_{e\in E} w(e)x(e)$ subject to

\[\begin{array}{llll}
x(\delta(v))+2x(\gamma(v))&=&b(v)&\hbox{for every } v\in V\\
x(\gamma(B))&\le& \f{b(B)\over 2}&\hbox{for every } B\con V\\ 
x(e)&\ge& 0&\hbox{for every }  e\in E
\end{array}
\]

\hskip60pt minimize 
$\sum_{v\in V} b(v)y(v) +  \sum_{B\con V} \f{b(B)\over2}\, z(B) $
subject to

\[\begin{array}{llll}
\H{yz}(e) &\ge& w(e)&\hbox{for every } e\in E\\
z(B)&\ge& 0&\hbox{for every }  B\con V
\end{array}
\]

The complementary slackness conditions
are essentially the same as ordinary matching:

$x(e)>0 \imp e$ is tight.

$z(B)>0 \imp$ $x(\gamma(B))=
\f{|b(B)|\over2}$.

As mentioned in Section \ref{bBlossomSec}
complementary slackness requires that a blossom
$B$ with $z(B)>0$
has precisely one incident matched edge,
i.e., \eqref{CSforBMatchingEqn} holds.
Let us review this fact. Our LP
constraint
$x(\gamma(B))\le\f{b(B)/2}$ 
is redundant if $b(B)$ is even (since
$2x(\gamma(B))\le x\set{\delta(v)}{ v\in B }
+2x\set{\gamma(v)}{v\in B}
=b(B)$).
So we can assume $b(B)$ is odd. Now equality 
in the constraint amounts 
to \eqref{CSforBMatchingEqn}.

The dual adjustment step differs from
ordinary matching only in allowing
a  loop to cause a blossom (Fig.\ref{DualbMatch}).
Like ordinary matching, the numerical quantities in our algorithm
are always half-integers. More precisely assume all given weights $w(e)$
are integral.
Assume 
either every initial $y$-value is integral or
every initial $y$-value is integral plus $1/2$; furthermore
every initial $z$-value is integral.
This  assumption holds for common initializations, 
e.g., $y\equiv \max_{e\in E}w(e)/2$
and $z\equiv 0$. 
It also holds for the initialization in our strongly polynomial
algorithm, Section \ref{bStrongSec}. 
(Note the $y$-values for $BG$, i.e., the transportation problem,
are integral-valued. So 
\eqref{yDefnTransportationEqn} gives integral $y$-values
for our algorithm assuming we double the given weight function.)
We will show that throughout the algorithm 
\begin{equation}
\label{yzIntegralEqn}
(\forall v^{\in V})(y(v)\in \mathbb{Z}/2)
\hbox{ and }
(\forall B^{\con V})(z(B)\in \mathbb{Z}).
\end{equation}
To prove \eqref{yzIntegralEqn} assume it holds before a dual adjustment. 
Examining the changes of Fig.\ref{DualbMatch} 
shows it suffices to prove $\delta$ is a half-integer.
Clearly $\delta_1$ and $\delta_3$ are half-integers.
We will 
show any edge joining two vertices of \S. has integral $y$-value.
This makes $\delta_2$ half-integral and completes the proof.

Any tight edge has $\H{yz}(e)=w(e)$. 
So \eqref{yzIntegralEqn} (specifically the integrality
of $z$) implies $y(e)\in \mathbb{Z}$. Any vertex
$v$ 
in \S. is joined to a free vertex $x$ by a path $P$ of tight edges.
Thus $y(v) +2y\set{u}{u\in P-v-x}+y(x)\in \mathbb{Z}$, i.e.,
$y(v) +y(x)\in \mathbb{Z}$.
Taking any other vertex $v'$ of \S. with similar relation
$y(v') +y(x')\in \mathbb{Z}$ gives
$y(v) +y(v')+y(x)+y(x')\in \mathbb{Z}$.
A free vertex is always outer, so its $y$-value always decreases by $\delta$.
So the initialization implies  $y(x)+y(x')\in \mathbb{Z}$.
Thus
$y(v) +y(v')\in \mathbb{Z}$ as desired.

The magnitude of numbers computed by the algorithm can be bounded as follows.
Let $W$ be the largest magnitude of an edge weight.
Assume all initial $y$ values are $\le W$ and $z\equiv 0$. 
We claim the largest value of $\Delta$ is $\le W b(V)$.
Clearly this implies every $y$ and $z$ value is $\le 2Wb(V)$.

To prove the claim consider  any point in the algorithm.
Let $b'(v)$ be the remaining degree requirement at
 $v$, i.e., $b'(v)=b(v)-d(v,M)$ for $M$ the current matching.
Since every matched edge is tight,
\begin{equation}
\label{DualToMEqn}
w(M)=\sum_{e\in M} \H{yz}(e)=
\sum_{v\in V} d(v,M)y(v)+\sum_{B\in\B.} \f{|B|/2}z(B).
\end{equation}
Thus we can rewrite the current value of the dual objective function as
$\sum_{v\in V} b'(v)y(v)+ w(M)$.
The dual adjustment 
preserves tightness of the edges of $M$. So
\eqref{DualToMEqn} holds and
the updated dual objective can be rewritten the same way.
 Thus the dual adjustment 
decreases the dual objective  value by
$b'(V)\delta\ge 2\delta$. The initial dual objective is $\le b(V)W$.
The final objective is the weight of a maximum $b$-matching, which is
$\ge -Wb(V)/2\ge -Wb(V)$. So we always have $\Delta=\sum \delta\le b(V)W$.

\begin{algorithm}[h]
\DontPrintSemicolon

\def\Or{\KwSty{or}\ }
$\delta_1\gets\min \set{y(e)-w(e)}{e=uv\notin M \mbox{ with $B_u$ outer, } B_v\notin \S.}$\;
$\delta_2=\min \set{(y(e)-w(e))/2}{e=uv\notin M \mbox{ with $B_u,B_v$ outer,  
either $B_u\ne B_v$ or $u=v$ atomic}}$\;
$\delta_3=\min \set{z(B)/2}{B \mbox{ an inner blossom of }\os.}$\;
$\delta=\min \{\delta_1, \delta_2, \delta_3\}$\;

\lFor{every vertex $v\in \S.$}\\
\Indp\lIf{$B_v$ is inner}{$y(v)\gets y(v)+\delta$}
\lElse{$y(v)\gets y(v)-\delta$}\;
\Indm\lFor{every blossom $B$ in \os.}\\
\Indp\lIf{$B$ is inner}{$z(B)\gets z(B) -2\delta$}
\lElse{$z(B)\gets z(B) +2\delta$}\;

\caption{Dual adjustment step for $b$-matching.}
\label{DualbMatch}
\end{algorithm}

As with ordinary matching, other versions of weighted $b$-matching
have LPs that are minor modifications of the original. Correspondingly, minor
modifications of our algorithm find such matchings. We illustrate with
maximum cardinality maximum weight $b$-matching
(defined in Section \ref{bMAnalSec}).
It is convenient to treat the more general problem of finding
a $b$-matching of maximum weight subject to the constraint
that it contains exactly $k$ edges.

The primal LP relaxes the vertex degree constraint to
\[
x(\delta(v))+2x(\gamma(v))\le b(v)\hskip 20pt \hbox{for every } v\in V
\]
and adds the cardinality constraint
\[
x(E)= k.
\]
The dual problem has a variable $c$ for the cardinality constraint,
the left-hand side of the dual edge constraint 
changes from $\H{yz}(e)$ to  $\H{yz}(e)+c$,
and the nonnegativity constraint
$y(v)\ge 0$ is added. The additional complementary slackness constraint
is
\[
y(v)>0 \imp x(\delta(v))+2x(\gamma(v))= b(v)\hskip20pt \hbox{for every } v\in V.\]

To find such a matching
we initialize our algorithm 
using a common value for every $y(v)$.
The algorithm halts after the search that increases the 
matching size to $k$. For maximum cardinality maximum weight $b$-matching,
this is the first time a search fails. To get an optimal LP solution,
let $Y$ be the common final value for $y(v)$, $v$ free,
or 0 if no such vertex exists.
(Fig.\ref{DualbMatch} implies that throughout the algorithm
all free vertices have the same $y$-value, and this value is 
the minimum $y$-value.)
Decrease all $y$ values by $Y$ and set $c=2Y$.
This solves the new LP. 
(In the dual edge constraint the new $y$-values decrease 
$\H{yz}(e)$ by $2Y$, which is balanced by
the new LP term $c=2Y$.) 
We conclude that our algorithm is correct.
It also proves the LP formulation is correct.

\bigskip

The LPs for $f$-factors incorporate limits
on the number of
copies of an edge as well as  $I(B)$ sets of blossoms.
(The graph may have parallel edges, so wlog we 
allow only 1 version of each copy to be in the $f$-factor.)

\hskip60pt maximize $\sum_{e\in E} w(e)x(e)$ subject to

\[\begin{array}{llll}
x(\delta(v))+2x(\gamma(v))&=&f(v)&\hbox{for every } v\in V\\
x(\gamma(B) \cup I )&\le& \f{f(B)+|I|\over 2}&\hbox{for every } B\con V,\,
I\con \delta(B)\\ 
x(e)&\le& 1&\hbox{for every }  e\in E\\
x(e)&\ge& 0&\hbox{for every }  e\in E
\end{array}
\]

The dual LP uses dual functions
$y:V \to \mathbb {R}$,
$z:2^V \times 2^E\to \mathbb {R_+}$.
Define
$\H{yz}:E\to \mathbb {R}$ by
\begin{equation}
\label{fHyzEqn}
\H{yz}(e) = y(e)  + z\set {(B,I)} {e \in \gamma(B)\cup I}.
\end{equation}

\hskip60pt minimize 
$\sum_{v\in V} f(v)y(v) +  
\sum_{B\con V,I\con \delta(B)} \f{f(B)+|I|\over2}\, z(B,I) +u(E)$
subject to

\[\begin{array}{llll}
\H{yz}(e) +u(e)&\ge& w(e)&\hbox{for every } e\in E\\
u(e)&\ge& 0&\hbox{for every } e\in E\\
z(B,I)&\ge& 0&\hbox{for every }  B\con V,\,I\con \delta(B)
\end{array}
\]

In
our algorithm
every nonzero $z$ value has the form $z(B,I(B))$ for $B$ a mature blossom.
So we  use the notation $z(B)$ as a shorthand for $z(B,I(B))$.

Say that $e$ is {\em dominated, tight,} or {\em underrated}
depending on whether
$\H{yz}(e)$ is  $\ge w(e)$, $= w(e)$, or $\le w(e)$, respectively;
{\em strictly dominated} and {\em strictly underrated} refer to the possibilities  $>w(e)$ and $< w(e)$ respectively.
The complementary slackness conditions for optimality
can be written with $u$ eliminated as

$x(e)>0 \imp e$ is  underrated

$x(e)=0 \imp e$ is  dominated

$z(B)>0 \imp$ $x(\gamma(B) \cup I(B) )= \f{f(B)+|I(B)|\over 2}$.

The numbers computed by the algorithm are analyzed similar to $b$-matching.
The same argument applies to show the algorithm always works with half-integers.
The same bound holds for the magnitude of numbers. The only addition to the
analysis is to account for the term $u(E)$ in the dual objective function.
Clearly the optimum $u$ function is defined by setting
$u(e)$ equal to the slack in $e$, $w(e)-\H{yz}(e)$, for 
every edge $e\in M$.
So \eqref{DualToMEqn} has the analog,
$w(M)= \sum_{e\in M} \H{yz}(e)+u(e)=
\sum_{v\in V} d(v,M)y(v)+\sum_{B\in\B.} \f{\frac{f(B)+I(B)}{2}}z(B)+u(E)$.
This equation holds both before and after the dual adjustment.
(Note the dual adjustment will change $u$ values also, and
each $u(e)$ may increase.) The
dual objective function can be rewritten just as before, as
$\sum_{v\in V} f'(v)y(v)+ w(M)$, both before and after the adjustment
step.
The rest of the analysis is identical to $b$-matching.

Similar to $b$-matching our algorithm extends to
variants of the maximum $f$-factor problem.
We again illustrate with
maximum cardinality maximum weight partial $f$-factors.
The LP is modified exactly as in $b$-matching.
Our modified algorithm and 
the definition of new LP variables is exactly the same.
The only difference in the analysis is 
that the new complementary slackness conditions
for edges are

$x(e)>0 \imp \H{yz}(e)+c \le w(e)$ 

$x(e)=0 \imp \H{yz}(e)+c \ge w(e)$. 

\noindent
As before the quantity $\H{yz}(e)+c$ equals the algorithm's
value of $\H{yz}(e)$, so these conditions are equivalent to the original ones.

\section{Grow/Expand steps}
\label{GrowExpandSection}
\def\zI{Z\_TO\_I} 
\def\ZY{DEL} 
We give a simple data structure to handle grow and expand steps.
First consider ordinary matching.
At any point in a search, for any vertex $v\in V$ define
$slack(v)$ to be the smallest slack in an unmatched edge
from an outer node to $v$. If $v\notin \S.$ 
and $slack(v)<\infty$, 
dual adjustments
reduce $slack(v)$. When $slack(v)$ becomes 0 a grow step can
be performed to make $B_v$ inner. But if $B_v$ is a blossom,
it may become inner before $slack(v)$ becomes 0. This blossom
may later get expanded, and $v$ may leave \S.. If not some smaller
blossom containing $v$ may get expanded causing $v$ to leave \S..
Continuing in this fashion $v$ may oscillate in and out of \S., becoming eligible and
ineligible for grow steps. 
This makes tracking potential 
grow steps nontrivial.
Note there is no such complication for grow steps using a matched
edge to add a new outer node, since matched edges are always tight
and outer nodes never leave \S..

The same overview applies to $b$-matching. $f$-factors are more general,
since matched edges need not be tight.
We first present the algorithm 
that applies to ordinary matching and
$b$-matching. Then we extend the algorithm  to $f$-factors.

\paragraph*{Data structures}
As in Section \ref{EdAlgSec}  for ordinary matching and
\ref{bBlossomSec} for $b$-matching and $f$-factors,
we use a tree representing
the laminar structure of blossoms. Specifically 
at the start of a search the current blossoms (from previous searches)
form a tree $\B.$.
The root of \B.  
corresponds to $V$, and each leaf corresponds to a vertex of $G$. 
The children of $B$ in \B. are the blossoms and atoms
in the cycle $C(B)$ forming
$B$. 
The subtree of a blossom $B$ has size $O(|V(B)|)$, as in Sections \ref{EdAlgSec} 
and \ref{bBlossomSec}.

Recall (Section \ref{TBMAlgSec})
the {rank} of a \B.-node $B$ is
$r(B)=\f{\log |V(B)|}$.
A \B.-child of $B$ is 
{\em small} if it has rank $<r(B)$, else {\em big}.
Clearly $B$ has at most one big child.
So the rank $r(B)$ descendants of $B$ form a path $P$
starting at $B$. Each node 
on $P$ except $B$ 
is the big child of its parent.%
\footnote{$P$ is a slight variant of the ``heavy path''
of  \cite{HT, T79}.} 
The data structure marks each node as big or small.

We also 
use this notion:
A child of a node on the above path $P$ is
a {\em small component} of $B$. 
Clearly a small component of $B$ is a small child of its parent.
If $B$ is a blossom then 
$V(B)=\cup \set{V(A)}{A \text{ a small component of $B$}}$.
(This fails if $B$ is a leaf of \B.. Such a $B$ has no children or components.)

The main task for the data structure is tracking  $slack(v)$ values.
Obviously this requires tracking $B_v$
(as usual $B_v$ denotes the currently maximal blossom or atom containing $v$).
The values $node(v)$ defined below allow identifying $B_v$ in $O(1)$ time.
$node(v)$ values are also used in blossom and augment steps to
compute paths in \os..

Recall the data structure for numerical quantities 
given in the last subsection of Section \ref{bMAnalSec}, in particular these
definitions:
$\Delta$  is the sum of all dual adjustment quantities $\delta$
in the current search.
Any outer vertex $v$ has a  quantity $Y(v)$, such that
the current value of $y(v)$ is $Y(v)-\Delta$.
A global Fibonacci heap \F. has entries for candidate
grow, blossom, and expand steps, with key equal to
the value of $\Delta$ 
when the corresponding edge becomes tight.

To compute current $y$ and $z$ values for nonouter nodes,
we
use an auxiliary quantity 
$\ZY(B)$ that tracks
$z$-values of expanded blossoms that have been converted into $y$-values.
To define this quantity let $y_0$ and $z_0$ denote the dual functions
at the start of the current search.
The algorithm stores the quantity
\[Y(v)=y_0(v).\]
Every node $B$ of \B. is labelled with the quantity
\begin{equation}
\ZY(B)=\mbox{\small{$\frac{1}{2}$}}\,z_0\set {A} {A \text{ a proper ancestor of $B$ in \B.}}.
\end{equation}
Observe 
that when $B$ is a maximal blossom, 
$\ZY(B)$ is equal to the total of all dual adjustments made while
$B$ was properly contained in an inner blossom.
At any point in time current $y$ values are computed by
\begin{equation}
\label{yYZtoYeqn}
y(v)=\begin{cases}
Y(v)+\ZY(B_v)&B_v \text{ not in \os.}\\
Y(v)+\ZY(B_v)+\Delta-\Delta_0(B_v)&B_v \text{ an inner node}
\end{cases}
\end{equation}
where 
$\Delta_0(B)$ denotes 
the value of $\Delta$ when blossom $B$ became an inner node
(blossom or atom).
We will  compute $y(v)$ in $O(1)$ time when it is needed.
To do this we must identify $B_v$ in $O(1)$ time.
This is done using the pointer $node(v)$, as we will describe below.

We track the best candidate edges for grow steps 
from outer nodes using a system
of Fibonacci heaps.
At any point in the algorithm every maximal nonouter  blossom $B$ 
has a Fibonacci heap $\F._B$.
The nodes of  $\F._B$ are the 
small components of $B$.
Thus if $B$ is not a node of \os.,
the smallest slack of an unmatched edge for a grow step to $B$
is the smallest value $slack(v)$, $v$ a vertex in $V(A)$, $A$ a blossom
or atom with a node in $\F._B$. 

The data structure must also handle  maximal nonouter atoms $B$.
For uniformity we assume atoms are handled like  blossoms -- they
have a Fibonacci heap of one node, the atom itself. 
We will not dwell on this case, the reader can make the
obvious adjustments for maximal nonouter atoms.

Returning to the general case, the data structure does not
explicitly store values $slack(v)$, 
since they change with every dual adjustment.
Instead
we store offsetted versions of related quantities as follows.

Observe that whenever
$B_v$ is not in \os.,
the slack in an unmatched edge $uv$ with
$B_u$ outer is 
\[y(u)+y(v)-w(uv)=(Y(u)-\Delta)+(Y(v)+\ZY(B_v))-w(uv).
\]
(Note this relation holds regardless of prior history, i.e.,
when $u$ was first in an outer node or the pattern of  $v$'s movement
 in and out of \S..)
So the data structure stores the quantity
\[
SLACK(v)=\min \set{Y(u)+Y(v)-w(uv)}{B_u \text{ outer}, uv \in E-M}
\]
for every vertex $v$ where $B_v$ is not outer. 
Note that the expression for a given edge $uv$ never changes in value, 
even as $B_u$ changes. The data structure
also records
the minimizing edge $uv$.
$SLACK(v)$ and its minimizing edge are updated as new outer nodes are created. 
At any point in time
when $v$ is not in \S.,
the current value of $slack(v)$ is
\begin{equation}
\label{SlackEqn}
slack(v)= SLACK(v)-\Delta+\ZY(B_v).
\end{equation}
The key of a node $A$ in $\F._B$ is
\begin{equation}
\label{AKeyEqn}
key(A,\,\F._B)=\min \set{SLACK(v)}{v\in V(A)}.
\end{equation}
At any point in time when $B$ is not in \os.,
the current smallest slack of an unmatched grow step
edge to $B$ is
$find\_min(\F._B ) -\Delta+\ZY(B)$.
Thus a grow step for $B$ can be done when
$\Delta=find\_min(\F._B )+\ZY(B)$.
So the key of $B$ in the global heap \F.
 is $find\_min(\F._B )+\ZY(B)$, if $B$ is not a node of \os..

For every vertex $v\in V$, $node(v)$ is  the unique  
ancestor of $v$ that is currently a node of some heap $\F._B$.
$node(v)$ is used in (\ref{AKeyEqn}) to 
maintain keys in $\F._B$ (i.e., $node(v)$ gives $A$ in (\ref{AKeyEqn})). 
$node(v)$ is also used 
in \eqref{yYZtoYeqn} to determine the current blossom $B_v$.
Specifically $node(v)$ is in the heap $\F._{B_v}$.

\paragraph*{Algorithms}
When a new outer node $B$ is created, 
every unmatched edge $uv$ ($u\in B$) is
examined. $SLACK(v)$ is decreased if appropriate. 
This may trigger a
$decrease\_key$ for $node(v)$ in $\F._{B_v}$. 
This may in turn
trigger a
$decrease\_key$ for $B_v$ in $\F.$,
if $B_v$ is currently not in \os..

When a grow step adds a blossom $B$ to \os.,
the node for $B$ in \F.  
is deleted. 
Note that whether $B$ becomes inner or outer,
it never gets reinserted in \F. in this search.
If $B$ becomes inner the value  $\Delta_0(B)$ is recorded.
If $B$ becomes outer, the values $y(v), v\in V(B)$ are required
to redefine $Y(v)$ (recall from Section \ref{bMAnalSec}).
This is done using the first alternative of \eqref{yYZtoYeqn}.
If $B$ becomes inner and later becomes outer in a blossom step,
$Y(v)$ is redefined using the second alternative of \eqref{yYZtoYeqn}.

Consider an expand step for an inner blossom $B$.
The \B.-children of $B$ (i.e., the nodes of $C(B)$)
become maximal blossoms or atomic, 
and we must update the data structure for them.
Let $B'$ be the big \B.-child  of $B$, if it exists.
For every \B.-child $A\ne B'$ of $B$, delete the node $A$ of $\F._B$.
Initialize a new F-heap $\F._A$ as follows
(modifying appropriately if $A$ is atomic):

\b

{\narrower

{\parindent=0pt

For each small component $D$ of $A$,
create a node in $\F._A$. For every $v\in V(D)$ update 
$node(v)$ to $D$.
Assign $key(D,\F._A)\gets\min\set{
SLACK(v)}{v\in V(D)}$.

}}

\b

\noindent
Let  the new heap $\F._{B'}$ be the (updated) heap $\F._B$.
Insert the \B.-children 
of $B$ that are no longer in \S. as entries in \F..
For the \B.-children that are inner nodes of \os. 
record their $\Delta_0$ value.
Process \B.-children that are outer nodes of \os. 
as above.

\bigskip

The main observation for correctness of the expand procedure is
that $\F._{B'}$ is the desired heap for $B'$. This follows since
the small components of $B'$ are those of $B$ minus the small children of $B$. 

It is easy to see the total time used in
the course of an entire search  is $O(m+n\log n)$.
When a small child $A$ becomes maximal it is charged $O(\log n)$ to account
for its
deletion from $\F._B$. For $D$ a small component of $A$,
each vertex $v\in V(A)$ 
is charged $O(1)$ for resetting
$node(v)$ and examining $SLACK(v)$. 
(The new $node(v)$ values are easily found by traversing
the subtree of $A$ in the blossom tree \B..
The traversal uses time proportional to the number of leaves,
i.e., $O(1)$ time for each vertex $v$.)
$v$ moves to a new small component
$O(\log n)$ times so this charge totals $O(n\log n)$.
Finally and most importantly, $decrease\_key$ uses $O(1)$ amortized time
in a Fibonnaci tree.

\paragraph*{$f$-factors}
Two new aspects of $f$-factors are that matched edges needn't be tight and
edges can be in $I$-sets. 
We will use some simple facts about $I$-sets.

\begin{lemma}
\label{IBconIALemma}
Consider
blossoms $A,B$ with $V(A)\con V(B)$, and edge $e\in \delta(A)\cap \delta(B)$.

\i $e=\eta(A)\iff e=\eta(B)$.

\ii $e\in I(A) \iff e\in I(B)$.
\end{lemma}

\begin{proof}
\i  Consider three cases for $A$. 

\case{$A\not\con \alpha(B)$} This makes $\eta(A)\in \gamma(B)$. 
So $e\in \delta(B)$ implies $e\ne \eta(A)$. Also
$e\in \delta(A)$ implies 
$e\ne \eta(B)$.

\case{$A= \alpha(B)$} This makes  $\eta(A)= \eta(B)$. Hence $e=\eta(A)$ iff
$e= \eta(B)$.

\case{$A\pcon \alpha(B)$} 
Edge $e$ of the hypothesis is in $\delta(A)\cap \delta(\alpha(B))$. By induction 
$e=\eta(A)\iff e=\eta(\alpha(B))$.
Since $ \eta(\alpha(B))=\eta(B)$ this implies \xi.

\b

\ii By \i there are two possibilities:

\case{$e\ne \eta(A),\eta(B)$} $e\in I(A)\iff e\in M \iff e\in I(B)$.

\case{$e=\eta(A)=\eta(B)$} $e\in I(A)\iff e\notin M \iff e\in I(B)$.
\end{proof}

Now observe
an edge $e=uv\in I(B_v)$ has
\begin{equation}
\label{zContributionEqn}
z_0\set{A}{V(A)\con V(B_v),\, e \in I(A)}=
z_0\set{A}{v\in V(A)\con V(B_v)}=2(\ZY(v)-\ZY(B_v)).
\end{equation}
The second equation is trivial and the first follows immediately part \ii
of the lemma.

The analog of the previous definition of 
$slack$ is
\begin{equation}
\label{fFactorSlackDefnEqn}
slack(v)=\min\set {|\H{yz}(uv)-w(uv)| }{uv\in E \mbox{ eligible at $u$}}.
\end{equation}
(Recall Lemma \ref{AlwaysEligible} and its terminology.)
As in Lemma \ref{fDualsPreservedLemma} 
define a sign $\sigma$ as $-1$ if $uv\in M$ else $+1$,
so any edge $uv$ has $|\H{yz}(uv)-w(uv)| =\sigma(\H{yz}(uv)-w(uv))$.

The highest level outline of the data structure is as before:
We track $slack$ by maintaining
the 
invariant \eqref{SlackEqn},
where the stored quantity $SLACK(v)$ will be defined below.
We define keys in $\F._B$ and $\F.$ exactly as before,
e.g., (\ref{AKeyEqn}). The invariant implies that
for any blossom $B$ not in \os.,
the current smallest $slack$ of a grow step
edge to $B$ is $find\_min(\F._B ) -\Delta+\ZY(B)$.
So the data structure gives the correct value for the next dual adjustment.

Our definition of $SLACK(v)$ involves two quantities
$IU(uv)$ and $IV(uv)$ 
that account for the contributions of
$I$-edges to the slack of $uv$, $IU$ at the $u$ end and $IV$ at the $v$ end.
We will define $IU$ and $IV$ to be fixed, stored quantities  
so the following relations hold.
At any time when $v\notin \S.$,
and $B_v$ is the maximal blossom/vertex currently containing $v$,
\begin{equation}
\label{vSlackInvariantEqn}
y(v)+z\set{A}{v\in V(A),\, uv \in I(A)}=Y(v)+IV(uv)+\sigma\ZY(B_v).
\end{equation}
At any time after $uv$ becomes eligible at $u$, 
\begin{equation}
\label{uSlackInvariantEqn}
y(u)+z\set{A}{u\in V(A),\, uv \in I(A)}=Y(u)+IU(uv)-\sigma\Delta.
\end{equation}
We reiterate that the only terms on the right-hand side of these
two equations that change with time are $\ZY(B_v)$ and $\Delta$.

Now define
\[SLACK(v)= \min \set{\sigma(Y(u)+Y(v)+IU(uv)+IV(uv)-w(uv))}
{uv\in E  \text{ eligible at }u}.\]
Let us show the above relations imply 
the desired invariant \eqref{SlackEqn} for $SLACK$.
Adding the two equations
and multiplying by $\sigma$ implies that 
at any point in time when $uv$ is eligible and $v\notin S$,
\begin{equation*}
\label{vNotinSEqn}
|\H{yz}(uv)-w(uv)|=
\sigma(Y(u)+IU(uv)+Y(v)+IV(uv)-w(uv))-\Delta+\ZY(B_v).
\end{equation*}
Applying this for every edge $uv$ in the definition of $SLACK$ gives
\eqref{SlackEqn} as desired.

It remains to give $IV$ and $IU$.
The contribution at the nonouter end $v$ is
defined by \[IV(uv)=\begin{cases}
0&uv\notin M\cup\eta(B_v)\\ 
2\ZY(v)&uv\in M-\eta(B_v)\\ 
2\ZY(B_v)&uv=\eta(B_v)\in M\\
2(\ZY(v)-\ZY(B_v))&uv=\eta(B_v)\notin M.
\end{cases}
\]

To discuss this definition we will use  the following terminology.
Recall that the algorithm computes $IV(uv)$ when $uv$ becomes eligible at $u$.
$IV(uv)$ is defined using the blossom/vertex $B_v$ at that time.
However we must verify
\eqref{vSlackInvariantEqn} whenever  $v\notin \S.$, so $B_v$ may change.
To keep the two cases straight say the {\em defining $B_v$}
is used to compute $IV(uv)$, and a {\em useful $B_v$} is one that
may be required later on in \eqref{vSlackInvariantEqn} 
to establish the invariant \eqref{SlackEqn}
for the algorithm. The defining $B_v$ is useful iff 
 $v\notin \S.$ when $IV(uv)$ is computed.
Clearly a useful $B_v$ is a subset of the defining $B_v$, but we shall see that
not every such $B_v$ is useful.

To
prove the definition is correct we will analyze each of its four cases
separately.
We will show
that if the defining $B_v$ is in that case, so is every useful $B_v$.
Then we will show 
 \eqref{vSlackInvariantEqn} is satisfied for every useful
$B_v$. 
To do this we will compute
the value of the left-hand side of \eqref{vSlackInvariantEqn}
and deduce the correct value of $IV(uv)$ by comparing to the right-hand
side.

To begin the analysis,
note that whenever $v\notin \S.$  the current value of $y(v)$ is 
\[Y(v)+\ZY(B_v)\]
since every dual adjustment increases $y(v)$ by $\delta$.
Also when $uv\in I(B_v)$ the $z$ contribution to  the left-hand side of
\eqref{vSlackInvariantEqn} is 
\[z_0\set{A}{v\in V(A)\con V(B_v)}=2(\ZY(v)-\ZY(B_v)),\]
by \eqref{zContributionEqn}.


\case{$uv\notin M\cup\eta(B_v)$}
We assume this case holds for the defining $B_v$.
So for any useful $B_v$, say $B$, $uv$ is an unmatched edge
and $uv\ne \eta(B)$
(by Lemma \ref{IBconIALemma}\xi). So this case holds for every useful $B$.

Now we establish \eqref{vSlackInvariantEqn} for any useful $B_v$.
The contribution to the left-hand side of \eqref{vSlackInvariantEqn}
 is $y(v)=Y(v)+\sigma\ZY(B_v)$.  This follows since this case has
$uv\notin I(B_v)$ (so there is no $z$ contribution) and
$\sigma=1$ (since $uv\notin M$).  
Comparing to the right-hand side of
\eqref{vSlackInvariantEqn} shows
$IV(uv)=0$, as desired.

\case{$uv\in M-\eta(B_v)$}
We assume this holds for the defining $B_v$.
So any useful $B_v$ has $uv$ matched and not its base edge
(by Lemma \ref{IBconIALemma}\xi).
Thus this case holds for any useful $B_v$.

Now consider any useful $B_v$.
If $B_v$ is a blossom then $uv\in I(B_v)$. So the $z$ contribution is
$2(\ZY(v)-\ZY(B_v))$. This also holds if $B_v$ is atomic, since
the $z$ contribution is 0. 
Since $uv\in M$,  $\sigma=-1$.
Adding the $y$ and $z$ 
contributions  to the left-hand side of \eqref{vSlackInvariantEqn}
gives total contribution
\[(Y(v)+\ZY(B_v))+2(\ZY(v)-\ZY(B_v))=
Y(v)+2\ZY(v) +\sigma\ZY(B_v).\]
Thus $IV(uv)=2\ZY(v)$, independent of $B_v$.

\b

The next two cases have $uv=\eta(B_v)$ for the defining $B_v$.
If $v\in \S.$ at this point then wlog $B_v$ is inner. 
Since $v=\beta(B_v)$, $v$  will remain in \S. for the rest of the search.
So $uv$ is irrelevant to the data structure.  If $v\notin \S.$ then
$B_v$ is itself the first useful $B_v$. The first time this $B_v$ becomes a node
of  \os., the preceding argument applies. It shows there are no other useful
$B_v$'s. In summary we have shown for the next two cases, every useful $B_v$
belongs to the same case.

\case{$uv=\eta(B_v)\in M$}
Since $uv\notin I(B_v)$ there is no $z$ contribution
(by Lemma \ref{IBconIALemma}\xii). So 
the total contribution is
$y(v)=Y(v)+\ZY(B_v)=Y(v)+2\ZY(B_v) +\sigma\ZY(B_v)$.
Thus $IV(uv)=
2\ZY(B_v)$.

\case{$uv=\eta(B_v)\notin M$}
This makes $uv\in I(B_v)$ so there is a $z$ contribution.
The total contribution is 
\[(Y(v)+\ZY(B_v))+2(\ZY(v)-\ZY(B_v))=
Y(v)+2(\ZY(v)-\ZY(B_v))+\sigma\ZY(B_v).\]
Thus $IV(uv)=2(\ZY(v)-\ZY(B_v))$.

\bigskip

\begin{figure}[t]
\centering
\input{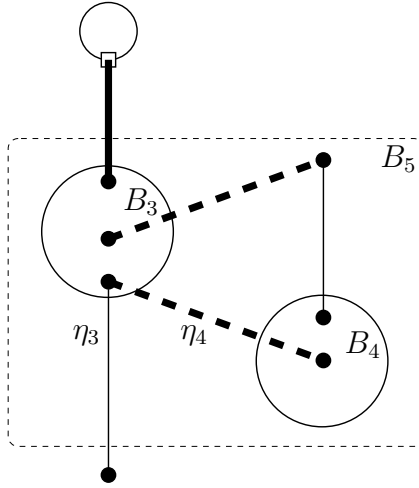}
 \caption{Precursor to structure of Fig.\ref{fBlossomFig}.}
 \label{fBlossomFigB4}
 \end{figure}

\remark {It might seem that the cases for $uv=\eta(B_v)$
are subject to a simplification because this edge is often tight. Specifically
if $B_v$ was not a maximal blossom at the beginning of the 
current search then $\eta(B_v)$
is tight when the search starts. So $\eta(B_v)$ will be tight 
when $B_v$ becomes maximal.
However this need not be the case when $\eta(B_v)$ becomes eligible.
For instance 
suppose a search starts out with the structure of
Fig.\ref{fBlossomFigB4}.
Then the inner blossom $B_5$ gets expanded
to give part of 
Fig.\ref{fBlossomFig}, where $\alpha_2=\eta_4=\eta(B_4)$. 
As mentioned 
(in the Examples section after Fig.\ref{fAlgExFig})
a dual adjustment makes $\alpha_2$ strictly underrated.
A subsequent expansion of $B_3$ may make $\alpha_2$ eligible, but still underrated.}

\b

The contribution at the \os. end $u$ is
\[IU(uv)=\begin{cases}
\ZY(B_u)-\Delta_0(B_u)&B_u \text{ inner}, uv\in M\\
2\ZY(u)-\ZY(B_u)+\Delta_0(B_u)&B_u \text{ inner}, uv=\eta(B_u)\notin M\\
0&B_u \text{ outer}, uv\notin M\\
2\big(\ZY(u)-\ZY(B_u)-2\Delta_0(O_u)+\Delta_0(B_u)\big)&B_u \text{ outer}, uv\in M.
\end{cases}
\]
$O_u$ is defined below.

To verify correctness
let $\Delta_0$ be the value of $\Delta$ when $uv$ first becomes eligible
for (any) $B_u$. We will show \eqref{uSlackInvariantEqn} holds at that point.
Thereafter, $uv$ remains eligible (Lemma \ref{AlwaysEligible}), so
\eqref{DdeMeqn} shows
the left-hand side of \eqref{uSlackInvariantEqn}
changes by $-\sigma\delta$ in every dual adjustment,
as does the right-hand side. Thus \eqref{uSlackInvariantEqn}
continues to hold in every dual adjustment.

\case{$B_u \text{ inner}, uv\in M$}
This makes $uv\notin I(B_u)$. 
(There are two cases: If $B_u$ is a blossom
then $uv=\eta(B_u)$ since $uv$ is eligible.
If $B_u$ is atomic then $I(B_u)=\emptyset$.)
Thus the contribution is
\[y(u)=Y(u)+\ZY(B_u)              
=Y(u)+ \ZY(B_u)    -\Delta_0(B_u)-\sigma\Delta_0.\]
Thus $IU(uv)=\ZY(B_u)-\Delta_0(B_u)$.

\case{$B_u \text{ inner}, uv=\eta(B_u)\notin M$}
This makes $B_u$ a blossom and $uv\in I(B_u)$. 
The contribution for $y(u)$ is the same as the previous case. 
The contribution for $z$ is $2( \ZY(u)-\ZY(B_u))$.
The total contribution is
$(Y(u)+\ZY(B_u))+2(\ZY(u)-\ZY(B_u))
= Y(u)+2\ZY(u)-\ZY(B_u)+\Delta_0(B_u)
-\sigma\Delta_0$.
Thus $IU(uv)=2\ZY(u)-\ZY(B_u)+\Delta_0(B_u)$.

\b

We are left with the case
where $uv$ first becomes eligible
when $u$ enters an outer node.
Furthermore
$uv\ne \eta(B_u)$ when $B_u$ is a blossom.
To prove the latter,
the preceding two cases apply
if blossom $B_u$ enters \os. as inner.
If $B_u$ enters as outer clearly $\eta(B_u)=\tau(B_u)\in \os.$.

Let $O_u$ be the first  outer node that contains $B_u$.
Let $\Delta_0(O_u)$ be the value of $\Delta$ when $O_u$ is formed.
So $\Delta_0=\Delta_0(O_u)$.
Recall that when $O_u$ is formed we redefine
$Y(u)$ to be the current value of $y(u)$ plus $\Delta_0(O_u)$. 
Hence at any time after $O_u$ is formed $B_u$ is outer and
\[y(u)=Y(u)-\Delta.\]
 Also the only $z$ contribution comes from $B_u$ (since
we assume $\Delta=\Delta_0$).

\case{$uv$ becomes eligible for $O_u$,  $uv\notin M$}
There is no $z$ contribution. (This is by definition if $B_u$ is atomic.
If $B_u$ is a blossom we have noted $uv\ne \eta(B_u)$.)
So  the total contribution is
$y(u)=Y(u)-\Delta_0(O_u)= Y(u)-\sigma\Delta_0$.
Thus $IU(uv)=0$.

\case{$uv$ becomes eligible for $O_u$,  $uv\in M$}
First suppose $B_u$ is a blossom.
This case makes $uv\in I(B_u)$. 
When $B_u$ becomes an \os.-node
(outer or inner) the $z$ contribution is 
\[z_0\set{A}{u\in V(A)\con V(B_u)}.\]
If $B_u$ enters as an inner node 
and is later absorbed in  an outer node, 
this $z$ contribution decreases by
\[2(\Delta_0(O_u)-\Delta_0(B_u)).\]
 This also holds
if $B_u$ enters as outer. (The latter may occur 
in a grow step that adds $B_u=O_u$, or in an expand step
that makes $B_u$ maximal and outer.)

It is possible that $B_u$ is an atom.
We must have $B_u$ outer, by the first case.
An atom has no $z$ contribution. This is consistent with the
two displayed $z$ contributions, since they are both 0 for an atom $B_u$
($B_u=O_u$).
 
So in all cases, the left-hand side of \eqref{uSlackInvariantEqn}
is 
\begin{eqnarray*}
&&(Y(u)-\Delta_0(O_u)) +
2( \ZY(u)-\ZY(B_u)-(\Delta_0(O_u)-\Delta_0(B_u)) )\\
&=&Y(u) +2(\ZY(u)-\ZY(B_u)-2\Delta_0(O_u)+\Delta_0(B_u))
-\sigma\Delta_0.
\end{eqnarray*}
Thus 
$IU(uv)
=2(\ZY(u)-\ZY(B_u)-2\Delta_0(O_u)+\Delta_0(B_u))
$.

\b

The only changes to the algorithm are the obvious ones for examining edges:
Matched edges must be examined and added to the data structure. $IU$ and
$IV$ quantities must be computed.
It is easy to see the latter uses $O(1)$ time per edge. So the timing estimate is not
affected.


\fi 
\end{document}